\documentclass[rmp,aps,reprint,superscriptaddress,floatfix,nofootinbib,showkeys,showpacs,nolongbibliography,author-year]{revtex4-2}

\pdfoutput=1

\usepackage{tabularx}
\usepackage{xspace}
\usepackage{listings}
\usepackage{lineno}

\usepackage[dvipsnames,svgnames]{xcolor} 
\usepackage{graphicx}
\usepackage{comment}

\usepackage{hyperref}
\usepackage{amsmath}
\usepackage{enumitem}
\setlist{wide, labelwidth=!, labelindent=0pt}
\usepackage{multirow}
\hypersetup{    
  colorlinks      = true,
  linkcolor       = {blue},
  linkbordercolor = {white},
  citecolor       = {blue},
  citebordercolor = {white},
  urlcolor        = {blue},
  urlbordercolor  = {white},
}

\usepackage{lipsum}
\usepackage{aas_macros}
\usepackage{orcidlink}

\usepackage{amsmath}
\usepackage{amssymb}
\usepackage{xspace}
\usepackage{xifthen}
\usepackage{eso-pic}

\definecolor{forestgreen}{HTML}{228B22}
\definecolor{urlblue}{HTML}{000000}

\newcommand{\NEW}[1]{{#1}} 

\mathchardef\mhyphen="2D

\newlength{\dhatheight}

\newcommand{\var}[1]{%
 \ensuremath{\texttt{\MakeUppercase{#1}}}\xspace
}
\newcommand{\bandvar}[2][]{%
  \ifthenelse{\isempty{#1}}{\var{#2}}{\var{#2\_#1}}%
}

\newcommand{\LCDM}{\ensuremath{\rm \Lambda CDM}\xspace}

\providecommand\physrep{\ref@jnl{Phys.~Rep.}}%
\providecommand\apjs{\ref@jnl{ApJS}}%
\providecommand{\jcap}{\ref@jnl{JCAP}}%

\begin{document}

\title{Dark Matter Constraints from Small-Scale Cosmic Structure}

\author{Ethan~O.~Nadler\orcidlink{0000-0002-1182-3825}}
\email[]{enadler@ucsd.edu}
\affiliation{Department of Astronomy \& Astrophysics, University of California, San Diego, La Jolla, California 92093, USA}

\author{Keir~K.~Rogers\orcidlink{0000-0003-1601-8144}}
\email[]{keir.rogers@kcl.ac.uk}
\affiliation{Department of Physics, King’s College London, Strand, London WC2R 2LS, UK}
\affiliation{Department of Physics, Imperial College London, Blackett Laboratory, Prince Consort Road, London, SW7 2AZ, UK}

\author{Alex~Drlica-Wagner\orcidlink{0000-0001-8251-933X}}
\email[]{kadrlica@fnal.gov}
\affiliation{Fermi National Accelerator Laboratory, P.O. Box 500, Batavia, IL 60510, USA}
\affiliation{Department of Astronomy \& Astrophysics, University of Chicago, 5640 S Ellis Avenue, Chicago, IL 60637, USA}
\affiliation{Kavli Institute for Cosmological Physics, University of Chicago, Chicago, IL 60637, USA}
\affiliation{NSF-Simons AI Institute for the Sky (SkAI),172 E. Chestnut St., Chicago, IL 60611, USA}

\begin{abstract}
Small-scale cosmic structure provides a powerful test of the fundamental nature of dark matter. A wide range of dark matter models impact matter clustering on small scales, including warm, fuzzy, and (self-)interacting dark matter. In these scenarios, dark matter physics such as free-streaming, wave interference, and self/Standard Model interactions alter the abundance and internal structure of dark matter halos. Cosmological and astrophysical probes of small-scale structure---including dwarf galaxies, strong lensing, the Lyman-$\alpha$ forest, stellar streams, and high-redshift galaxies---are therefore sensitive to these effects. Here, we review dark matter constraints provided by small-scale structure, focusing on observables that probe scales smaller than $\sim 1~\mathrm{Mpc}$, which define the frontier of current measurements. We summarize how these constraints have been translated to limits on microphysical dark matter models, and we discuss key modeling uncertainties and observational systematics. Finally, we highlight the growing importance of probe combination and simulation-based inference for this field, and we overview upcoming observational facilities that will sharpen small-scale structure tests of dark matter physics.
\keywords{Cosmology, Dark matter, Particle astrophysics, Particle dark matter}
\end{abstract}

\maketitle

\tableofcontents

\section{Introduction}
\label{sec:intro}

It has been nearly a century since \citet{Zwicky:1933gu} referred to ``dark matter'' (DM) in the Coma cluster. Since then, astrophysical and cosmological data have provided overwhelming evidence that most of matter in the universe is invisible and non-baryonic (see \citealt{Bertone:2016nfn} for a historical overview). This evidence includes observations of galaxy rotation curves~\citep{Rubin:1970zza}, the large-scale structure of the universe (LSS; \citealt{Blumenthal:1984bp}), and the cosmic microwave background (CMB; \citealt{WMAP:2003elm}), among many other cosmic probes of DM.

Despite this progress, the nature of DM remains unknown. In the standard $\Lambda$CDM cosmological model, cold, collisionless dark matter (CDM) makes up $\approx 27\%$ of the cosmic energy budget but has unspecified microphysical properties. Critically, determining the fundamental properties of DM is guaranteed to reveal physics beyond $\Lambda$CDM and, likely, beyond the Standard Model (SM) of particle physics (e.g., see \citealt{Bertone:2004pz} for an overview). \NEW{Since} cosmic probes provide the only positive, \NEW{definitive} detection of DM to date, these observables represent a compelling avenue to study DM microphysics. 

Studies of the formation and evolution of cosmic structure have played a key role in the development of the CDM paradigm (e.g., see \citealt{Frenk:2012ph} for a review). For example, early work showed that hot dark matter (HDM) is incompatible with the large-scale clustering of galaxies (e.g., \citealt{White:1983fcs}). These results ruled out the hypothesis that DM is composed of standard neutrinos (e.g., \citealt{Doroshkevich:1980vy}) and bolstered interest in CDM. In parallel, theoretical efforts were undertaken to evade these constraints, resulting in the development of models like sterile neutrino DM~\citep{Dodelson:1993je,Shi:1998km}.

Modern LSS surveys have further strengthened our understanding of DM. In particular, a wide variety of DM particle properties---including its decay lifetime, SM interactions, and primordial velocity distribution---are now tightly constrained (e.g., \citealt{Schneider11120330,Wang:2012eka,Cyr-Racine:2013fsa,Kunz:2016yqy,DES:2020mpv,Bartlett:2022ztj,Gluscevic:2017ywp,Boddy:2018kfv}). These results are broadly consistent with CDM predictions, although certain tensions in LSS data may be resolved by new DM physics (e.g., \citealt{Amon:2022azi,He:2023dbn,He:2025npy,Buen-Abad:2025bgd,Rogers:2023ezo}).

Deviations from CDM are generally allowed to be largest on smaller length scales, which probe DM microphysics~\citep{Buckley171206615}. \emph{Small-scale cosmic structure}, defined here as structure on comoving scales that are nonlinear today (see Section~\ref{sec:small-scale}), is thus a key test of DM microphysics. In recent decades, measurements of the smallest observable cosmic structures deep into the nonlinear regime have been the arena for several ``challenges'' to CDM (e.g., see \citealt{Bullock:2017xww} and \citet{Tulin:2017ara} for reviews). In parallel, model-building efforts have broadened beyond the weakly-interacting-massive-particle (WIMP) paradigm given null results from terrestrial experiments, and cosmological predictions now exist for many DM models (e.g., \citealt{Marsh:2015xka,Abazajian:2017tcc}).

There have recently been major strides in testing DM models using probes of cosmic structure on the smallest accessible scales. On one hand, more robust theoretical predictions have been developed, including growing bodies of cosmological simulations that self-consistently compare multiple models beyond CDM (e.g., \citealt{ForouharMoreno2022GalacticSatelliteSystems,Stucker:2021vyx,Rose:2024xcb,Nadler:2025fcv,Despali:2025koj,Shen:2023lsf}), semi-analytic techniques that target small-scale structure (e.g., \citealt{Lacey2016UnifiedMultiwavelength,Benson:2010kx,Jiang:2020rdj,Kravtsov:2022}), and models for the interplay between baryonic and DM physics on small scales (see \citealt{Sales:2022ich} for a review). On the other hand, observations of the Lyman-$\alpha$ forest~\citep{Weinberg:2003eg}, faint dwarf galaxies~\citep{Simon190105465}, stellar streams~\citep{Bonaca:2024dgc}, strong gravitational lenses~\citep{Vegetti:2023mgp}, and the high-redshift universe~\citep{Ellis:2007ds} have become increasingly sensitive to structure on smaller scales.

Thus, the smallest cosmic structures now provide stringent constraints on many fundamental DM properties. These limits have been achieved using a diverse set of observables (spanning high-redshift probes to measurements of the very local universe) and modeling techniques (spanning full cosmological hydrodynamic simulations, SAMs, and effective theories). In preparation for the rapid influx of small-scale structure data expected from new facilities in the coming years (e.g., \citealt{Chakrabarti:2022cbu}), it is timely to summarize the status of DM constraints from small-scale structure.

This review focuses on constraints from \emph{frontier probes}, i.e., observables that statistically constrain DM properties on the smallest accessible scales. Within this space, we summarize constraints derived from: \begin{enumerate}
    \item \emph{Established probes}---dwarf galaxies, strong lensing, and the Lyman-$\alpha$ forest---which are relatively mature and currently drive the tightest limits on DM models;
    \item \emph{Emerging probes}---stellar streams, high-redshift galaxies, and 21-cm cosmology---which are rapidly gaining sensitivity and theoretical interest. 
\end{enumerate} 
We also briefly discuss how \emph{precision probes} of the matter distribution on larger (but still nonlinear) scales, such as weak lensing, complement these efforts, but we do not comprehensively summarize these results.

Given the breadth of the field, this review focuses on constraints that apply to microphysical DM properties (free-streaming, wave interference, non-gravitational interactions, etc.) set by the statistical properties of the smallest observable cosmic structures. We also discuss how these constraints have been translated to limits on specific particle models of DM where applicable. Note that we do not cover constraints from CMB spectral distortions, stellar heating in dwarf galaxies, or other probes of minihalos or DM/inflationary scenarios with extreme primordial power spectrum enhancement; these are covered in other reviews (e.g., \citealt{Chluba:2015bqa,Allahverdi:2020bys}). We also do not cover direct or indirect detection constraints \citep{Gaskins:2016cha}, primordial black hole DM~\citep{Carr:2021bzv}, or composite DM scenarios~\citep{Cline:2021itd}. Our goal is to complement these efforts by providing a unified treatment of the probes most sensitive to DM microphysics through its effects on the statistics of small-scale structure.

Our review suggests that DM constraints from small-scale structure are poised to enter a new phase. Historically, small-scale structure observables have largely been studied in isolation, with each probe providing independent constraints on DM physics. However, different probes ultimately measure aspects of the same underlying matter distribution and are governed by a common set of physical processes. As a result, the most powerful and robust constraints are expected to arise from joint analyses across observational probes. To achieve such constraints, it is important to develop a unified modeling framework in which multiple small-scale structure observables are statistically combined within a common modeling framework. We discuss how such a framework may be enabled through advances in simulation, machine-learning, and semi-analytic modeling techniques; this effort represents a central challenge and opportunity for the field over the coming decade.

This review is organized as follows. In Section~\ref{sec:small-scale}, we define ``small-scale structure'' and argue that upcoming observations will begin an era of high-precision cosmology at the small-scale frontier, analogous to the era of ``precision cosmology'' that has arisen from observations of the CMB and LSS in recent decades. In Section~\ref{sec:scenarios}, we discuss the classes of DM models covered in this review, including warm, interacting, self-interacting, decaying, and fuzzy DM. In Section~\ref{sec:constraints}, we summarize DM constraints from established probes of small-scale structure: dwarf galaxies, strong gravitational lensing, and the Lyman-$\alpha$ forest; in Section~\ref{sec:emerging}, we summarize DM constraints from emerging probes: stellar streams, high-redshift galaxies, weak lensing, and 21-cm cosmology. In Section~\ref{sec:combination_constraints}, we summarize early results from small-scale probe combination. We discuss the theoretical and observational outlook for DM constraints from small-scale structure in Section~\ref{sec:outlook}, and we conclude in Section~\ref{sec:conclusion}.

\NEW{Readers interested in the key small-scale structure predictions of different DM scenarios should refer to the paragraphs at the end of each subsection in Section~\ref{sec:scenarios}. The summaries in Sections~\ref{sec:constraints}, \ref{sec:emerging}, and \ref{sec:current_status} contain high-level takeaways regarding DM constraints split by small-scale structure probe and DM scenario, including systematics and important directions for future work. The main body of Sections~\ref{sec:constraints} and \ref{sec:emerging} is intended as a comprehensive summary of the recent literature.}

Throughout, we adopt Planck 2018 cosmological parameters of $\Omega_{\mathrm{dm}} h^2 = 0.1207$, $\Omega_{\mathrm{b}} h^2 = 0.02237$, $n_s = 0.96$, $h=0.6736$, and $\sigma_8 = 0.81$~\citep{Planck:2018vyg}. However, many of the constraints we summarize adopt different cosmological parameters; these differences should be accounted for when directly comparing literature results.

\section{The Small-Scale Structure Frontier}
\label{sec:small-scale}

\subsection{Defining ``Small-Scale Structure''}

We define ``small-scale structure'' based on the standard formulation of the matter power spectrum and the growth of structure in the universe.
The dimensionless matter power spectrum is given by
\begin{equation}
\Delta^2(k,z)\equiv k^3P(k,z)/2\pi^2,
\end{equation}
where $k$ is the wavenumber and $P(k,z)$ is the linear matter power spectrum, extrapolated from early times to redshift $z$ using the linear growth factor. Note that $\Delta^2(k,z)$ represents the contribution to the variance of density fluctuations per logarithmic wavenumber interval $\mathrm{d}\ln k$.

The nonlinear wavenumber $k_{\mathrm{nl}}(z)$ is defined by the scale at which dimensionless density fluctuations are order unity in CDM~\cite{Cabass:2022avo},
\begin{equation}
    \Delta^2(k_{\mathrm{nl}}(z),z)\equiv 1.\label{eq:knl_def}
\end{equation}
Near $z=0$, this yields
\begin{equation}
    k_{\mathrm{nl}}(z)\approx 0.17~\mathrm{Mpc}^{-1}\times\left[\frac{\sigma_8}{0.81}\,\frac{D(z)}{D(0)}\right]^{-2},\label{eq:k_nl}
\end{equation}
where $D(z)$ is the linear growth factor~\citep{Carroll:2000fy}. Equation~\ref{eq:k_nl} is accurate at the $\approx 10\%$ level for $z\lesssim 1$. At higher redshifts, $\Delta^2(k)$ flattens and $k_{\mathrm{nl}}$ rises much more steeply, reaching $\approx 1.6$, $5$, and $18~\mathrm{Mpc}^{-1}$ at $z=2$, $3$, and $4$; we solve Equation~\ref{eq:knl_def} numerically when plotting $k_{\mathrm{nl}}(z)$. Because $\Delta^2(k)$ grows only logarithmically at high $k$, $k_{\mathrm{nl}}$ exceeds $10^3~\mathrm{Mpc}^{-1}$ by $z\approx 7$.

\NEW{Throughout, we define ``small-scale structure'' as structure on scales that are nonlinear today:}
\begin{equation}
    {k>k_{\mathrm{nl}}(z=0)\approx 0.17~\mathrm{Mpc}^{-1},}\label{eq:small_scale}
\end{equation}
\NEW{i.e., scales smaller than $\lambda_{\mathrm{nl}}(z=0)=2\pi/k_{\mathrm{nl}}(z=0)\approx 40~\mathrm{Mpc}$. At high redshift, part of this range is still linear as discussed above. High-redshift probes (the Lyman-$\alpha$ forest, high-$z$ galaxies, and 21-cm cosmology) therefore access small-scale structure while it is still (quasi-)linear, which makes them relatively clean tests of \emph{ab initio} DM physics. Collapsed halos nonetheless exist at these times because they form from rare, high peaks of the linear density field. For example, the $10^9$--$10^{11}~M_\odot$ halos that host galaxies at $z\approx 6$--$10$ are $\nu\equiv\delta_c/\sigma(M,z)\approx 2$--$5$ peaks, so their abundance is exponentially sensitive to $P(k)$.}

Frontier probes, which are the focus of this review, typically measure scales of $\lambda\lesssim 1~\mathrm{Mpc}$, or $k\gtrsim 6~\mathrm{Mpc}^{-1}$. Note that substructure within DM halos is manifestly nonlinear, since even the largest halos that exist today have sizes that are much smaller than $\lambda_{\mathrm{nl}}(z=0)$.

\begin{figure*}[t!]
\centering
\includegraphics[trim={0cm 0.25cm 0 0},width=\textwidth]{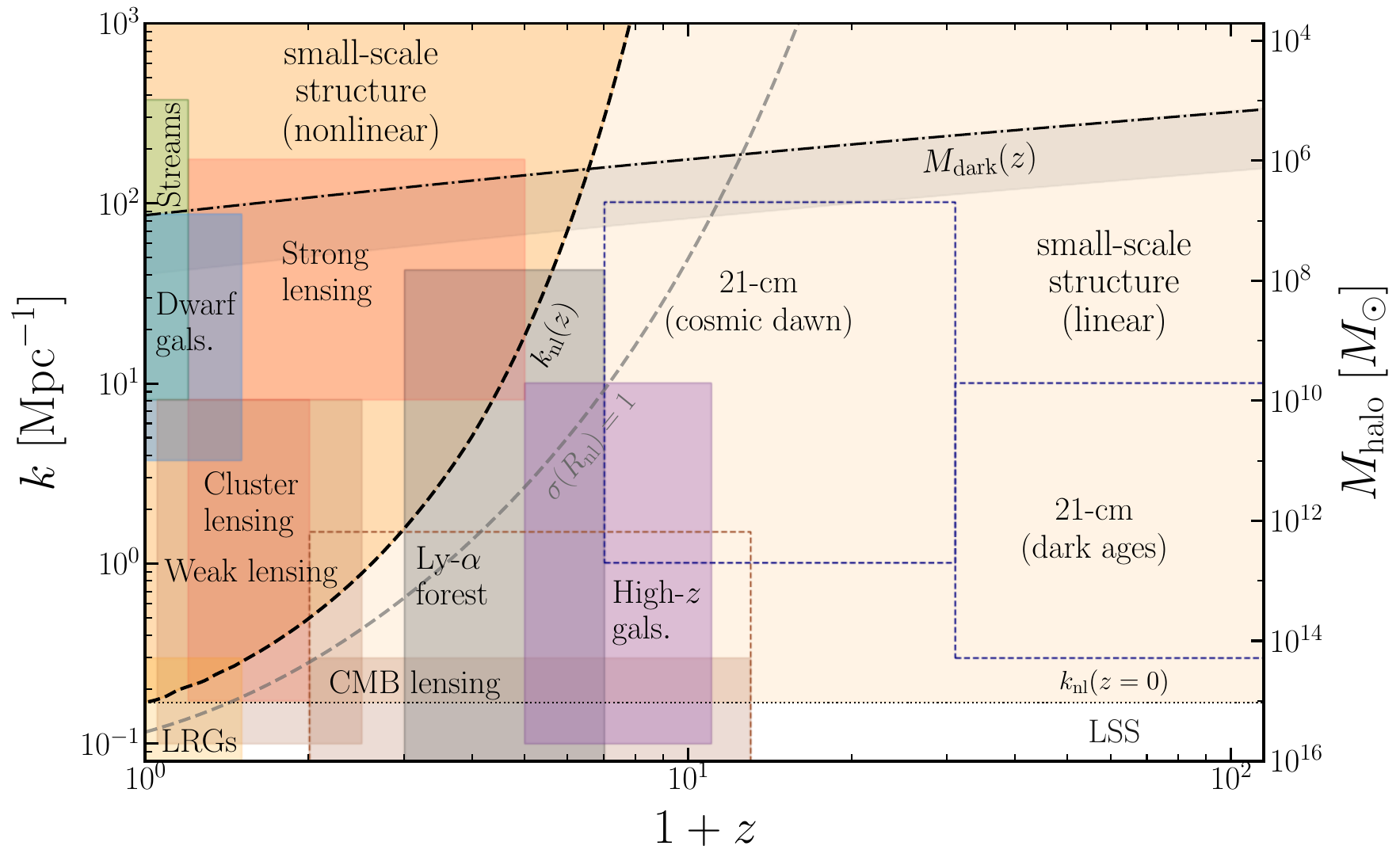}
    \caption{Wavenumber and mass scales probed by small--scale cosmic structure observables as a function of redshift. Filled boxes show probes with current data, and unfilled boxes show future datasets. \NEW{The gold shaded region shows small-scale structure, defined as comoving scales that are nonlinear today, $k>k_{\mathrm{nl}}(z=0)\approx 0.17~\mathrm{Mpc}^{-1}$ (thin dotted line; Equation~\ref{eq:small_scale}). The thick black dashed line shows the redshift-dependent nonlinear wavenumber $k_{\mathrm{nl}}(z)$, obtained by solving Equation~\ref{eq:knl_def} with the linear CDM power spectrum; small scales are nonlinear to the left of this line (dark gold) and linear to the right of this line (light gold). The gray dashed line shows an alternative definition of the nonlinear scale, $R_{\mathrm{nl}}$, at which the linear variance is unity.} The dot-dashed black line, $M_{\rm dark}(z)$, shows a conservative lower limit on the \NEW{peak mass of the smallest} halo to host a luminous galaxy, and the gray shaded region indicates the theoretical uncertainty on this lower limit~\citep{Benitez-Llambay:2020zbo,Nadler:2025mnz}. Wavenumbers are associated with peak halo masses via Equation~\ref{eq:m_k}. Stellar streams and dwarf galaxies are slightly extended in $z$ for visual clarity. This review focuses on DM constraints from measurements that probe $\lambda\lesssim 1~\mathrm{Mpc}$, or $k\gtrsim 6~\mathrm{Mpc}^{-1}$. Adapted from \citet{Sabti:2021unj} and \citet{Boddy:2022knd}, with the addition of LRGs~\citep{Reid:2009xm}.}
    \label{fig:k_z}
\end{figure*}

Many of the small-scale structure probes we review trace the abundance and internal structure of DM halos, which provide a crucial link between models, simulations, and observables. We relate peak halo mass and wavenumber through \citep[e.g.,][]{Nadler:2025fcv}:
\begin{align}
    M_{\mathrm{halo}}(k) &\equiv \frac{4\pi}{3}\Omega_m\rho_{\mathrm{crit},0}\left(\frac{\pi}{k}\right)^3 \nonumber \\
    &\approx 5.2\times 10^9~M_{\mathrm{\odot}}\times\left(\frac{\Omega_m h^2}{0.143}\right)\times \left(\frac{k}{10~\mathrm{Mpc}^{-1}}\right)^{-3},
    \label{eq:m_k}
\end{align}
where $M_{\mathrm{halo}}(k)$ is the mass enclosed within a volume $4\pi/3 (\pi/k)^3$ and $\Omega_m\rho_{\mathrm{crit},0}$ is the mean matter density of the universe at $z = 0$; here, $\Omega_m = \Omega_{\mathrm{DM}}+\Omega_b$ is the total matter density in units of the critical density. Thus, Equation~\ref{eq:m_k} gives a characteristic mass scale for isolated halos at $z=0$, \NEW{and for subhalos before infall}. \NEW{We refer to these as ``peak'' masses, which we distinguish from the ``stripped'' masses of subhalos after accretion. As discussed below, peak masses are relevant for galaxy--halo connection modeling and probed by dwarf galaxy abundance measurements, while stripped masses are probed by dynamical tracers (e.g., stars in dwarf galaxies or stellar streams) or gravitational lensing; we refer the reader to \citet{Zavala:2019gpq} for a more general review of DM halo properties.} Figure~\ref{fig:k_z} shows the wavenumber and mass scales probed by the small-scale structure observables  we will discuss as a function of redshift.

\subsection{Toward Precision Cosmology on Small Scales}

The era of ``precision cosmology'' emerged over the last three decades as a result of two core advances: (1) cosmological models that made well-defined predictions, and (2) experimental measurements that were sufficiently precise to distinguish between these models.
On the modeling side, a major milestone was the construction of the \LCDM model.
Observationally, the precision era was enabled by measurements of the CMB~\citep{Hu:2001bc}, Type Ia supernovae~\citep{Filippenko:2003ta}, galaxy clustering~\citep{Reid:2009xm}, and baryon acoustic oscillations (BAOs; \citealt{SDSS:2005xqv}). 
Precision measurements have also come from LSS, including galaxy clusters~\citep{Allen:2011zs}, weak lensing~\citep{Kilo-DegreeSurvey:2023gfr}, and cross-correlations between probes and wavelengths~\citep{DES:2026fyc}. 
At the frontier of precision cosmology today is the use of higher-order, non-Gaussian information in full-shape and field-level inferences of the galaxy power spectrum (see \citealt{Leclercq:2025ywu} for a review).

Cosmological probes on small spatial scales have lagged their large-scale counterparts for two main reasons:
(1) making quantitative model predictions on small, nonlinear scales is significantly more challenging and computationally expensive, and
(2) observations are statistically limited by the relatively small number of systems that are accessible due either to the rarity of their occurrence (e.g., quadruply-lensed quasars; \citealt{Oguri:2010ns}) or their intrinsic faintness (e.g., ultra-faint dwarf galaxies; \citealt{Simon190105465}).
However, major technological advances in nonlinear modeling, computing, and observational sensitivity are now beginning to enable precision measurements on small scales.

\NEW{The numerical tools connecting primordial DM physics to $P(k)$ and thereby setting the initial conditions for structure formation are well validated in CDM (e.g., \citealt{Lesgourgues:2011rh,Lewis:2026mif}) and are becoming increasingly accurate for non-CDM models~\citep{Kamionkowski:2021njk,Ji:2022iji,Vogel:2022odl,Lee:2025zym,Lee:2025vgv}}. Meanwhile, large suites of cosmological simulations, which model the nonlinear physics that shapes small-scale structure, are being developed (see \citealt{Banerjee:2022qcb} for a review). On the observational side, astronomical surveys are discovering the rare and/or faint objects needed to map out the DM distribution on small scales, while follow-up observations using large telescopes and specialized instruments are providing measurements that probe DM physics beyond gravity (see \citealt{Chakrabarti:2022cbu} for a review).

Thus, the study of small-scale structure is entering a regime that is no longer restricted to the study of a handful of extreme systems, but is instead focused on statistically rigorous population-level analyses. 
At the same time, the field faces an exciting new observational frontier: sensitivity to completely dark halos with $M_{\rm halo} \lesssim 10^7~M_{\mathrm{\odot}}$.  Table~\ref{tab:probes} lists observational probes of small-scale structure along with their corresponding redshifts, wavenumbers, and \NEW{peak halo mass ranges}.\footnote{Cluster cosmology, although it probes nonlinear scales, is primarily sensitive to DM models through the profiles of high-mass halos and is covered by dedicated reviews \citep[e.g.,][]{Allen:2011zs}; we briefly discuss cluster-scale SIDM constraints in Section~\ref{sec:sidm}.}

\begin{table*}[t!]
\caption{Observational probes of small-scale structure (first column) and associated redshift (second column), wavenumber (third column), and \NEW{peak halo mass} (fourth column) ranges. Wavenumbers and halo masses are related via Equation~\ref{eq:m_k}; however, note that certain observables (e.g., the Lyman-$\alpha$ forest) do not probe collapsed objects. \NEW{Small-scale structure corresponds to $k>k_{\mathrm{nl}}(z=0)\approx 0.17~\mathrm{Mpc}^{-1}$ (Equation~\ref{eq:small_scale}); at high redshift, some of these scales are still linear (Figure~\ref{fig:k_z}).}
\label{tab:probes}}
\begin{ruledtabular}
\begin{tabular}{lccc}
Probe & Redshift range & Wavenumber range [$\mathrm{Mpc}^{-1}$] & \NEW{Peak halo mass range} [$M_\odot$] \\
\hline
\multicolumn{4}{l}{\emph{Current data (precision probes)}} \\[2pt]
Galaxy clustering (LRGs)             & $0$--$0.5$     & $0.03$--$0.3$   & $2\times 10^{14}$--$2\times 10^{17}$ \\
Weak lensing (CMB)              & $1$--$12$     & $0.01$--$0.3$   & $2\times 10^{14}$--$5\times 10^{18}$ \\
Weak lensing (galaxy) & $0.05$--$1.5$ & $0.1$--$8$      & $10^{10}$--$5\times 10^{15}$ \\
\hline
\multicolumn{4}{l}{\textit{Current data (frontier probes)}} \\[2pt]
Strong lensing (cluster)          & $0.2$--$1$       & $0.17$--$8$      & $10^{10}$--$10^{15}$ \\
High-$z$ galaxies   & $4$--$10$      & $0.1$--$10$       & $5\times10^{9}$--$5\times10^{15}$ \\
Lyman-$\alpha$ forest    & $2$--$6$       & $0.07$--$42$     & $7\times10^{7}$--$2\times10^{16}$ \\
Dwarf galaxies           & $\lesssim 0.1$     & $4$--$86$       & $10^{7}$--$10^{11}$ \\
Strong lensing (galaxy) & $0.2$--$4$ & $8$--$170$     & $10^{6}$--$10^{10}$ \\
Stellar streams          & $\approx 0$     & $8$--$370$     & $10^{5}$--$10^{10}$ \\
\hline
\multicolumn{4}{l}{\textit{Future data}} \\[2pt]
Weak lensing (CMB)              & $1$--$12$     & $0.01$--$1.5$   & $2\times 10^{12}$--$5\times 10^{18}$ \\
21-cm (dark ages) & $30$--$200$ & $0.3$--$10$ & $5\times10^{9}$--$2\times10^{14}$ \\ 21-cm (cosmic dawn) & $6$--$30$ & $1$--$100$  & $5\times10^{6}$--$5\times10^{12}$
\end{tabular}
\end{ruledtabular}
\end{table*}

\subsection{Connecting Theory to Observables}
\label{sec:modeling_connection}

DM constraints from small-scale structure are achieved primarily through the following modeling techniques, ordered from highest to lowest physical complexity: simulations (cosmological or controlled/hybrid), SAMs, and empirical models. These techniques are not mutually exclusive; for example, semi-analytic or empirical models are often evaluated on cosmological simulation snapshots.

Cosmological simulations have served as a foundation for LSS studies, and there has recently been substantial progress in large suites of high-resolution cosmological simulations targeting small-scale structure (e.g., see \citealt{Angulo:2021kes} for a review). Cosmological N-body simulations have established key predictions of CDM, including a halo mass function that rises steeply toward low masses~\citep{Tinker:2008ff}, universal halo structure often modeled by the Navarro--Frenk--White profile (NFW; \citealt{Navarro:1996gj}), and the mass--concentration--formation time relation~\citep{Wechsler:2001cs,Diemer:2018vmz}. 

\NEW{In CDM, cosmological N-body simulations of LSS are well established (see \citealt{Angulo:2021kes} for a review). On smaller scales, zoom-in simulations have continued to push to higher resolution (see \citealt{Nadler:2022dvo} for a compilation of existing zoom-in simulations). This effort has led to convergent predictions for certain subhalo population statistics, e.g., the present-day subhalo mass function (SHMF); however, important uncertainties related to artificial disruption and subhalo finding/tracking remain. Meanwhile, hydrodynamic simulations have demonstrated the ability to produce realistic galaxy populations (see \citealt{Vogelsberger:2019ynw} for a review of galaxy formation simulations). However, the impact of different subgrid physics prescriptions represents an important systematic: several discrepancies between hydrodynamic codes and models persist on small scales, particularly for low-surface-brightness galaxies and galaxy rotation curve diversity more broadly~\citep{Sales:2022ich}.} 

Moving beyond CDM, recent simulation suites have begun systematically mapping from DM particle physics parameters to small-scale structure observables in models beyond CDM (e.g., ETHOS; \citealt{Vogelsberger:2015gpr,Lovell:2017eec,Lovell:2018gap,Bose:2018juc,Bohr:2020yoe}). Several suites that include models beyond CDM now simultaneously include several different DM models (e.g., COZMIC; \citealt{Nadler:2025fcv,Nadler:2025jwh,An:2025gju}), and others span different baryonic feedback prescriptions (e.g., DREAMS; \citealt{Rose:2024xcb,Rose:2025}). Simulations of small-scale structure in both CDM and alternative models face important challenges, which we discuss in Section~\ref{sec:theory_discussion}.

\begin{figure*}[t!]
\centering
\includegraphics[trim={0cm 0.25cm 0 0},width=\textwidth]{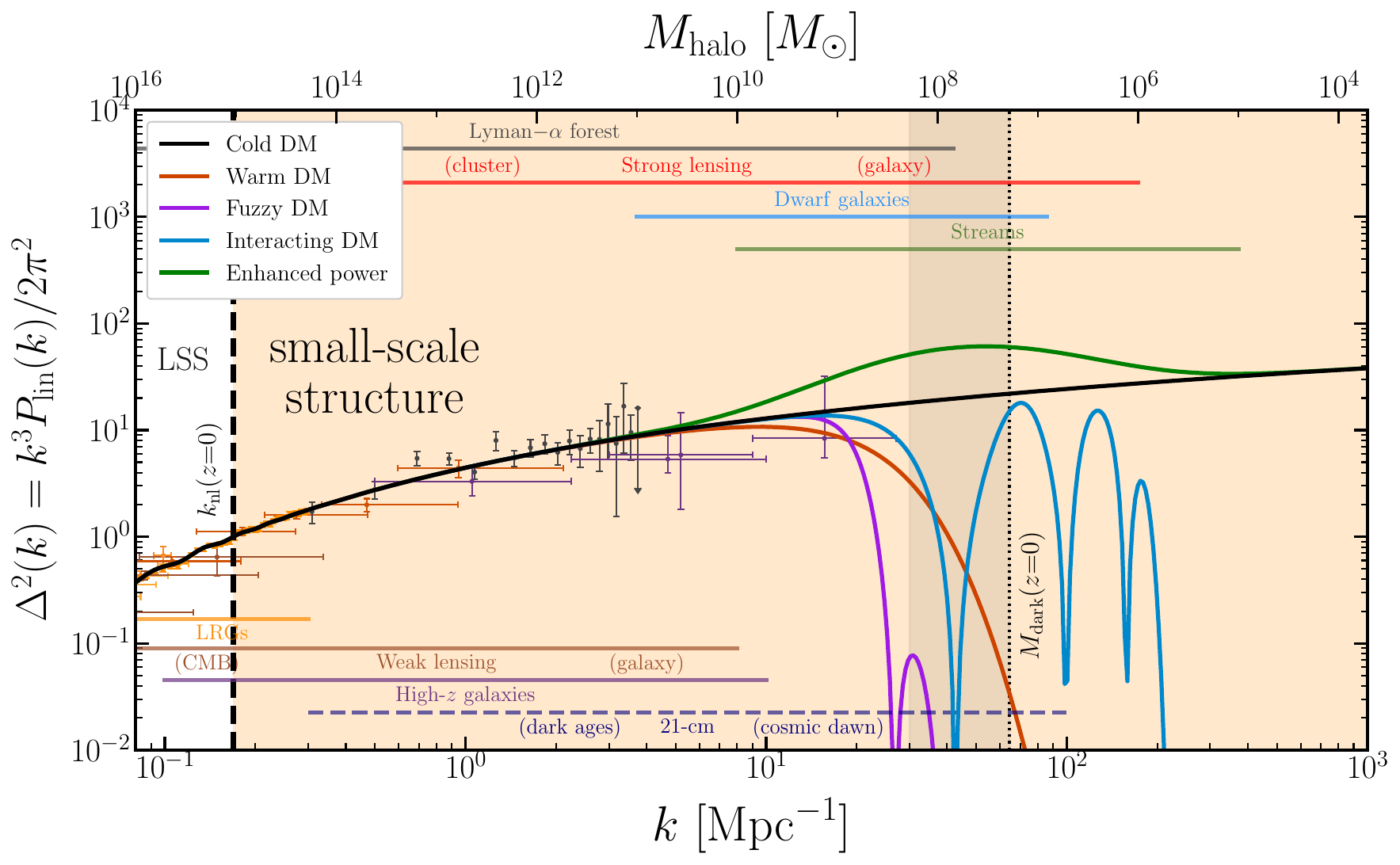}
    \caption{Dimensionless linear matter power spectrum, $\Delta^2(k)$, extrapolated linearly to $z=0$, as a function of wavenumber. \NEW{Small-scale structure corresponds to scales where $\Delta^2(k)>1$ today in CDM, i.e., $k>k_{\mathrm{nl}}(z=0)\approx 0.17~\mathrm{Mpc}^{-1}$ (gold shaded region; Equation~\ref{eq:small_scale}). The dashed vertical line shows $k_{\mathrm{nl}}(z=0)$.} The dotted vertical line shows an estimate for the \NEW{peak mass of the smallest luminous halos} at $z=0$, and the gray shaded region indicates the theoretical uncertainty on this lower limit. Error bars show $P(k)$ constraints from \citet{Chabanier:2019eai}, with the addition of high-$z$ galaxy UV luminosity function measurements from \citet{Sabti:2021unj} and \citet{Munoz:2026diy}, and horizontal lines show wavenumber ranges corresponding to the probes from Figure~\ref{fig:k_z}. Power spectrum predictions are shown for the following DM models: CDM (black), WDM (red; $m_{\mathrm{WDM}}=3.5~\mathrm{keV}$ using the \citet{Vogel:2022odl} transfer function), FDM (purple; $m_{\mathrm{FDM}}=10^{-21}~\mathrm{eV}$ using the \citet{Passaglia:2022bcr} transfer function), IDM (blue; DM--proton scattering with $\sigma_0=10^{-22}~\mathrm{cm}^2$ and $\sigma\sim v^4$, using the \citet{Boddy:2018kfv} transfer function), and a phenomenological model with enhanced power (green; an illustrative Gaussian enhancement of the power ratio with $k_0=41.8~\mathrm{Mpc}^{-1}$, $A=2$, and $\sigma_k=0.5$, following the functional form of \citealt{Tkachev:2024qua} and \citealt{Nadler:2025crd}). Adapted from \citet{Bechtol:2022koa}.}
    \label{fig:p_k}
\end{figure*}

Semi-analytic models (SAMs) solve simplified versions of the physical equations underlying structure and galaxy formation (see \citealt{Baugh:2001ha} for an overview). SAMs are often evaluated on merger trees, which can be extracted from cosmological simulations or generated semi-analytically, e.g., using extended Press--Schechter theory (ePS; \citealt{Press:1973iz,Zentner:2006vw}). SAMs are now commonly used in small-scale structure studies. As we will discuss, recent analyses of several small-scale DM probes including stellar streams (e.g., \citealt{Dropulic:2024tsk,Menker:2024koc,Adams:2024zhi}), dwarf galaxies (e.g., \citealt{Dekker:2021scf,Dekker:2024nkb,Esteban:2023xpk,Ahvazi,Ando:2025qtz,Newton:2024jsy}), and strong lensing (e.g., \citealt{Gilman:2019nap,Gilman:2025fhy,Keeley:2024brx,Keeley:2025oig}) employ SAMs. These models can be orders-of-magnitude more efficient than cosmological simulations, and can also be applied to cosmological simulation outputs, e.g., as a means to model galaxy populations~\citep{Helly:2002rk,Benson:2011bq,Weerasooriya}. A growing body of work aims to incorporate non-gravitational DM physics directly into SAMs by calibrating them on simulations in alternative DM models (e.g., \citealt{Ando:2024kpk,Lonergan:2025atf,Ono:2025jpo,Nadler:2025yni,Benson:2026kdp}).

Finally, empirical models predict halo and/or galaxy populations without explicitly describing the underlying physical processes (e.g., see \citealt{Lapi:2026} for a review). Empirical models often aim to capture the galaxy--halo connection, i.e., the statistical relation between the properties of DM halos and the galaxies that they host, e.g.\ through the stellar mass--halo mass (SMHM) relation (see \citealt{Wechsler:2018pic} for a review). Techniques in this category that are well-established for LSS studies---including abundance matching or halo occupation distribution (HOD) modeling---have recently been extended to model structure on smaller scales (e.g., \citealt{Read:2018gpi,Nadler:2018iux,Kado-Fong:2025,Xu:2026efb}). Other empirical models describe halo and/or galaxy properties through parametric relations that are constrained by data or calibrated to simulations (e.g., \citealt{Moster:2018,Behroozi:2019kql}). Again, these techniques have been extended to study small-scale structure, including populations of low-mass satellite galaxies~\citep{OLeary:2023,Wang:2024jsi}.

\section{Dark Matter Scenarios}
\label{sec:scenarios}

\subsection{\emph{Ab Initio} vs.\ \emph{In Situ} DM Physics}

We focus on two broad classes of deviations from the CDM paradigm when summarizing DM constraints, following \citet{Bechtol:2022koa}:
\begin{itemize}
    \item \emph{Ab initio} deviations: DM properties alter linear structure relative to CDM, typically before matter--radiation equality; then, nonlinear structure formation proceeds as in CDM (i.e., gravity is the only relevant force experienced by DM and wave behavior is unimportant). 
    \item \emph{In situ} deviations: DM properties dynamically alter nonlinear structure formation, over the course of cosmic history, often (but not exclusively) through new force(s) experienced by DM particles.
\end{itemize}

\emph{Ab initio} effects are often parameterized by the transfer function $T(k)$; when squared, this quantity represents the ratio of $P(k)$ to that in CDM:
\begin{equation}
    T^2(k)\equiv \frac{P(k)}{P_{\mathrm{CDM}}(k)}.
\end{equation}
We further define the half-mode wavenumber $k_{\mathrm{hm}}$ via
\begin{equation}
    T^2(k_{\mathrm{hm}})\equiv 0.25.
\end{equation}
The corresponding half-mode mass $M_{\mathrm{hm}}$ is given by $M_{\mathrm{hm}}\equiv M_{\mathrm{halo}}(k_{\mathrm{hm}})$ using Equation~\ref{eq:m_k} \NEW{and should thus be interpreted as a peak halo mass scale}. We discuss the shape of the transfer function for various DM models below; Figure~\ref{fig:p_k} shows several representative examples.

\emph{In situ} effects, in contrast, are more difficult to parameterize in a unified way. The couplings between DM and itself and/or SM particles determine the magnitude of \emph{in situ} deviations from CDM, but effects on nonlinear structure are model-dependent; in addition, other DM properties (e.g., the DM particle mass in the ultralight DM regime) can dynamically affect structure formation.

Many particle DM models produce both \emph{ab initio} and \emph{in situ} deviations from CDM. The two types of deviation need not arise from the same component; for example, multi-component dark sectors can feature mixed behavior. Even single-component DM models often introduce both effects in practice; for example, particle constructions of SIDM generically alter both $P(k)$ and halo evolution~\cite{Cyr-Racine:2015ihg,Nadler:2025wra}, though SIDM is usually simulated by adding self-interactions to CDM-like initial conditions~\cite{Tulin:2017ara,Adhikari:2022sbh}. For clarity, we discuss \emph{ab initio} and \emph{in situ} constraints separately where possible, and we highlight the signatures of both effects for each small-scale probe and beyond-CDM scenario.

In this section, we review the main classes of DM models covered throughout, focusing on their implications for small-scale structure. We refer the reader to reviews that cover each scenario in more detail: see \citet{Tulin:2017ara} and \citet{Adhikari:2022sbh} for self-interacting DM, \citet{Abazajian:2017tcc} for warm DM, \citet{Hui179504} for fuzzy DM, and \citet{Hambye:2010zb} for decaying DM. We do not directly review DM scenarios that lead to $P(k)$ enhancement, which can include various axion DM production mechanisms (e.g., \citealt{Arvanitaki:2019rax,Amin:2022nlh,Sobotka:2024tat,Co:2025lrd}).

\subsection{Warm Dark Matter}

Warm dark matter (WDM) refers to a class of models in which DM has a non-negligible free-streaming length imprinted during the radiation--dominated epoch. Free streaming arises from the velocity dispersion of the WDM particles at early times and erases density perturbations on scales smaller than the free-streaming length $\lambda_\mathrm{fs}$. This effect suppresses $P(k)$ on small scales, preventing structure formation below a characteristic mass scale and delaying structure formation relative to CDM even above this scale~\citep{Schneider11120330}.

From a particle physics standpoint, WDM is often realized by thermally-produced particles that decouple in the radiation-dominated epoch while (mildly) relativistic. A canonical example is a sterile neutrino produced via mixing with active neutrinos \citep[e.g., the Dodelson–Widrow mechanism;][]{Dodelson:1993je}; this production mechanism is now largely ruled out by a combination of structure formation and indirect detection data~\citep{Shakya:2015xnx}. This motivated the study of other production mechanisms---including resonant production in the presence of a lepton asymmetry \citep{Shi:1998km} and production through the decay of heavier particles~\citep{Shaposhnikov:2006xi}---since the primordial WDM momentum distribution can be colder in these cases while avoiding indirect detection constraints. Light gravitinos in models that feature gauge-mediated supersymmetry breaking provide another concrete example of WDM~\citep{Gorbunov:2008ui}. This section focuses on thermal-relic WDM; our review of DM constraints covers both thermal and non-thermal models.

A particle with mass $m_\mathrm{WDM}$ produced in thermal equilibrium becomes nonrelativistic at a temperature $T_\mathrm{nr} \sim m_\mathrm{WDM}/3$. For WDM masses of observational interest ($m_\mathrm{WDM} \sim 1$–$10\,\mathrm{keV}$), this occurs deep in the radiation-dominated epoch, well before matter--radiation equality but after Big Bang Nucleosynthesis (BBN). The free-streaming length accumulated before matter domination sets the scale below which $P(k)$ is suppressed. In particular, for wavenumbers $k \ll k_\mathrm{fs}$, WDM perturbations grow identically to CDM, while for $k \gg k_\mathrm{fs}$ perturbations are exponentially suppressed.

\begin{figure*}[t!]
\centering
\includegraphics[width=\textwidth]{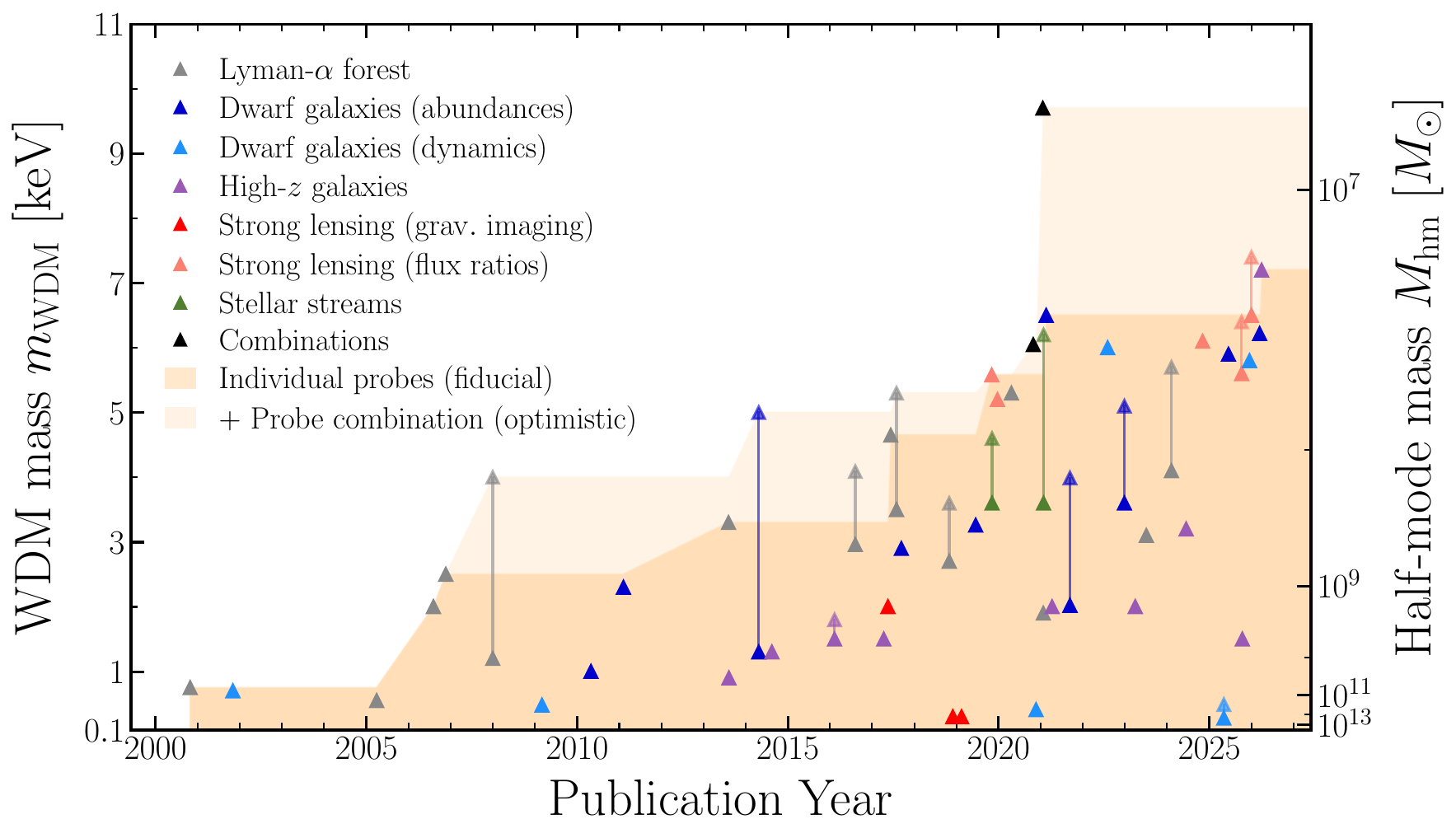}
    \caption{Summary of lower limits on the thermal-relic WDM mass as a function of publication date. Constraints are colored by analyses of the Lyman-$\alpha$ forest (gray), dwarf galaxies (dark blue for limits from abundances and blue for limits from profiles), strong gravitational lensing (red for gravitational imaging and light red for flux ratio anomalies), high-$z$ galaxies (purple), stellar streams (green), and combinations of small-scale structure probes (black). Results that report both a fiducial and optimistic limit are shown by opaque arrows (placed at the fiducial constraint with lower $m_{\mathrm{WDM}}$) connected to transparent arrows (placed at the more optimistic limit). The dark shaded region shows the aggregate WDM limit from small-scale structure over time, based on the fiducial bounds from all individual probes; the light region shows the cumulative bound based on the optimistic limits and including probe combination results. \NEW{The MW satellite plus strong lensing flux ratio limit from \citet{Nadler:2021dft} currently provides the strongest lower bound, while the dark shaded band shows the envelope of limits from all individual probes. Note that dwarf galaxies and stellar streams have common systematics related to the MW subhalo population and host halo properties, while strong lensing, the Lyman-$\alpha$ forest, and high-$z$ galaxies have distinct systematics.} See Appendix~\ref{sec:wdm_summary} for details on the references, statistical interpretations, and WDM transfer function assumptions.}
    \label{fig:wdm_timeline}
\end{figure*}

The WDM linear power spectrum is typically calculated with Boltzmann codes such as \textsc{CLASS} \citep{Lesgourgues:2011rh} or \textsc{CAMB} \citep{Lewis:1999bs} using an appropriate primordial phase-space distribution to determine the initial conditions. The WDM transfer function resulting from this procedure can be written following \citet{Bode:2000gq} and \citet{Viel:2005qj},
\begin{equation}
    T^2_{\mathrm{WDM}}(k,m_{\mathrm{WDM}}) = \left[1+(\alpha(m_{\mathrm{WDM}})\times k)^{2\nu}\right]^{-10/\nu},\label{eq:transfer_wdm}
\end{equation}
where recent fits to transfer functions from linear Boltzmann solvers yield $\nu=1.049$ and
\begin{equation}
\alpha(m_{\mathrm{WDM}})=a \left(\frac{m_{\mathrm{WDM}}}{1~\mathrm{keV}}\right)^b \left(\frac{\omega_{\mathrm{WDM}}}{0.12}\right)^{\eta}\left(\frac{h}{0.6736}\right)^{\theta}h^{-1}~\mathrm{Mpc},\label{eq:alpha_wdm}
\end{equation}
with $\omega_{\mathrm{WDM}}\equiv \Omega_{\mathrm{WDM}}h^2$, $a=0.0437$, $b=-1.188$, $\theta=2.012$, and $\eta=0.2463$ for spin-$1/2$ particles~\citep{Vogel:2022odl}. The scale $\alpha$ determines the cutoff wavenumber, with $k_\mathrm{fs} \sim 1/\alpha$. The half-mode wavenumber $k_{\mathrm{hm}}$ is smaller than $k_{\mathrm{fs}}$ by about one order of magnitude due to the integrated nature of free-streaming.

The WDM transfer function yields a half-mode mass
\begin{align}
    M_{\mathrm{hm}}(m_{\mathrm{WDM}}) &\approx 4\times 10^8~M_{\mathrm{\odot}}\times 
    \left(\frac{m_{\mathrm{WDM}}}{3~\mathrm{keV}}\right)^{-3.564}\times& \nonumber \\
    &\ \ \ \left(\frac{\omega_{\mathrm{DM}}}{0.12}\right)^{1.582}\times 
    \left(\frac{h}{0.6736}\right)^{3.036}.&
    \label{eq:Mhm_mwdm}
\end{align}
Structure formation is strongly suppressed relative to CDM on scales $M\lesssim M_\mathrm{hm}$; for $m_\mathrm{WDM} = 10\,\mathrm{keV}$, this corresponds to $M_\mathrm{hm} \approx 5\times 10^6~M_\odot$ in our fiducial cosmology, characteristic of the halos below the galaxy formation threshold that will be probed by upcoming strong gravitational lensing and stellar stream data.

Through its effect on $P(k)$, WDM leads to a suppression of the halo and subhalo mass functions across cosmic time. Fitting functions for this suppression as a function of $m_{\mathrm{WDM}}$ have been measured in simulations (e.g., \citealt{Lovell:2013ola,Stucker:2021vyx,Nadler:2025fcv}), and this mass function suppression is the primary signature used to constrain WDM with small-scale structure. However, WDM also gives rise to other effects on structure. For example, WDM's non-negligible primordial velocity dispersion sets a minimum \NEW{peak halo mass} below which virialized objects cannot dynamically form~\citep{Tremaine:1979we}. In addition, WDM halo concentrations are reduced relative to CDM (even for halos with $M>M_{\mathrm{hm}}$) because structure formation is delayed in these models as a result of $P(k)$ suppression (e.g., \citealt{Bose:2015mga}).

Recent work has improved the accuracy of WDM transfer function calculations~\citep{Vogel:2022odl}, extended structure formation constraints to mixed cold-plus-warm models~\citep{Tan:2024cek,An:2025gju}, and incorporated WDM into semi-analytic predictions~\citep{Benson:2012su,Dekker:2021scf}. Meanwhile, sterile neutrino model building has broadened beyond the Dodelson--Widrow and Shi--Fuller scenarios to address structure formation and other astrophysical constraints, particularly from X-ray indirect detection (e.g., see \citealt{DeGouvea:2019wpf,Johns:2019cwc,Astros:2023xhe,Fuller:2024noz}). Given the range of production mechanisms available and the sensitivity of upcoming surveys, WDM continues to serve as a benchmark for a broader class of DM models that suppress small-scale structure.

Figure~\ref{fig:wdm_timeline} summarizes the evolution of small-scale structure WDM constraints since the turn of the century. The typical sensitivity of WDM limits from small-scale probes has steadily improved with time due to increasing data quality and more accurate models of nonlinear structure. However, constraints from different analyses of each probe vary, even when similar datasets are analyzed.

In summary, the primary observational signature of WDM is a suppression of the halo and subhalo mass functions \NEW{for systems with peak masses} near and below $M_{\mathrm{hm}}$, which is set by $m_{\mathrm{WDM}}$ via Equation~\ref{eq:Mhm_mwdm}. Currently, the strongest lower limits on $m_{\mathrm{WDM}}$ from individual small-scale structure probes are set by MW satellite abundances (Section~\ref{sec:dwarf_abundances}), stellar stream perturbations (Section~\ref{sec:streams}), strong gravitational lensing flux ratio anomalies (Section~\ref{sec:galaxy_sl}), and high-resolution Lyman-$\alpha$ forest measurements (Section~\ref{sec:lyman_alpha}), and high-$z$ galaxies (Section~\ref{sec:high_z}). The most optimistic joint analyses of these probes now yield $m_{\mathrm{WDM}}\gtrsim 10~\mathrm{keV}$ (\citealt{Nadler:2021dft}; Section~\ref{sec:combination_constraints}).

\subsection{Fuzzy Dark Matter}

Fuzzy dark matter (FDM) refers to the class of ultralight DM models in which the wave-like nature of the DM affects cosmological structure formation \citep{Hu008506, Hui179504, Antypas:2022asj}. Following the uncertainty principle, to achieve an astrophysically-sized de Broglie wavelength $(\lambda_\mathrm{dB} \gtrsim 1\,\mathrm{pc})$,  DM particles must have low momentum. Given the lowest velocity dispersion inferred from the most compact ultra-faint galaxies, \(v \sim 1\,\mathrm{km}\,\mathrm{s}^{-1}\), this (approximately) determines the particle mass for which the wave-like behavior of DM is relevant:
\begin{equation}
m_\mathrm{FDM} \approx 2 \times 10^{-18}\,\mathrm{eV}\times\left(\frac{\lambda_\mathrm{dB}}{\mathrm{pc}}\right)^{-1}\times \left(\frac{v}{1\ \mathrm{km}\,\mathrm{s}^{-1}}\right)^{-1}.
\label{eq:FDM_dB}
\end{equation}
\NEW{Here, $\lambda_\mathrm{dB}\equiv\hbar/(m_\mathrm{FDM}v)$ is the reduced de Broglie wavelength.} It follows that the lighter the particle mass, the larger the wavelength. An FDM particle with \(m_\mathrm{FDM} \sim 10^{-22}\,\mathrm{eV}\) will have a wavelength that would produce a \(\sim \mathrm{kpc}\) core in a dwarf galaxy (\(\sim 20\,\mathrm{km}\,\mathrm{s}^{-1}\)), while FDM with \(m_\mathrm{FDM} \sim 10^{-25}\,\mathrm{eV}\) will affect galaxy cluster scales. Note that the FDM field is coherent on this length-scale. An important time-scale is the time to traverse such a coherent granule:
\begin{equation}
\tau_\mathrm{coh} = 2\pi\,\frac{\lambda_\mathrm{dB}}{2v} \approx 6\,\mathrm{Myr}\times \left(\frac{m}{10^{-18}\,\mathrm{eV}}\right)^{-1}\times \left(\frac{v}{1\ \mathrm{km}\,\mathrm{s}^{-1}}\right)^{-2}.
\label{eq:FDM_time_coherence}
\end{equation}

Given their extremely low mass, FDM particles must be bosons with high occupation number  to constitute a sizable fraction of the DM relic density.\footnote{Fermions cannot comprise all of the DM for particle masses \(\lesssim 1\,\mathrm{keV}\) given the Pauli exclusion principle~\citep{Tremaine:1979we} unless there is a vast number of particle species~\citep{Davoudiasl:2020uig}.} It follows that FDM can consist of ultralight (pseudo-) scalars, vectors, tensors or even higher-spin particles \citep{Jain:2021pnk}. Given the differing degrees of freedom, each of these models will affect structure formation slightly differently.

\begin{figure*}[t!]
\centering
\includegraphics[trim={0cm 0.25cm 0 0},width=\textwidth]{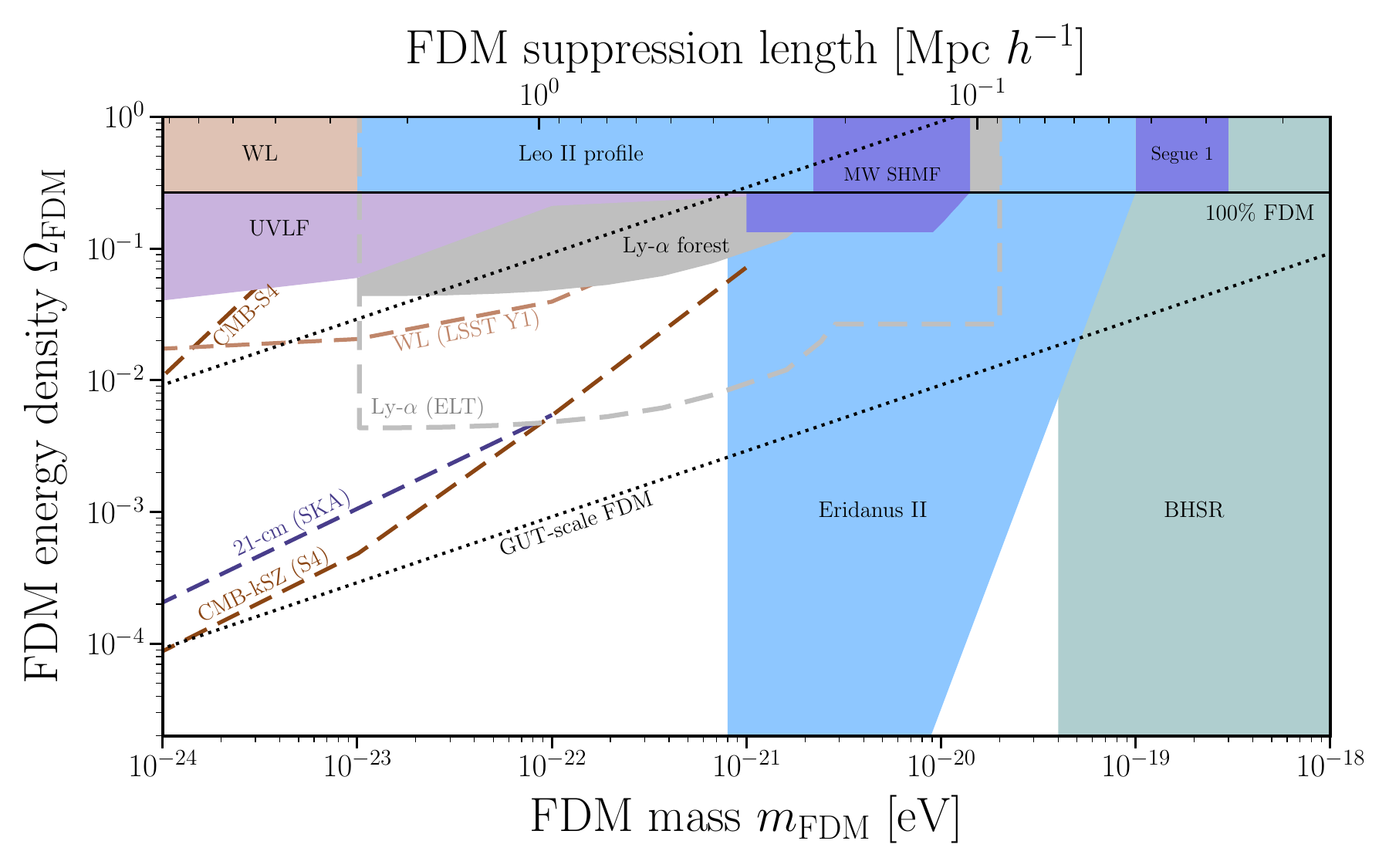}
    \caption{Summary of constraints on the FDM cosmic energy density, $\Omega_{\mathrm{FDM}}$, as a function of the FDM particle mass, \(m_\mathrm{FDM}\), with the equivalent FDM Jeans suppression wavelength, \(\lambda_\mathrm{J}\), shown at the top. \NEW{Shaded regions show current} constraints from analyses of dwarf galaxy abundances (MW SHMF; \citealt{Crumrine:2026rnp}), \NEW{the dynamics of} individual dwarf \NEW{galaxies} (Leo II profile; \citealt{Zimmermann:2024xvd}), FDM dynamical heating in ultra-\NEW{faint} dwarf galaxies (Eridanus II, \citealt{Marsh:2018zyw}; Segue 1 and 2, \citealt{Dalal:2022rmp}), the Lyman-$\alpha$ forest (Ly-\(\alpha\) forest; \citealt{Kobayashi:2017jcf,Rogers:2020ltq}), the high-\(z\) galaxy UV luminosity function (UVLF; \citealt{Winch:2024mrt}), galaxy weak lensing (WL; \citealt{Dentler:2021zij}) and black hole superradiance (BHSR; \citealt{Mehta:2020kwu}). \NEW{Dashed lines show forecast} sensitivities \NEW{of} future analyses of \NEW{lensing from CMB-S4-class experiments} (CMB-S4; \citealt{Dvorkin:2022bsc}), galaxy weak lensing (WL (LSST Y1); \citealt{Preston:2025tyl}), 21-cm intensity mapping (21-cm (SKA);  \citealt{Bauer:2020zsj}), the CMB kinetic Sunyaev--Zeldovich effect (CMB-kSZ (S4); \citealt{Farren:2021jcd}) and the Lyman-\(\alpha\) forest (Ly-\(\alpha\) (ELT); \citealt{Grin:2019mub}). \NEW{The black dotted lines bound the theoretically preferred range in which} the axion decay constant, \(f_\mathrm{a}\), is at the grand unified scale (GUT-scale FDM; \citealt{Marsh:2015xka}). \NEW{The solid horizontal line marks the density at which FDM constitutes all of the dark matter.} \NEW{Figure adapted from} code maintained (and updated) at \url{https://github.com/keirkwame/DarkMatterLimits}.}
    \label{fig:fdm_limits}
\end{figure*}

A concrete pseudo-scalar example is the axion \cite{Peccei1977CP,Preskill_1983,Abbott_1983,Dine_1983,Weinberg1978LightBoson,Wilczek1978ProblemOf, 1992SvJNP..55.1063B}. The axion field acquires an angular degree of freedom in the early universe at a high-energy scale, \(f_\mathrm{a}\), the axion decay constant. This symmetry breaks at a lower energy, leading to damped harmonic oscillations from an initial misalignment angle, \(\theta_\mathrm{a}\), in the presence of Hubble friction. When the field oscillates, the axion behaves like a pressureless DM candidate in a time-averaged sense. At early times, when Hubble friction initially dominates, the field slowly rolls and the axion can drive accelerated expansion if its energy density is sufficient. For \(m_\mathrm{FDM} \lesssim 10^{-27}\,\mathrm{eV}\), the matter-like oscillations begin after matter--radiation equality, and the dark energy-like phenomena become observationally relevant. We only consider heavier axions here. While the axion was originally postulated as a solution to both the strong CP and DM problems, a broader class of axion-like particles is a typical product of high-energy theories, including string theory \citep{Svrcek:2006yi,PhysRevD.81.123530}. In string theory, the compactification of extra space-time dimensions leads to the formation of tens, or even hundreds, of axions in a scenario dubbed the ``axiverse.'' Recent work has systematically studied axiverse models that produce FDM \citep{Sheridan:2024vtt} and connected them to cosmological observations \citep{Jain:2025vfh}.

In the axion model, the energy density \(\Omega_\mathrm{FDM}\) (\(\omega_\mathrm{FDM} \equiv \Omega_\mathrm{FDM} h^2\)) relates to the decay constant and initial misalignment via
\begin{equation}
    \omega_\mathrm{FDM} \approx 0.12\,\theta_\mathrm{a}^2\times\left(\frac{m_\mathrm{FDM}}{4.4 \times 10^{-19}\,\mathrm{eV}}\right)^\frac{1}{2} \times\left(\frac{f_\mathrm{a}}{10^{16}\,\mathrm{GeV}}\right)^2.
    \label{eq:FDM_density}
\end{equation}
Given the large number of axions in the axiverse and the large decay constants needed to produce a cosmologically-significant abundance, it is typical to expect each axion to be some fraction of the DM. Cosmological observations therefore typically constrain both \(m_\mathrm{FDM}\) and \(\Omega_\mathrm{FDM}\), which can be related to fundamental parameters of the axion model by Equation~\ref{eq:FDM_density}.

FDM inherently produces \textit{in situ} deviations from standard CDM, although many aspects of the model can be observationally constrained from its \textit{ab initio} effects on the $P(k)$. \emph{In situ} dynamics arise since the FDM field can be described by the wave-like Klein-Gordon equation. It is typical to take the limit of a single, non-relativistic, classical, scalar field (although see discussion of multi-species and higher spin FDM below). This simplifies the equations of motion to the coupled Schr\"{o}dinger-Poisson equations \citep[e.g.,][]{Ferreira:2020fam}:
\begin{equation}
    i\dot{\psi} = -\frac{3}{2}iH\psi - \frac{\nabla^2 \psi}{2a^2m_\mathrm{FDM}} + \frac{g}{8m_\mathrm{FDM}^2}|\psi|^2\psi + m_\mathrm{FDM}\Phi\psi,
    \label{eq:FDM_Schrodinger}
\end{equation}
\begin{equation}
    \nabla^2\Phi = 4\pi G \delta\rho,
    \label{eq:Poisson}
\end{equation}
where \(\psi\) is the complex axion field, \(\Phi\) is the gravitational potential, \(\delta \rho\) is the matter density perturbation, \(H\) is the Hubble parameter, \(a\) is the scale factor, and \(g\) is the axion self-coupling. In most studies discussed in the literature, \(g\) is set to zero. When \(g\) is positive (negative), this leads to repulsive (attractive) non-gravitational self-interactions; these interactions can change the dynamics significantly, even for very small \(|g|\) (e.g., \citealt{Glennon:2020dxs,Kirkpatrick:2021wwz}).

A striking \textit{in situ} consequence of Equations~\eqref{eq:FDM_Schrodinger} and \eqref{eq:Poisson} is the existence of wave-like DM structures within collapsed objects like halos and filaments. The ground state excitation of the FDM wavefunction \(\psi\) (the ``soliton'') has a central cored density profile, in contrast to the cusps that form in CDM-only simulations. In the halo outskirts, like CDM, FDM relaxes to an NFW density profile, but with additional density interference fringes and, on the smallest scales, time-oscillating DM granules. The coherence length- and time-scales of FDM granules are determined by Equations~\eqref{eq:FDM_dB} and \eqref{eq:FDM_time_coherence}. These oscillations are a source of dynamical heating to any orbiting stars \citep{Marsh:2018zyw,El-Zant:2019ios,Dalal:2022rmp}, though the strength of this effect strongly depends on the initial conditions and tidal history of the halos and the galaxies that they host \citep{Eberhardt:2025lbx}. 

Numerical simulations have been run to account for FDM self-interactions (both repulsive and attractive), vector, tensor and higher-spin fields, multi-field models, mixtures of CDM and FDM \citep{Lague:2023wes}, and the inclusion of baryonic physics, each causing their own distinct signatures; these are active areas of study (e.g., see \citealt{Eberhardt:2025caq} and \citealt{Schive:2025bcm} for recent reviews). In addition to the effects of FDM on halo structure, the impact of FDM can vary nontrivially with larger-scale environment; for example, \citet{Zimmermann:2024vng} showed that axion wave dynamical effects can both suppress and enhance clustering on different scales in cylindrical filaments of the cosmic web.

Even before the cosmic web forms, there are \textit{ab initio} effects on the linear matter power spectrum from FDM. The linear perturbations to the FDM density, in a comoving gauge, have the equations of motion of a damped harmonic oscillator with a scale-dependent dispersion relation, \(\omega_k^2\), or sound speed \citep{Frieman:1995pm,Amendola:2005ad,Marsh:2010wq}. This leads to a transition when \(\omega_k(k_\mathrm{J}) = 0\), where \(k_\mathrm{J}\) is the FDM Jeans wavenumber, and the comoving FDM Jeans wavelength is given by
\begin{equation}
\begin{split}
    \lambda_\mathrm{J} = \frac{2 \pi}{k_\mathrm{J}}&\approx 9.4\,\mathrm{Mpc}\times (1 + z)^\frac{1}{4}\times\\
    & \ \ \ \ \left(\frac{\omega_\mathrm{FDM}}{0.12}\right)^{-\frac{1}{4}} \times\left(\frac{m_\mathrm{FDM}}{10^{-26}\,\mathrm{eV}}\right)^{-\frac{1}{2}}.
\end{split}
\end{equation}
For \(\lambda > \lambda_\mathrm{J}\), perturbations grow like CDM; for \(\lambda < \lambda_\mathrm{J}\), perturbations oscillate and are suppressed. Thus, FDM suppresses $P(k)$ below the Jeans wavelength. If there is a mixture of CDM and FDM, the CDM component is also partially suppressed. The full effect is calculated in modified Boltzmann codes such as \textsc{AxionCAMB}~\citep{Hlozek:2014lca}, \textsc{AxiECAMB}~\citep{Liu:2024yne}, and \textsc{AxiCLASS}~\citep{Poulin:2018dzj,Smith:2019ihp}. \citet{Winch:2023qzl} additionally consider large initial misalignment angles dubbed ``extreme axions.''

The one-component FDM transfer function can be approximated following \citet{Passaglia:2022bcr}:
\begin{equation}
    T^2_{\mathrm{FDM}}(k,m_{\mathrm{FDM}}) = \left[\frac{\sin(x^p)}{x^p(1+Bx^{6-p})}\right]^2,\label{eq:transfer_fdm}
\end{equation}
where $p=5/2$, $x = Ak/k_{\mathrm{J,eq}}$, and $k_{\mathrm{J,eq}}\equiv 9\,m_{\mathrm{FDM,22}}^{1/2}~\mathrm{Mpc}^{-1}$ is the comoving Jeans wavenumber at matter--radiation equality, with
\begin{equation}
\begin{split}
A &= 2.22\left(m_{\mathrm{FDM,22}}\right)^{1/25-\ln\left(m_{\mathrm{FDM,22}}\right)/1000},\\
B &= 0.16 \left(m_{\mathrm{FDM,22}}\right)^{-1/20},
\end{split}
\end{equation}
and $m_{\mathrm{FDM,22}}\equiv m_{\mathrm{FDM}}/10^{-22}~\mathrm{eV}$. This yields
\begin{equation}
    M_{\mathrm{hm}}(m_{\mathrm{FDM,22}}) \approx 4.5\times 10^{10}\,M_{\mathrm{\odot}} \times m_{\mathrm{FDM,22}}^{-1.41}\label{eq:mhm_mfdm}.
\end{equation}
This form agrees with the earlier \citet{Hu008506} fit to within $\approx 20\%$ in $M_{\mathrm{hm}}$, predicting slightly more suppression for $m_{\mathrm{FDM}}\sim 10^{-22}~\mathrm{eV}$ and slightly less suppression for $m_{\mathrm{FDM}}\gtrsim 2\times 10^{-21}~\mathrm{eV}$, where current bounds lie; note that the \citet{Hu008506} fit is still used to derive many FDM constraints.

Figure~\ref{fig:fdm_limits} summarizes current constraints in the $m_{\mathrm{FDM}}$--$\Omega_{\mathrm{FDM}}$ plane. For pure FDM \citep[i.e., $\Omega_{\mathrm{FDM}}=0.265$;][]{Planck:2018vyg}, lower limits now typically reach $m_{\mathrm{FDM}}\gtrsim 10^{-21}$ to $10^{-20}~\mathrm{eV}$. In addition to the observational probes of small-scale structure discussed in this review, we also show bounds from two effects that are specifically sensitive to FDM. First, the formation of clouds of ultralight bosons around supermassive black holes (BHs) would slow down the black hole spin through the superradiance effect \citep{PhysRevD.81.123530}. Thus, lower limits on supermassive black hole spins can be translated to exclusions on the existence of bosons in a particular particle mass regime (\(4 \times 10^{-20}\,\mathrm{eV} \leq m_\mathrm{FDM} \leq 4 \times 10^{-17}\,\mathrm{eV}\)), regardless of their DM density \citep{Mehta:2020kwu}. However, it is argued that other astrophysical phenomena can spin up black holes thereby counteracting the spin-down effect of ultralight bosons \citep{Hoof:2024quk,Sarmah:2024nst}. Second, we also show the forecasted sensitivity to FDM from the Ostriker-Vishniac effect in \NEW{measurements of the kinetic Sunyaev--Zeldovich (kSZ) effect from potential future CMB experiments}~\citep{Farren:2021jcd}. Here, sensitivity arises from FDM suppression of small-scale structure growth.

In summary, FDM produces both \emph{ab initio} suppression of the halo and subhalo mass functions, with a characteristic scale set by $M_{\mathrm{hm}}(m_{\mathrm{FDM}})$ via Equation~\ref{eq:mhm_mfdm}, along with \emph{in situ} modifications to halo density profiles due to wave dynamics. The former effect is constrained by the same probes as WDM---dwarf galaxy abundances (Section~\ref{sec:dwarf_abundances}), stellar stream perturbations (Section~\ref{sec:streams}), strong lensing measurements of halo abundances (Section~\ref{sec:galaxy_sl}), the Lyman-$\alpha$ forest (Section~\ref{sec:lyman_alpha}), and the high-\(z\) galaxy UV luminosity function (Section~\ref{sec:high_z})---while the latter effect is often constrained by the internal kinematics of individual dwarf galaxies and dynamical heating of stellar populations (Section~\ref{sec:dwarf_profiles}).

\subsection{Interacting Dark Matter}

Non-gravitational interactions between DM and SM particles---or between DM and other dark sector particles---affect small-scale structure (see \citealt{Gluscevic:2019yal} for a review). If DM--SM interactions are efficient before recombination, they lead to momentum and heat transfer from SM to DM particles. This can yield a suppression of $P(k)$ due to collisional damping on scales smaller than the size of the cosmological horizon when interactions freeze out~\citep{Boehm:2000gq,Boehm:2004th,McDermott:2010pa}. For the purposes of this review, we refer to such models as ``interacting dark matter'' (IDM).

\begin{figure*}[t!]
\centering
\includegraphics[trim={0cm 0.25cm 0 0},width=0.9\textwidth]{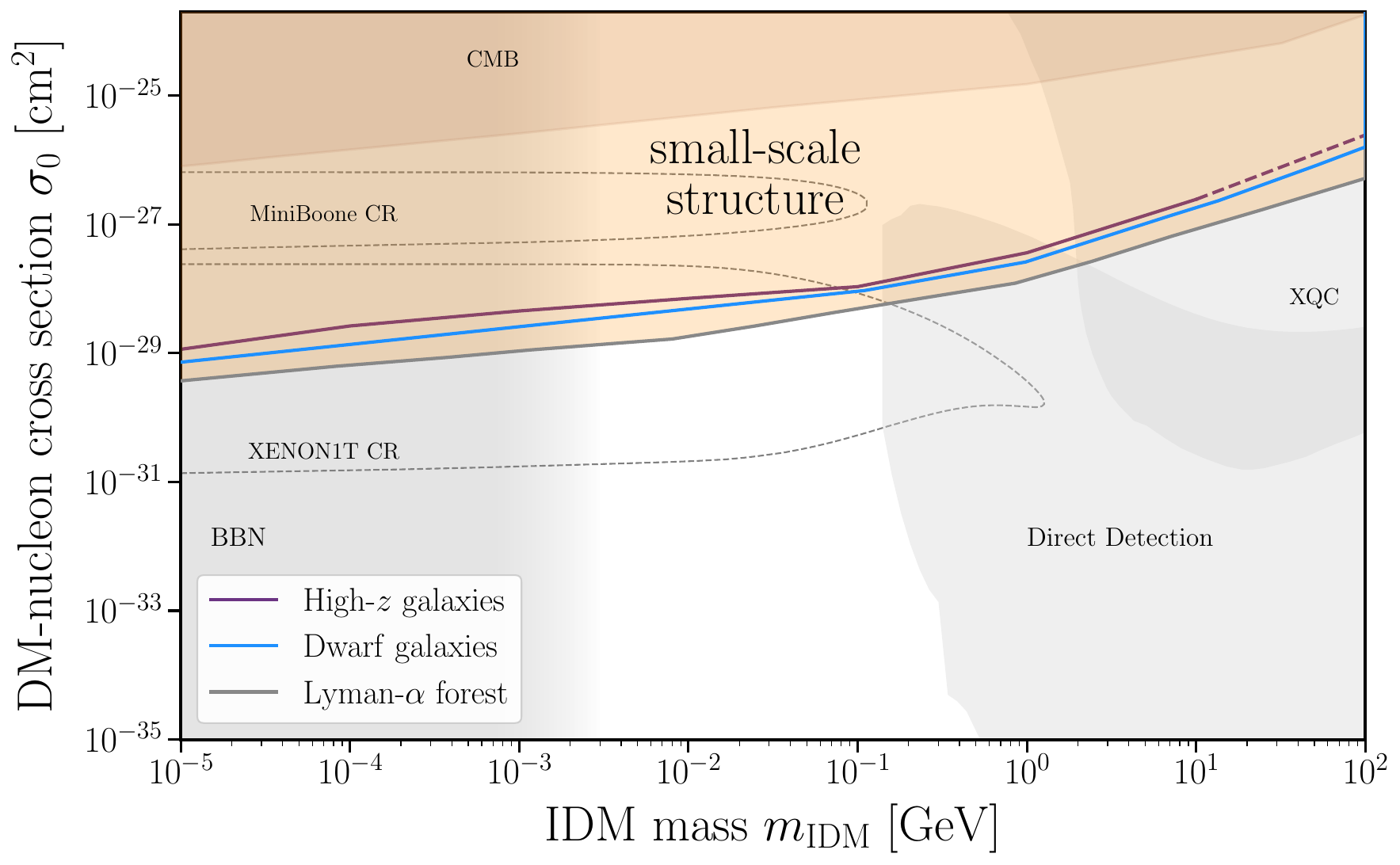}
    \caption{Representative constraints on the velocity-independent DM--nucleon scattering cross section as a function of the DM particle mass. Upper limits from small-scale structure are shown by the gold shaded region, including limits from high-$z$ galaxies (purple; \citealt{Lazare:2025gha}), dwarf galaxies (blue; \citealt{DES:2020fxi}), and the Lyman-$\alpha$ forest (black; \citealt{Rogers:2021byl}). LSS limits from CMB data are shown in brown~\citep{Gluscevic:2017ywp}. IDM mass limits from Big Bang Nucleosynthesis (BBN) are shown by the vertical gray shaded region~\citep{Krnjaic:2019dzc,An:2022sva}. Filled gray contours show direct detection \NEW{and X-ray Quantum Calorimeter (XQC)} bounds~\citep{Mahdawi:2017cxz,Emken:2018run,Akerib:2022ort}, and unfilled gray contours show cosmic ray up-scattering constraints (e.g., \citealt{Alvey:2022pad}). Adapted from \citet{Drlica-Wagner:2022lbd}.}
    \label{fig:idm}
\end{figure*}

WIMPs provide a classic example of collisional damping. For weak-scale interactions, the \NEW{peak mass of the smallest halos that can form given} the small-scale $P(k)$ cutoff is $M_{\mathrm{min}}\sim 10^{-6}~M_{\mathrm{\odot}}$~\citep{Green:2003un,Loeb:2005pm}. Note that $M_{\mathrm{min}}$ depends on the IDM cross section and can be significantly larger in models with higher-amplitude DM--SM cross sections. Here, we parameterize DM--SM interactions in an effective theory framework (following, e.g., \citealt{Gluscevic:2017ywp,Boddy:2018kfv}). Specifically, we write the cross section for interactions between DM particle $\chi$ and SM particle $i$ as
\begin{equation}
    \sigma_{\chi\mathrm{-i}} \equiv \sigma_{0,\chi\mathrm{-i}}v_{\chi\mathrm{-}i}^{n},\label{eq:sigma_idm}
\end{equation}
where $\sigma_{0,\chi\mathrm{-i}}$ is the scattering amplitude, $v$ is the relative velocity, and $n$ is a constant power-law index. Particle physics models of IDM can feature interactions that map to multiple power-law operators, but we treat each $n$ separately for simplicity. Note that IDM models featuring DM--radiation scattering are commonly parameterized by replacing velocity with temperature in Equation~\ref{eq:sigma_idm}. 

The effect of IDM on $P(k)$ depends on $n$. For velocity-independent scattering ($n=0$), the IDM power spectrum is very similar to thermal-relic WDM. For DM--proton scattering with IDM masses $m_{\mathrm{IDM}}\gg m_p$ (where $m_p$ is the proton mass) and $n=0$ \citep{Nadler:2019zrb}, 
\begin{equation}
    M_{\mathrm{hm}}(\sigma_0,m_{\mathrm{IDM}}) \approx 3\times 10^{10}~\left(\frac{\sigma_0/m_{\mathrm{IDM}}}{10^{-26}~\mathrm{cm^2/{GeV}}}\right)^{1.92}~M_{\mathrm{\odot}}.
\end{equation}

This similarity between $P(k)$ for velocity-independent DM--proton scattering and thermal-relic WDM extends to some other IDM models~\citep{Boehm:2001hm,Nguyen:2021cnb,Crumrine:2024sdn}. However, IDM models with $n>0$ often lead to prominent dark acoustic oscillation (DAO) features in $P(k)$, precluding a general mapping between small-scale structure in IDM and WDM cosmologies. Meanwhile, momentum transfer increases over time rather than freezing out in the early universe for models with $n<-2$, complicating effects on small-scale structure~\citep{Dvorkin:2013cea,Boddy180800001,Ali-Haimoud:2023pbi}. For other IDM models, the \emph{in situ} effects of DM--SM interactions are expected to be small because interactions freeze out in the early universe. However, DM--baryon scattering rates can be enhanced near the centers of galaxies today~\cite{Fischer:2025snw,Hainje:2026}, or in regions where a local overdensity of DM scattering partners are produced (e.g., near a supernova in DM--radiation scattering models; \citealt{Heston:2024ljf}). These kinds of \emph{in situ} effects have not widely been used to set IDM constraints beyond analyses of a few specific systems~\citep{Bhoonah:2018wmw,Wadekar:2019mpc}.

Other classes of IDM models feature interactions between DM and a low-mass mediator (e.g., \citealt{Dvorkin:2020xga}) or dark radiation~\citep{Feng:2009mn}. Many such models are captured within the Effective Theory of Structure Formation (ETHOS) framework (\citealt{Cyr-Racine:2015ihg}; also see \citealt{Cyr-Racine:2012tfp}). In these scenarios, interactions transfer heat and momentum to the DM, resulting in a small-scale $P(k)$ cutoff that depends on the dark radiation temperature; this cutoff is often parameterized by the decoupling temperature $T_{\mathrm{kd}}$ between the DM and dark radiation. For example, in the warm SIDM model of \citet{Nadler:2025wra} ($m_\chi=0.1~\mathrm{GeV}$, $\sigma_0/m=147.1~\mathrm{cm^2~g^{-1}}$, and $w=24.33~\mathrm{km\ s^{-1}}$), a decoupling temperature of $T_{\mathrm{kd}}=2.32~\mathrm{keV}$ (corresponding to an initial dark-to-visible sector temperature ratio of $0.11$) leads to a $P(k)$ cutoff with a half-mode scale comparable to $10~\mathrm{keV}$ thermal-relic WDM; this mapping depends on the particle parameters. In such models, $P(k)$ also generally inherits small-scale DAOs, with positions and peak heights that depend on the underlying particle parameters~\citep{Bohr:2020yoe}; recent work has also quantified these effects in DM--SM scattering and atomic DM models~\citep{Barron:2025dys}. Many models with dark radiation scattering also feature DM self-interactions, which we discuss below; these scenarios also generally lead to CMB signals through their impact on the number of relativistic degrees of freedom at early times (e.g., \citealt{Huo:2017vef}).

Figure~\ref{fig:idm} summarizes bounds on velocity-independent DM--nucleon scattering, illustrating how small-scale structure complements direct detection and early-universe probes. In particular, BBN places an upper limit on the IDM mass through the contribution of IDM to relativistic degrees of freedom at early times. This bound depends on the specific DM--SM coupling, but typically excludes $m_{\mathrm{IDM}}\lesssim 1~\mathrm{MeV}$ (e.g., \citealt{Krnjaic:2019dzc,An:2022sva,An:2024nsw}). Meanwhile, direct detection experiments set upper limits on the cross section for $m_{\mathrm{IDM}}\gtrsim 1~\mathrm{GeV}$ (see \citealt{Akerib:2022ort} for a review); reinterpretations of direct detection limits for cosmic ray-upscattered IDM probe DM masses (e.g., \citealt{Alvey:2022pad}). At intermediate IDM masses, small-scale structure sets stringent constraints on DM--SM interactions.

In summary, many IDM models produce $P(k)$ suppression, making dwarf galaxy abundances (Section~\ref{sec:dwarf_abundances}), the Lyman-$\alpha$ forest (Section~\ref{sec:lyman_alpha}), and the high-$z$ galaxy UV luminosity function (UVLF; Section~\ref{sec:high_z}) powerful probes of this physics. For velocity-independent DM--proton scattering, upper limits on the cross section from small-scale structure probes are now at the level of $\sigma_0\lesssim 10^{-28}~\mathrm{cm}^2$ for $m_{\mathrm{IDM}}=1~\mathrm{GeV}$. On the other hand, IDM models with velocity-dependent scattering or dark radiation interactions can produce oscillatory features in $P(k)$ that differ from WDM suppression, and the sensitivity of different probes to these features varies substantially; we discuss this probe-by-probe in Section~\ref{sec:constraints}.

\subsection{Self-interacting Dark Matter}
\label{sec:sidm}

Self-interacting dark matter (SIDM) refers to a class of models featuring non-gravitational interactions between DM particles~\citep{Spergel:1999mh,Tulin:2017ara,Adhikari:2022sbh}. A simple, well-studied case is that of elastic, velocity-independent interactions between DM particles of a single species, typically parameterized by the cross section per unit DM particle mass, $\sigma/m$. In this scenario, the SIDM scattering rate in a DM halo is then
\begin{equation}
\begin{aligned}
    &R_{\mathrm{scat}}t_{H} \approx 15\times \\
    &\left(\frac{\rho_{\mathrm{DM}}}{0.1~M_{\mathrm{\odot}}~\mathrm{pc}^{-3}}\right)\times\left(\frac{v_{\mathrm{rel}}}{50~\mathrm{km\ s}^{-1}}\right)\times \left(\frac{\sigma/m}{1~\mathrm{cm^2~g^{-1}}}\right),
\end{aligned}
\label{eq:rscat}
\end{equation}
where $t_H$ is the Hubble time, $\rho_{\mathrm{DM}}$ is the local DM density within the halo, and $v_{\mathrm{rel}}$ is the scattering velocity. Equation~\ref{eq:rscat} implicitly defines a characteristic radius, $r_1$, within which DM particles scatter at least once per Hubble time on average, i.e., $R_{\mathrm{scat}}(r_1)t_H=1$~\citep{Kaplinghat:2015aga}. This radius roughly separates the isothermal SIDM core from the CDM-like outer halo. Thus, for a $10^9~M_{\mathrm{\odot}}$ halo, $r_1\sim 1~\mathrm{kpc}$ for a cross section of $\sigma/m\sim 1~\mathrm{cm^2~g}^{-1}$ at $v\sim 20$--$30~\mathrm{km\ s}^{-1}$.

A classic \emph{in situ} signature of SIDM on cosmic structure is the formation of density cores in halo centers. The core sizes of simulated SIDM halos are well predicted by the isothermal Jeans model, in which the core radius is somewhat smaller than $r_1$~\citep{Robertson:2020pxj}. In the presence of baryons, the scale radius of the baryon distribution sets the core size, and for sufficiently steep baryonic density profiles, cores can be erased (e.g., \citealt{Kaplinghat:2013xca,Kamada:2016euw,Jiang:2022aqw}). In the subsequent phase of SIDM halo evolution, the heat flow reverses and halos undergo gravothermal core collapse, in which their central regions collapse and reach extremely high densities in a runaway fashion~\citep{Balberg:2002ue,Koda:2011yb}. Core collapse can be achieved within the age of the universe for high cross sections ($\sigma/m\gtrsim 10~\mathrm{cm^2~g^{-1}}$ on the relevant velocity scales), and is achieved more quickly in high-concentration halos~\citep{Essig:2018pzq,Kaplinghat:2019svz}; tidal stripping can also accelerate~\citep{Sameie:2019zfo,Nishikawa:2019lsc} or delay~\citep{Zeng:2021ldo,Zeng:2023fnj} collapse. In addition, the interplay between baryonic feedback and collapse is under active study, with current results indicating that the process is slowed but not stopped by potential fluctuations~\citep{Kong:2026piq}. Beyond these imprints on density and velocity dispersion profiles, self-interactions tend to sphericalize halos' shapes and isotropize their velocity distributions relative to CDM~\citep{Rocha:2012jg,Peter:2012jh}.

Observations of galaxy clusters set stringent upper limits of $\sigma/m\lesssim 1~\mathrm{cm^2\ g}^{-1}$ on the cross section at large scattering velocities, $v\sim 1000~\mathrm{km\ s}^{-1}$~\citep{Kaplinghat:2015aga,Andrade:2020lqq,Sagunski:2020spe}. Thus, in order to appreciably affect small-scale structure, the SIDM cross section must rise with decreasing scattering velocity. A well-studied example is a DM sector with a light particle, $\phi$, that mediates interactions between heavy DM particles, $\chi$, through a Yukawa potential, leading to a differential cross section in the weakly-coupled Born limit of~\citep{Feng:2009hw,Tulin:2013teo}
\begin{equation}
    \frac{\mathrm{d}\sigma}{\mathrm{d}\cos\theta} = \frac{\sigma_0w^4}{2\left[w^2+v^2\sin^2(\theta/2)\right]^2},\label{eq:xsec}
\end{equation}
where $w$ is the velocity scale at which the SIDM cross section transitions from a $v^{-4}$ to $v^0$ scaling and $\theta$ is the scattering angle \citep{Ibe:2009mk,Yang:2022hkm}. Recent work has shown that halo gravothermal evolution is captured by the effective cross section~\citep{Yang:2022hkm,Yang:2022zkd,Outmezguine:2022bhq}
\begin{equation}
\label{eq:eff}
\sigma_{\rm eff} = \frac{2 \int v^2\mathrm{d} v \mathrm{d} \cos\theta \frac{\mathrm{d} \sigma}{\mathrm{d} \cos\theta} \sin^2\theta v^5 \exp \left[-\frac{v^2}{4\nu_{\rm eff}^2}\right]}{\int v^2\mathrm{d} v \mathrm{d} \cos\theta \sin^2\theta v^5  \exp \left[-\frac{v^2}{4\nu_{\rm eff}^2}\right]}, 
\end{equation}
where $\nu_{\rm eff}\approx0.64V_{\rm max}$ is the characteristic velocity dispersion of an NFW halo with maximum circular velocity $V_{\rm max}$. Nonetheless, the velocity and angular dependence of the underlying differential cross section should be accounted for when making detailed predictions, and the mapping from $\sigma_{\mathrm{eff}}$ to microscopic cross sections is an active area of study~\citep{Ramos:2025lvk}.

\begin{figure}[t!]
\hspace{-4mm}
\includegraphics[trim={0 0cm 0 0cm},width=0.5\textwidth]{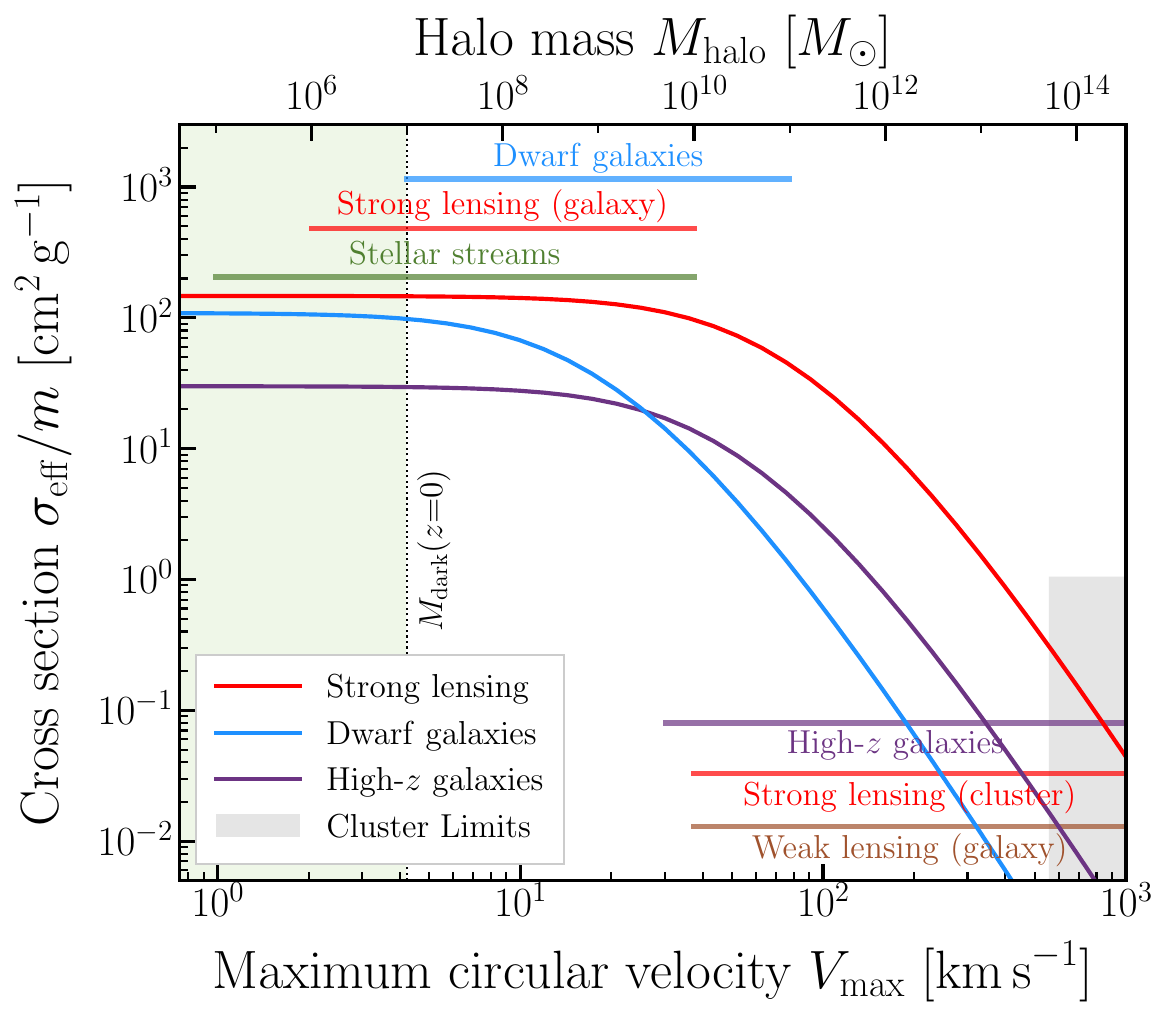}
   \caption{Effective SIDM cross section as a function of halo maximum circular velocity. Horizontal lines show velocity and \NEW{peak halo mass} ranges corresponding to the small-scale structure probes from Figure~\ref{fig:k_z}. Approximate cross section upper limits from clusters are shaded gray, and the velocity range probed by dark halos is shaded green. Solid lines show models motivated by observations of dwarf galaxies (blue; \citealt{Correa:2020qam}), strong-lensing substructure (red; \citealt{Nadler:2023nrd}), and Little Red Dots (purple; \citealt{Jiang:2025jtr}; also see \citet{Feng:2025rzf}). We refer the reader to \citet{Fischer:2023lvl} and \citet{Almeida:2025cee} for more detailed figure summaries of SIDM constraints.}
    \label{fig:xsec}
\end{figure}

Figure~\ref{fig:xsec} shows three example cross sections that affect the halo profiles of dwarf galaxies ($\sigma_0=108.9~\mathrm{cm^2~g^{-1}}$, $w=29.4~\mathrm{km\ s}^{-1}$; \citealt{Correa:2020qam}), the population of dense strong lensing perturbers ($\sigma_0=147.1~\mathrm{cm^2~g^{-1}}$, $w=120~\mathrm{km\ s}^{-1}$; \citealt{Nadler:2023nrd}), and the abundance of Little Red Dots ($\sigma_0=30~\mathrm{cm^2~g^{-1}}$, $w=80~\mathrm{km\ s}^{-1}$; \citealt{Jiang:2025jtr}). Thus, halos of different masses probe the SIDM cross section at different velocity scales. This follows because the velocity dispersion of particles within a halo is set by the halo mass. On the other hand, SIDM subhalos are subject to both internal scattering and collisions between subhalo and host halo particles; the latter probes the SIDM cross section at the subhalo's orbital velocity~\citep{Nadler:2020ulu,Klemmer:2026ivv}. We note that, while the models in Figure~\ref{fig:xsec} are useful benchmarks, a wide variety of cross section shapes can arise from particle models of SIDM (e.g., \citealt{Tulin:2013teo,Colquhoun:2020adl}).

Because SIDM primarily alters halo profiles rather than abundances, the tightest constraints are typically set by probes of the internal structure of halos across a wide halo mass range. At high velocities ($v\approx 1000~\mathrm{km\ s}^{-1}$), upper limits on the SIDM cross section of $\sigma/m\lesssim 0.1$ to $1~\mathrm{cm^2~g}^{-1}$ have been derived from weak and strong lensing measurements of cluster shapes~\citep{Peter:2012jh}, mergers~\citep{Randall:2008ppe,Bradac:2008eu,Wittman:2017gxn,Jee:2026qjg}, profiles~\citep{Kaplinghat:2015aga,Elbert:2016dbb,Sagunski:2020spe,Andrade:2020lqq,ODonnell:2025pkw}, BCG properties~\citep{Harvey:2018uwf}, and combinations of weak lensing and galaxy counts around clusters~\citep{Adhikari:2024aff}. At intermediate velocities ($v\approx 100~\mathrm{km\ s}^{-1}$), SIDM-only explanations of galaxy rotation curve diversity favor $\sigma/m \approx \mathcal{O}(10)~\mathrm{cm^2~g}^{-1}$~ (e.g., \citealt{Kamada:2016euw,Ren:2018jpt,Roberts:2024uyw,Jia:2026ocr}); however, baryonic feedback can produce similar effects~\citep{Zentner:2022xux}.

Meanwhile, lower relative velocity scales ($v\lesssim 100~\mathrm{km\ s}^{-1}$) are probed by dwarf galaxy kinematics (Section~\ref{sec:dwarf_profiles}) and the density profiles of strong lensing perturbers in galaxies (Section~\ref{sec:galaxy_sl}) and clusters (Section~\ref{sec:cluster_sl}), \NEW{including a recent gravitational imaging detection of an extremely compact substructure~\citep{Powell:2025rmj,Vegetti:2026} and hints from flux ratio anomaly measurements that (sub)halo populations are denser than predicted by CDM~\citep{Gilman:2026wul}}. Low-amplitude cross sections that uniformly core low-mass halos are constrained at the level of $\sigma/m\lesssim 1~\mathrm{cm^2~g}^{-1}$ by cuspy dwarf galaxies like Draco (e.g., \citealt{Read:2018pft}), while cored systems like Fornax prefer larger cross sections if the cores are creating by SIDM rather than baryonic feedback (e.g., \citealt{Valli:2017ktb}). Meanwhile, high-amplitude cross sections with $\sigma/m\gtrsim 100~\mathrm{cm^2~g}^{-1}$ that place a majority of halos in the core collapse phase today may also be disfavored (e.g., \citealt{Ando:2025qtz,Almeida:2025cee,Fischer:2026ryr}). Models with $50\ \mathrm{cm^2~g}^{-1}\lesssim \sigma/m\lesssim100~\mathrm{cm^2~g}^{-1}$ lead to a mixture of cored (originally low-concentration) and core-collapsed (originally high-concentration) halos, potentially allowing these scenarios to match the observed diversity of satellite profiles (e.g., \citealt{Yang:2022mxl}). 

Thus, the SIDM cross section is bounded from both above and below over certain velocity ranges, leaving only a finite range of viable parameter space open if elastic DM self-interactions significantly impact small-scale structure. \NEW{As suggested by Figure~\ref{fig:xsec}, there is tentative evidence that the strong lensing anomalies mentioned above, along with similar hints of dense low-mass substructure from stellar streams (e.g., \citealt{Bonaca181103631,Zhang:2024fib,Nibauer:2025ezn}), are consistent with a single underlying velocity-dependent SIDM cross section that reaches $\sim 100~\mathrm{cm^2~g}^{-1}$ at $v\sim 10~\mathrm{km\ s}^{-1}$ (also see \citealt{Yu:2025tmp}).}

Finally, we note that this review focuses on the case of elastic DM self-interactions; we do not attempt to summarize the growing body of literature on models beyond elastic SIDM, including inelastic~\citep{Vogelsberger:2018bok}, dissipative~\citep{Essig:2018pzq,Huo:2019yhk,Shen:2021frv,Gemmell:2023trd,Roy:2024bcu,Schmidt:2026hmn}, and multi-state interactions~\citep{ONeil:2022szc,Leonard:2024mqo,Yang:2025dgl,Yang:2025xsp,Patil:2025nmj}, models where only a fraction of the total DM budget interacts~\citep{Roberts:2025bad}, or the long-range limit of DM self-interactions, i.e., dark forces with a scalar mediator \citep{Graham:2025gtd}.

\subsection{Decaying Dark Matter}

The phenomenological models discussed above assume that DM particles are stable on cosmological timescales. However, many DM particle models predict that the DM component can decay to a lower-mass species in the dark sector~\citep{Pagels:1981ke,Weinberg:1982zq,Berezinsky:1991sp,Covi:1999ty,Feng:2003xh,Kaplinghat:2005sy,Allahverdi:2014bva}. We refer to models where the decay process is relevant for structure formation as decaying dark matter (DDM). 

DDM alters small-scale structure relative to CDM because decays inject heat into halos. The most direct effect of this process is to core and reduce the inner densities of DM halos~\citep{Peter:2010jy}. In turn, these alterations to halo density profiles affect subhalo evolution, resulting in suppression of subhalo abundances and maximum circular velocity functions relative to CDM~\citep{Wang:2014ina,DES:2022doi,Nadler:2025yni}.

Although many different microphysical models can effectively behave like DDM, we focus on constraints on two-body DDM to illustrate how small-scale structure data constrains these scenarios. In the two-body model, a ``parent'' DM particle $\chi$ decays to a lower-mass daughter $\chi'$ plus massless dark radiation $\gamma'$~\citep{Sanchez-Salcedo:2003pym,Strigari:2006jf,Peter:2010au}:
\begin{equation}
    \chi \rightarrow \chi' + \gamma'.
\end{equation}
In the two-body DDM model, decays impart momentum on the daughter particles with a ``kick velocity'' $V_{\mathrm{kick}}=\epsilon\,c$, where $\epsilon\equiv (m_{\chi}-m_{\chi'})/m_{\chi}$ is the mass splitting between the parent and daughter DM particles. DDM models with $\tau\sim 10~\mathrm{Gyr}$ and $V_{\mathrm{kick}}\sim 40~\mathrm{km\ s^{-1}}$ can significantly impact the evolution of halos with \NEW{peak masses} below $\approx 10^{10}~M_{\mathrm{\odot}}$~\citep{Nadler:2025yni}.
We note that decays directly to dark radiation (or to SM particles) are generally more tightly constrained than two-body decays; thus, the two-body DDM model represents a conservative extension to CDM that may have observable imprints on small-scale structure.

In summary, the primary observational signature of DDM is a reduction in the central densities of low-mass halos, which dynamically leads to a suppression of the DDM subhalo mass function, with the magnitude of these effects set by $\tau$ and $V_{\mathrm{kick}}$; the DDM field halo mass function can also be suppressed due to dynamical variations in the threshold for halo collapse in these models~\citep{Montandon:2026jyl}. Dwarf galaxy abundances (Section~\ref{sec:dwarf_abundances}) and profiles (Section~\ref{sec:dwarf_profiles}) exclude DDM lifetimes of $\tau\lesssim 20$--$30~\mathrm{Gyr}$ for $V_{\mathrm{kick}} \approx 20$ to $40~\mathrm{km\ s}^{-1}$ (e.g., \citealt{Peter:2010au,DES:2022doi}); the Lyman-$\alpha$ forest provides similar lifetime sensitivity at kick velocities of $300~\mathrm{km\ s}^{-1}\lesssim V_{\mathrm{kick}}\lesssim 1500~\mathrm{km\ s}^{-1}$ (e.g., \citealt{Wang:2013rha,Fuss:2022zyt}; Section~\ref{sec:lyman_alpha}). These results are complemented by precision probes on larger scales, including CMB, weak lensing, and galaxy clustering, which set more stringent lifetime limits at even larger kick velocities (e.g., \citealt{FrancoAbellan:2021sxk,Simon:2022ftd}).\footnote{\NEW{Some models that couple DM and dark energy yield a DM particle mass that evolves with time \citep[e.g.,][]{2006PhRvD..73h3509D}. 
In these scenarios, the comoving DM energy density is not conserved. Models motivated by recent DESI observations typically require the DM mass to increase at late times, in contrast to two-body DDM. In particular, such models can produce an apparent phantom crossing consistent with the preference for evolving dark energy found when combining CMB and DESI BAO data \citep{Khoury:2025txd,Khoury:2026svx,Chakraborty:2025syu}; however, the nonlinear small-scale structure signatures of these models remain largely unexplored.}}

\section{Established Probes}
\label{sec:constraints}

This section summarizes constraints on DM properties from well-established probes of small-scale structure: dwarf galaxies (Section~\ref{sec:dwarfs}), strong lensing (Section~\ref{sec:strong_lensing}), and the Lyman-$\alpha$ forest (Section~\ref{sec:lyman_alpha}).

Throughout, we quote both the numerical values of the DM constraints and (where available) the statistical interpretation of these results. However, we emphasize that statistical assumptions vary between studies, particularly for constraints on models that only approach CDM in a specific limit. For example, in Bayesian analyses that adopt a prior on $m_{\mathrm{WDM}}$ or related quantities, the range of the prior affects the resulting constraints (e.g., see the discussion in \citealt{Jethwa161207834}, \citealt{Gilman:2019vca}, and \citealt{Straight:2026wts}). Thus, care should be taken when directly comparing the results of different analyses and when interpreting the constraining power of various datasets based on derived limits alone.

\subsection{Dwarf Galaxies}
\label{sec:dwarfs}

Dwarf galaxies have been used to constrain a plethora of DM properties. The most competitive constraints on \emph{ab initio} deviations have generally been derived from the MW satellite population, which includes systems down to ultra-faint dwarfs~\citep{Simon190105465}. Ultra-faints drive the strongest constraints on $P(k)$ modifications because fainter galaxies generally occupy lower-mass halos, which probe smaller scales and are thus more sensitive to deviations from the CDM paradigm. As a result, the ability to accurately model the specific satellite population of the MW, \NEW{accounting for relevant features like the existence of the LMC}, is an important consideration for all such analyses. \NEW{This is currently achievable for only relatively small numbers of hosts with zoom-in simulations (e.g., \citealt{Buch:2024ssx}), while the generation of constrained semi-analytic merger trees is an area of ongoing work (e.g., \citealt{Nadler:2022kmy}). Thus, further theoretical development is needed to generate realizations of MW-like systems at the scale needed for precision DM inference.}

Meanwhile, tests of \emph{in situ} DM physics benefit from measurements of low-mass halo profiles over a wide halo mass range; the DM profiles of brighter dwarfs within and beyond the MW are useful for this purpose. This section reviews constraints derived from dwarf galaxy abundances (e.g., total number counts or luminosity functions) and density profiles, which respectively (but not exclusively) inform \emph{ab initio} and \emph{in situ} DM models.

\subsubsection{Abundances}
\label{sec:dwarf_abundances}

Constraints on DM physics using dwarf galaxy abundances generally focus on the MW satellite population. These constraints have been derived using one (or more) of the following approaches:
\begin{enumerate}[label=(\roman*)]
    \item Comparing the total number of observed (or completeness-corrected) MW satellites to the total number of predicted subhalos in a given DM model;
    \item Forward-modeling the observable properties (e.g.\ luminosities or stellar velocity dispersions) of MW satellites in a given DM model, and comparing to observations; 
    \item Comparing the completeness-corrected radial distribution of MW satellites to DM model predictions.
\end{enumerate}

\textbf{Thermal-relic Warm Dark Matter}: \citet{Maccio:2009isa} used the \textsc{Galform} SAM to constrain WDM, finding that $m_{\mathrm{WDM}}>1~\mathrm{keV}$ is required to avoid underpredicting the luminosity function of classical and Sloan Digital Sky Survey (SDSS) satellites. \citet{Kennedy:2013uta} used a similar method, model, and dataset to derive WDM constraints; these authors also explored the degeneracy between the MW host halo mass and WDM limit, showing that MW host halo masses $<1.4\times 10^{12}~M_{\mathrm{\odot}}$ imply a limit of $m_{\mathrm{WDM}}>3.3~\mathrm{keV}$. This degeneracy is due to the fact that higher-mass halos host more subhalos, allowing for more severe WDM suppression when matching a fixed count of observed satellites.

Analyses based on WDM simulations complemented these early semi-analytic results. \citet{Polisensky:2010rw} used WDM zoom-in simulations to compare the total number of WDM subhalos to classical and SDSS satellite abundances, yielding $m_{\mathrm{WDM}}>2.3~\mathrm{keV}$. \citet{Lovell:2013ola} conducted a similar analysis, finding $m_{\mathrm{WDM}}>1.5~\mathrm{keV}$. The difference between the \citet{Polisensky:2010rw} and \citet{Lovell:2013ola} limits is mainly due to the higher host halo mass used in the latter study \NEW{($\sim 2\times 10^{12}~M_{\mathrm{\odot}}$ rather than a range from $1.4$ to $2\times 10^{12}~M_{\mathrm{\odot}}$)}, which results in weaker WDM constraints. \NEW{Both studies' values of host halo mass are within the spread of current MW halo mass estimates (see \citealt{Wang:2019ubx}), although recent measurements generally favor the lower-mass end of this range.}

Recent studies have drawn on the strengths of both SAM and simulation-based analyses by applying empirical galaxy--halo connection models to subhalo populations from DM--only zoom-in simulations. \citet{Jethwa161207834} and \citet{Nadler:2019zrb} used this approach to derive WDM constraints from the classical-plus-SDSS satellite luminosity function. \NEW{Crucially, these studies propagated Poisson error in a way that simultaneously accounts for the uncertainty in both predicted and observed satellite counts, which is necessary to place robust constraints}. \citet{Jethwa161207834} and \citet{Nadler:2019zrb} respectively obtained $m_{\mathrm{WDM}}>2.9~\mathrm{keV}$ and $m_{\mathrm{WDM}}>3.3~\mathrm{keV}$ at $95\%$ confidence. Using Bayesian forward-modeling frameworks, both of these studies marginalized over uncertainties in the faint end of the galaxy--halo connection and the efficiency of satellite disruption by the MW disk. Thus, these studies simultaneously constrained the galaxy--halo connection and DM properties; this approach to marginalizing over theoretical uncertainties has become increasingly common in satellite population modeling (e.g., \citealt{Nadler:2018iux,DES:2019ltu,Danieli:2023,Liu:2025vhk}).

\citet{DES:2020fxi} used this approach to model the fainter and more abundant population of satellites detected by Pan-STARRS1 (PS1) and the Dark Energy Survey (DES), as presented in \citet{DES:2019vzn}. In particular, \citet{DES:2020fxi} applied an empirical galaxy--halo connection model to zoom-in simulations that resemble the MW by hosting an LMC-like subhalo with a realistic mass, orbit, current distance, and infall time. Their analysis yielded $m_{\mathrm{WDM}}>6.5~\mathrm{keV}$ at $95\%$ confidence after analytically marginalizing over the MW halo mass, substantially improving on the limits from \citet{Nadler:2019zrb} due to the improved sensitivity of PS1 and DES data, which was incorporated in both the data vector of observed satellites and via the observational selection function from \citet{DES:2019vzn}. These analyses also revealed the importance of the LMC system for modeling the full satellite population of the MW, and particularly its spatial anisotropy relative to the DES and PS1 survey footprints.

\citet{Nadler:2025fcv} updated this bound to $m_{\mathrm{WDM}}>5.9~\mathrm{keV}$ at $95\%$ confidence using the same framework but updated WDM transfer functions, and \citet{Liu:2025vhk} used the same framework but jointly modeled MW and M31 satellites to obtain $m_{\mathrm{WDM}}>6.22~\mathrm{keV}$ at $95\%$ confidence. Both of these analyses also used updated WDM SHMF fits from the COZMIC simulation suite, which agree well with previous WDM SHMFs from \citet{Lovell:2013ola}. Note that the transfer function that is ruled out in the \citet{Nadler:2025fcv} analysis is nearly identical to the \citet{DES:2020fxi} analysis; thus, the mapping from $m_{\mathrm{WDM}}$ to the transfer function drives this difference relative to the previous constraints (also see \citealt{Vogel:2022odl}). \NEW{For the purpose of the WDM constraint, the \citet{Nadler:2025fcv} limit therefore supersedes the \citet{DES:2020fxi} result.}

In parallel, \citet{Newton:2020cog} used the inferred total number of MW satellites from \citet{Newton:2018}, which was based on SDSS and DES observations and predictions for the underlying satellite radial distribution from simulations, to constrain WDM. By marginalizing over the MW halo mass, \citet{Newton:2020cog} derived $m_{\mathrm{WDM}}>2.02~\mathrm{keV}$ at $95\%$ confidence based on subhalo counts alone, which strengthened to $m_{\mathrm{WDM}}>3.99~\mathrm{keV}$ when modeling galaxy formation in low-mass subhalos using the \textsc{Galform} SAM. Meanwhile, \citet{Dekker:2021scf} used the SASHIMI SAM to derive WDM constraints based on the completeness-corrected total count of PS1 and DES satellites, finding $m_{\mathrm{WDM}}>3.6~\mathrm{keV}$ ($m_{\mathrm{WDM}}>5.1~\mathrm{keV}$) for $M_{\mathrm{MW}}=2\times 10^{12}~M_{\mathrm{\odot}}$ ($M_{\mathrm{MW}}=0.6\times 10^{12}~M_{\mathrm{\odot}}$) at $95\%$ confidence level (C.L.).

\textbf{Sterile Neutrino Warm Dark Matter}: Several studies have constrained specific non-thermal WDM production mechanisms using MW satellites, often focusing on sterile neutrinos.\footnote{Constraints on other non-thermal WDM-like models have also been derived using MW satellites (e.g., \citealt{Miller:2019pss,Das:2020nwc,Dvorkin:2020xga,Banerjee:2023utz,DEramo:2025jsb}).} In these scenarios, transfer functions differ in detail from thermal-relic WDM. Thus, these analyses typically rely on dedicated simulations or construct a mapping between thermal-relic WDM constraints and the non-thermal model of interest.

\citet{Polisensky:2010rw} derived limits on sterile neutrinos produced via the Dodelson--Widrow~\citep{Dodelson:1993je} and Shi--Fuller~\citep{Shi:1998km} mechanisms by analytically scaling their thermal-relic WDM constraints to match a characteristic cutoff scale in the transfer function. This yielded $m_s>13.3~\mathrm{keV}$ and $m_s>8.9~\mathrm{keV}$, respectively. \citet{Horiuchi:2013noa} and \citet{Cherry:2017dwu} then used extended Press--Schechter (ePS) theory to predict SHMFs using sterile neutrino transfer functions. The latter study derived constraints on resonantly-produced sterile neutrinos based on completeness-corrected SDSS satellite counts. These constraints are expressed in the sterile neutrino mixing angle vs.\ mass plane and conservatively yield $m_{s}>7~\mathrm{keV}$ at $95\%$ confidence. \citet{Dekker:2021scf} updated these constraints using the SASHIMI SAM and completeness-corrected PS1 and DES satellite counts, which yield a stronger limit of $m_{s}>11.6~\mathrm{keV}$ owing to the use of deeper data. \citet{Schneider:2016uqi} placed limits on resonantly-produced sterile neutrinos by comparing ePS predictions to the inferred total MW satellite count, finding that models with $m_s=7~\mathrm{keV}$ and $10~\mathrm{keV}$ are excluded for some mixing angles, but that $m_s=15~\mathrm{keV}$ is consistent with the data for all mixing angles.

Using the empirical galaxy--halo connection approach described above, \citet{DES:2020fxi} translated their thermal-relic WDM constraints to Dodelson--Widrow and Shi--Fuller sterile neutrino models. To do so, these authors required that $P(k)$ for a sterile neutrino model must be strictly more suppressed than the ruled-out thermal-relic $P(k)$ on all scales relevant for MW satellites; this procedure yields more conservative constraints than matching the half-mode scale of non-thermal and thermal WDM models. Thus, \citet{DES:2020fxi} derived $m_s>50~\mathrm{keV}$ for Dodelson--Widrow sterile neutrinos, while for Shi--Fuller production, their constraints yield $m_s \gtrsim 17~\mathrm{keV}$ when marginalized over mixing angle.
Their constraints in the mass--mixing angle plane strongly disfavored the interpretation of the $3.5~\mathrm{keV}$ X-ray line~\citep{Boyarsky:2014jta} as sterile neutrino WDM decay; when combined with X-ray limits, these results ruled out nearly the entire parameter space for Dodelson--Widrow sterile neutrinos (assuming that they comprise all of the DM), highlighting the complementarity of cosmological and indirect detection constraints. Using the same inference pipeline, \citet{An:2023mkf} placed constraints on sterile neutrinos produced in the presence of SM neutrino self-interactions, finding that the MW satellite population excludes heavy ($\gtrsim 1~\mathrm{GeV}$) mediator masses and yields $m_s>37~\mathrm{keV}$ at $95\%$ confidence.

Recently, \citet{Newton:2024jsy} derived sterile neutrino WDM constraints using the \textsc{Galform} SAM. These limits are generally less stringent than those in \citet{DES:2020fxi}; for example, they only mildly disfavor the sterile neutrino interpretation of the $3.5~\mathrm{keV}$ line. As discussed in \citet{Newton:2024jsy}, the primary driver of this discrepancy with \citet{DES:2020fxi} is the mismatch between sterile neutrino phase-space distributions produced by the \textsc{resonance-dm}~\citep{Ghiglieri:2015jua} and \textsc{sterile-dm}~\citep{Venumadhav:2015pla} codes, as described in \citet{Bodeker:2020hbo}; also see \citet{Alvey:2020xsk} and \citet{Lovell:2023olv}. The sterile neutrino WDM limits discussed above can also be relaxed in models with large lepton flavor asymmetries but zero total lepton asymmetry; in this case, substantially smaller mixing angles than usually allowed can produce the correct relic abundance, opening up new regions of parameter space for $m_s\lesssim 60~\mathrm{keV}$~\citep{Akita:2025txo}.

\textbf{Fractional Warm Dark Matter}: Models in which a fraction, $f_{\mathrm{WDM}}$, of the total DM budget is warm also affect the MW satellite population. In particular, these models partially suppress $P(k)$ and thus reduce predicted MW satellite abundances. Fractional WDM can therefore be constrained using the same methods described above for full WDM scenarios. Note that $P(k)$ and the SHMF do not completely cut off in fractional scenarios; as a result, constraints on fractional WDM models are more sensitive to the detailed shape of the SHMF suppression than constraints on full WDM.

\citet{Anderhalden:2012jc} used cosmological simulations of fractional thermal-relic WDM to derive $m_{\mathrm{WDM}}>0.1~\mathrm{keV}$ for $f_{\mathrm{WDM}}=0.2$ and $m_{\mathrm{WDM}}>0.3~\mathrm{keV}$ for $f_{\mathrm{WDM}}=0.5$ based on the completeness-corrected abundance of classical and SDSS satellites. These results thus demonstrated a degeneracy between $m_{\mathrm{WDM}}$ and $f_{\mathrm{WDM}}$. Next, \citet{Diamanti:2017xfo} used a semi-analytic description of fractional thermal-relic WDM to predict that models with $m_{\mathrm{WDM}}\lesssim 1~\mathrm{keV}$ for $f_{\mathrm{WDM}}\gtrsim0.25$ yield too few satellites to match total MW satellite counts.

\citet{Tan:2024cek} and \citet{An:2025gju} derived limits on fractional thermal-relic WDM by applying an empirical galaxy--halo connection model to zoom-in simulations. In particular, these authors combined fractional WDM transfer functions with the MW satellite inference framework from \citet{DES:2020fxi} to derive constraints using DES and PS1 satellite observations. The main methodological difference is that \citet{Tan:2024cek} used a SAM to predict SHMFs, while \citet{An:2025gju} used cosmological zoom-in simulations of fractional non-cold DM. These studies yielded similar constraints, with \citet{An:2025gju} deriving limits ranging from $m_{\mathrm{WDM}}>3.6~\mathrm{keV}$ (for $f_{\mathrm{WDM}}=0.5$) to $m_{\mathrm{WDM}}>5.4~\mathrm{keV}$ (for $f_{\mathrm{WDM}}=0.9$). All of these limits are stronger than the previous generation of fractional WDM constraints, consistent with the improvement for full WDM scenarios discussed above. Note that \citet{Tan:2024cek} also placed constraints on fractional Shi--Fuller sterile neutrino WDM.

\textbf{Fuzzy Dark Matter}: Like WDM, ultralight FDM also suppresses $P(k)$. However, this suppression results from a different physical mechanism (i.e., Jeans suppression vs.\ free-streaming) and thus has a different shape and scaling with underlying DM particle parameters compared to WDM. We now discuss the \emph{ab initio} impact of FDM on dwarf galaxy abundances, and we discuss constraints based on \emph{in situ} effects due to wave interference using dwarf galaxies in a later subsection.

\citet{Hu008506} originally argued that a $P(k)$ cutoff at $k=4.5~h~\mathrm{Mpc}^{-1}$ (which, at the time, was thought to alleviate the missing satellites problem) could be produced by FDM models with $m_{\mathrm{FDM}}\sim 10^{-22}~\mathrm{eV}$. \citet{Hui179504} presented a more detailed calculation of the minimum soliton mass capable of surviving repeated orbits around a larger host, finding that $m_{\mathrm{FDM}}=10^{-22}~\mathrm{eV}$ corresponds to a minimum surviving soliton mass of $\approx 7\times 10^8~M_{\mathrm{\odot}}$, consistent with the mass function suppression seen in FDM simulations~\citep{Schive:2014dra}. However, modern analyses of the MW satellite population significantly strengthened these bounds. For example, \citet{Nadler:2019zrb} translated their MW satellite constraints---derived from forward-modeling classical and SDSS MW satellites---to FDM by matching $k_{\mathrm{hm}}$ for FDM models to the ruled-out $3.3~\mathrm{keV}$ WDM model from their analysis. This procedure yielded $m_{\mathrm{FDM}}>2.9\times 10^{-21}~\mathrm{eV}$ at $95\%$ confidence.\footnote{\citet{Marsh:2018zyw} derived $m_{\mathrm{FDM}}\gtrsim 8\times 10^{-22}~\mathrm{eV}$ based on predictions for the SHMF and the existence of the dwarf galaxy Eridanus II, noting that a comparable bound is obtained regardless of whether infall or stripped mass DM halo mass is used (however, also see the discussion in \citealt{Chiang:2021uvt}).}

\citet{Schutz:2020jox} subsequently noted that differences between the shapes of the WDM and FDM transfer functions may be important and was neglected from approaches that relied purely on matching the half-mode scale. 
Thus, it is preferable to use SHMFs derived using the FDM transfer function to derive constraints. However, existing studies reach different conclusions about the FDM SHMF. For example, the commonly-used \citet{Schive:2015kza} FDM SHMF fitting function differs substantially from semi-analytic predictions~\citep{Du:2018unj}. \citet{DES:2020fxi} integrated both of these SHMFs into their MW satellite forward model to derive $m_{\mathrm{FDM}}>2.9\times 10^{-21}~\mathrm{eV}$ based on DES and PS1 satellite data using the \citet{Du:2018unj} SHMF fit. Notably, this result did not strengthen the limit from \citet{Nadler:2019zrb}, despite using more sensitive data, because the \citet{Du:2018unj} model predicts fairly mild SHMF suppression. On the other hand, \citet{DES:2020fxi} showed that the same analysis yields $m_{\mathrm{FDM}}>9.1\times 10^{-21}~\mathrm{eV}$ when using the \citet{Schive:2015kza} FDM SHMF fit.

Next, \citet{Nadler:2025fcv} simulated FDM models as part of the COZMIC N-body zoom-in simulation suite. These simulations evolved FDM transfer functions using DM--only simulations, which is a reasonable approximation for the purposes of the SHMF because $k_{\mathrm{hm}}<k_J$ for the FDM models of interest (i.e., the SHMF suppression is set on scales larger than the FDM Jeans length); this conclusion is supported by full Schrödinger-Poisson simulations~\citep{May:2021wwp,May:2022gus}. \citet{Nadler:2025fcv} found that the FDM SHMF is significantly \emph{more} suppressed at low subhalo masses compared to previous mass function fits, and thus derived $m_{\mathrm{FDM}}>1.4\times 10^{-20}~\mathrm{eV}$ at $95\%$ confidence using the \cite{DES:2020fxi} modeling framework. \NEW{The COZMIC suite from \citet{Nadler:2025fcv} currently provides the most direct prediction for the SHMF based on $P(k)$ suppression in FDM, and the resulting limit supersedes earlier analyses based on semi-analytic fitting functions. Full Schrödinger--Poisson simulations of the low-mass MW subhalo populations will be needed to go beyond these conservative limits.}

\citet{Crumrine:2026rnp} extended the above constraints to fractional FDM models, finding $m_{\mathrm{FDM}}>9.1\times 10^{-21}~\mathrm{eV}$ for a $50\%$ FDM component, and that current MW satellite abundance data do not constrain lower FDM fractions, except in scenarios with extreme $P(k)$ suppression on large scales. Meanwhile, \citet{Liu:2025vhk} obtained $m_{\mathrm{FDM}}>1.75\times 10^{-20}~\mathrm{eV}$ at $95\%$ confidence by jointly fitting MW and M31 satellites in a similar modeling framework. Finally, \citet{Nadler:2026wtt} analyzed ultralight DM produced with a field power spectrum peaked at a subhorizon wavenumber $k_*$, finding that MW satellite abundances yield a lower limit on the DM particle mass of $6\times 10^{-18}~\mathrm{eV}\times (k_*/10^4~\mathrm{Mpc}^{-1})$ for $k_*>10^4~\mathrm{Mpc}^{-1}$ at $95\%$ confidence in this scenario.

\textbf{Interacting Dark Matter}: Constraints on DM--baryon scattering IDM using MW satellite have been derived by leveraging WDM limits. For example, \citet{Nadler:2019zrb} considered velocity-independent DM--proton interactions and showed that transfer functions in these models are nearly identical to thermal-relic WDM. Thus, they derived IDM constraints by matching the half-mode scale of their ruled-out $3.3~\mathrm{keV}$ thermal-relic WDM model based on forward modeling classical and SDSS satellites. This yielded $\sigma_{\mathrm{DM}-p}<8\times 10^{-28}~\mathrm{cm}^2$ for $m_{\mathrm{DM}}=1~\mathrm{GeV}$. This constraint scales linearly with $m_\mathrm{DM}$ for $m_{\mathrm{DM}}\gg 1~\mathrm{GeV}$, and as $m_{\mathrm{DM}}^{1/4}$ for $m_{\mathrm{DM}}\ll 1~\mathrm{GeV}$. \citet{DES:2020fxi} updated this constraint based on their DES and PS1 satellite analysis by mapping $k_{\mathrm{hm}}$ to their newly-constrained $6.5~\mathrm{keV}$ WDM model. This yielded $\sigma_{\mathrm{DM}-p}<2.7\times 10^{-28}~\mathrm{cm}^2$ for $m_{\mathrm{DM}}=1~\mathrm{GeV}$ and a similar scaling with $m_{\mathrm{DM}}$ compared to \citet{Nadler:2019zrb}, improving on the previous limits.

Although velocity-independent DM--baryon scattering yields transfer functions that are very similar to thermal-relic WDM, IDM transfer functions generally feature DAOs due to the coupling between DM and SM particles at early times. \citet{Maamari:2020aqz} constrained velocity-dependent IDM models based on the \citet{DES:2020fxi} WDM constraints by requiring that DM--proton scattering transfer functions are strictly more suppressed than the reference WDM model up to a critical wavenumber probed by MW satellites. This procedure conservatively ensured that DAOs do not affect the abundance of subhalos that host satellite galaxies and yielded upper limits of $\sigma_{\mathrm{DM}-p}<1.5\times 10^{-21}$, $5.5\times 10^{-16}$, and $6.2\times 10^{-10}~\mathrm{cm}^2$ for $n=2$, $4$, and $6$, respectively, for $m_{\mathrm{DM}}=1~\mathrm{GeV}$. \citet{Nguyen:2021cnb} extended this method to DM--electron scattering to obtain $\sigma_{\mathrm{DM}-e}<6.2\times 10^{-24}$, $3.9\times 10^{-21}$, and $3.0\times 10^{-18}~\mathrm{cm}^2$ for $n=2$, $4$, and $6$, respectively, for $m_{\mathrm{DM}}=1~\mathrm{GeV}$. The scaling of these limits with $m_{\mathrm{DM}}$ varies with $n$ and differs from the scaling obtained using half mode-matching. \citet{Crumrine:2026rnp} extended these cross section constraints to fractional DM--baryon scattering scenarios for IDM fractions down to $50\%$.

Meanwhile, \citet{Buen-Abad:2021mvc} derived constraints on DM--proton and electron-scattering IDM for $n=0$, $-2$, and $-4$ by mapping to the MW satellite WDM limits from \citet{DES:2020fxi} and using both transfer function constraint techniques. Their $n=0$ bounds are similar to the previous studies, while the constraints weaken relative to limits derived from other cosmic probes for $n=-2$ and $-4$. This occurs because the shape of the $P(k)$ suppression becomes flatter with decreasing $n$, leading LSS probes like the CMB to place more stringent limits than MW satellites for $n=-4$.

For DM--radiation scattering, \citet{Boehm:2014vja} performed cosmological N-body zoom-in simulations using IDM transfer functions for DM--photon and neutrino scattering models, which have transfer functions similar to thermal-relic WDM. By comparing the number of predicted subhalos to the completeness-corrected count of classical and SDSS satellites, they derived $\sigma_{\mathrm{DM}-\gamma}<5.5\times 10^{-9}\sigma_T\left(m_{\mathrm{DM}}/1~\mathrm{GeV}\right)$ at $2\sigma$, where $\sigma_T$ is the Thompson scattering cross section. In a follow-up study, \citet{Schewtschenko:2014fca} showed that halo concentrations and maximum circular velocities are reduced due to $P(k)$ suppression in DM--radiation scattering scenarios; note that this is also an \emph{in situ} effect.  \citet{Escudero:2018thh} then derived bounds on DM--photon scattering IDM by requiring that the total number of MW subhalos that host satellites exceeds the number of completeness-corrected DES and SDSS satellites derived in \citet{Newton:2018}. To predict the SHMF, these authors mapped IDM models to effective WDM models by matching their half-mode scales. This procedure yielded $\sigma_{\mathrm{DM}-\gamma}<8\times 10^{-10}\sigma_T\left(m_{\mathrm{DM}}/1~\mathrm{GeV}\right)$ at $95\%$ confidence.

\citet{Akita:2023yga} constrained DM--neutrino interactions based on the completeness-corrected count of DES and PS1 satellites from \citet{DES:2019vzn}, using a SAM to predict the SHMF. In particular, they derived a conservative limit of $\sigma_{\mathrm{DM}-\nu}<10^{-32}~\mathrm{cm}^2\left(m_{\mathrm{DM}}/1~\mathrm{GeV}\right)$ for energy-independent scattering, which strengthens when imposing an extra condition on the subhalos that are able to form satellite galaxies. These authors also derived constraints on energy-dependent scattering with power-law exponents of $n=2$ and $4$, as well as on several specific particle physics models that feature DM--neutrino interactions.

Recently, \citet{Crumrine:2024sdn} constrained both DM--neutrino and DM--photon scattering by requiring that their transfer functions are strictly more suppressed than the ruled-out WDM model from the \citet{DES:2020fxi} DES and PS1 satellite analysis. They derived $\sigma_{\mathrm{DM}-\gamma}<1.98\times 10^{-38}~\mathrm{cm}^2$ and $\sigma_{\mathrm{DM}-\nu}<3.16\times 10^{-38}~\mathrm{cm}^2$ for $m_{\mathrm{DM}}=1~\mathrm{MeV}$. These constraints scale linearly with $m_{\mathrm{DM}}$ for $m_{\mathrm{DM}}\gg 1~\mathrm{MeV}$ and strengthen at lower $m_{\mathrm{DM}}$ due to the effects of DM sound speed on linear matter perturbations. In addition, these authors showed that the photon and neutrino--scattering constraints are related by a constant factor at high $m_{\mathrm{DM}}$ derived from the ratio between neutrino and photon energy densities. \citet{Crumrine:2026rnp} considered fractional DM--radiation scattering scenarios in a similar framework, again extending cross section limits down to IDM fractions of $50\%$.

\subsubsection{Dynamics}
\label{sec:dwarf_profiles}

Constraints on DM physics from dwarf galaxy dynamics often use one of the following techniques:
\begin{enumerate}[label=(\roman*)]
    \item Comparing the stellar kinematics and sizes (and hence enclosed \NEW{dynamical masses} or densities) of specific dwarfs to theoretical predictions;
    \item Comparing the same quantities as (i), but on a population level; 
    \item Comparing the number or luminosity function of surviving satellite galaxies to theoretical predictions.
\end{enumerate}

\textbf{Warm Dark Matter}: \citet{Tremaine:1979we} noted that the primordial velocity dispersion of fermionic DM particles places an upper limit on the phase-space densities of DM halos today. Thus, the observed central densities and velocities dispersions of dwarf galaxies place lower limits on the mass of fermionic WDM. \citet{Dalcanton:2000hn} used this argument and the properties of classical MW satellites to derive $m_{\mathrm{WDM}}\gtrsim 0.7~\mathrm{keV}$. 

Subsequent work refined this limit and extended it to sterile neutrino WDM. For example,  \citet{Boyarsky:2008ju} used the properties of SDSS-discovered satellites to derive $m_{\mathrm{WDM}}>0.48~\mathrm{keV}$, and they translated this thermal-relic bound to several sterile neutrino production scenarios (also see \citealt{Gorbunov:2008ka}). \citet{Alvey:2020xsk} then refined these analyses to obtain $m_{\mathrm{WDM}}>0.41~\mathrm{keV}$ at $95\%$ C.L. Finally, \citet{Bezrukov:2025ttd} derived complementary limits of $m_{\mathrm{WDM}}>0.28$ or $0.49~\mathrm{keV}$ at $95\%$ C.L. depending on the approach used to infer dwarfs' phase-space densities.

Beyond using the observed central phase-space densities of individual dwarfs to constrain WDM, \citet{Kim:2021zzw} showed that the velocity dispersion function of MW satellites probes WDM due to the reduction of halo concentrations in these models, which is ultimately determined by $P(k)$ suppression. These authors found that the completeness-corrected velocity dispersion function can only be reproduced in models with $m_{\mathrm{WDM}}\gtrsim 6~\mathrm{keV}$. As noted by \citet{Delos:2023exh}, there is a competing effect in WDM, since its large free-streaming scale is expected to lead to the formation of sizable prompt cusps at halo centers, increasing predicted stellar velocity dispersions. \citet{Delos:2025ees} then implemented this effect in a SAM of MW satellite galaxies and showed that WDM can overpredict the stellar velocity dispersions of the faintest and most compact MW satellites, leading to a constraint of $m_{\mathrm{WDM}}>5.8~\mathrm{keV}$ at $95\%$ confidence.

\citet{Maccio:2012qf} presented a general ``catch-22'' argument regarding \emph{in situ} WDM constraints from MW satellites. In particular, these authors argued that WDM models which appreciably alter dwarfs' inner halo properties must severely suppress the abundance of these systems to begin with, implying that the satellite luminosity function (which probes the \emph{ab initio} effects of WDM) is generally more constraining than dwarfs' phase-space densities (which probe \emph{in situ} effects). This is consistent with the fact that current WDM constraints from dwarf galaxy abundances, discussed above, are typically stronger than constraints from dwarf dynamics summarized in this section. In this context, using the velocity dispersion function (following \citealt{Kim:2021zzw}) combines these types of constraints and should yield even stronger limits on WDM and similar models that suppress (or enhance) $P(k)$. \citet{Esteban:2023xpk} applied this idea to constrain DM models that suppress or enhance small-scale primordial curvature power spectra, and \citet{Dekker:2024nkb} placed limits on blue- and red-tilted primordial power spectra; both studies used SAMs to connect $P(k)$ to predictions for MW satellite galaxy and halo properties.

\NEW{We note that the observational selection effects involved in measuring the velocity dispersions of the MW satellite population are less well characterized than those associated with abundance measurements, which could lead to biases in the above results. This difference largely arises because velocity dispersions are measured on an individual basis, such that observational sensitivities and systematics vary system-by-system, while satellite abundances can be measured homogeneously by wide-area photometric surveys. As a result, DM constraints from satellite abundance measurements are currently more robust to observational systematics than those from velocity dispersions-based measurements. Future spectroscopic surveys with well-characterized completeness will be needed to place dynamical limits on an equal footing; we discuss relevant facilities in Section~\ref{sec:observation_future}.}

\textbf{Fuzzy Dark Matter}: Solitonic cores at the centers of FDM (sub)halos alter the inner density profiles of dwarf galaxies relative to CDM predictions, and this signature has been used to constrain FDM. In particular, \citet{Marsh:2015wka} compared analytic predictions for FDM halo density profiles to the cored central density profiles inferred for the Fornax and Sculptor dwarf galaxies. This yielded a preference for models with $m_{\mathrm{FDM}}\approx 6\times 10^{-23}~\mathrm{eV}$ over CDM.\footnote{Also see \citet{Schive:2014dra}, who compared a different FDM soliton model to kinematic data for a stellar subcomponent of Fornax, yielding a comparable best-fit value of $m_{\mathrm{FDM}}$.} Meanwhile, \citet{Gonzalez-Morales:2016yaf} derived an upper limit on $m_{\mathrm{FDM}}$ by analyzing the luminosity-averaged velocity dispersion of the individual chemodynamical components of Fornax and Sculptor, finding $m_{\mathrm{FDM}}< 0.4\times 10^{-22}~\mathrm{eV}$ at $97.5\%$ confidence. A subsequent analysis by \citet{Chen:2016unw} using eight classical satellites yielded a slightly weaker upper limit, with $m_{\mathrm{FDM}}<2.49\times 10^{-22}~\mathrm{eV}$ at the $2\sigma$ level based on kinematic data from \citet{Walker:2009zp}. Note that the effects of a central soliton on dwarf density profiles are partially degenerate with baryonic feedback, which can also produced cored central density profiles~\citep{Read:2018pft}. The above analyses do not model such baryonic effects; doing so would allow for higher FDM masses to fit the data.

As noted by \citet{Marsh:2015wka}, the upper limits of $m_{\mathrm{FDM}}\lesssim 10^{-22}~\mathrm{eV}$ discussed above are in tension with structure formation bounds. For example, FDM limits from MW satellites abundances strongly disfavor $m_{\mathrm{FDM}}<10^{-22}~\mathrm{eV}$. Population analyses showed that a related tension appears when considering the full range of MW satellites' inner densities. In particular, UFDs are generally found to occupy subhalos with relatively high density profile amplitudes compared to classical satellites~\citep{Kaplinghat:2019svz}. For example, \citet{Safarzadeh:2019sre} derived $m_{\mathrm{FDM}}\gtrsim 10^{-21}~\mathrm{eV}$ by comparing analytic soliton profiles to the enclosed \NEW{dynamical masses} that have been measured for UFDs. This is incompatible with the upper limits on $m_{\mathrm{FDM}}$ from classical satellites discussed in the previous paragraph. \citet{Zimmermann:2024xvd} then found $m_{\mathrm{FDM}}>2.2\times 10^{-21}~\mathrm{eV}$ at $95\%$ confidence by comparing an ensemble of reconstructed FDM wavefunctions to density profiles inferred from a Jeans analysis of the Leo~II dwarf spheroidal galaxy. More recently, \citet{Benito:2025xuh} derived $m_{\mathrm{FDM}}>2.2\times 10^{-21}~\mathrm{eV}$ at $95\%$ confidence by comparing analytically-generated soliton profiles to the profiles of 27 UFDs inferred using Eddington inversion.

In related work, \citet{Hayashi:2021xxu} reported $m_{\mathrm{FDM}}=1.1^{+8.3}_{-0.7}\times 10^{-19}~\mathrm{eV}$ based on a Jeans analysis of Segue 1, and \citet{Wardana:2026} extended FDM analyses of dwarf profiles to eight MW satellites, finding that only $-20.3<\log(m_{\mathrm{FDM}}/\mathrm{eV})<-19.2$ and $-22.1<\log(m_{\mathrm{FDM}}/\mathrm{eV})<-21.5$ are consistent with data. When interpreted as lower limits, these results imply even stronger bounds on $m_{\mathrm{FDM}}$ than the UFD limits discussed in the previous paragraphs. Most recently, \citet{Zhao:2026son} placed bounds on self-interacting FDM using similar techniques, finding lower limits on $m_{\mathrm{FDM}}$ ranging from $10^{-22}$ to $10^{-21}~\mathrm{eV}$ when self-interactions are weak by analyzing Leo II. The profiles of low surface brightness galaxies also offer insights into FDM; for example, \citet{Montes:2023ahn} found that the cored profile of the extremely diffuse SMC-mass galaxy Nube is well fit by an FDM model with $m_{\mathrm{FDM}}=0.8^{+0.4}_{-0.2}\times 10^{-23}~\mathrm{eV}$. Constraints from more massive dwarfs like Nube often differ by orders of magnitude from those set by the faintest dwarfs because solitonic cores grow with decreasing halo mass. \NEW{Taken together, despite different modeling assumptions and treatments of the data, recent studies generally imply that FDM must be heavier than $\sim 10^{-21}~\mathrm{eV}$ to reproduce the observed diversity of dwarf galaxy (and particularly MW satellite) inner densities.}

Complementary FDM constraints have been placed based on dynamical heating. In particular, density fluctuations in the FDM field can inject energy into the stellar populations of dwarf galaxies, causing their orbits to migrate outward and expanding their sizes.\footnote{Note that both tidal stripping~\citep{Errani:2022aru} and subhalo-driven heating~\citep{Penarrubia:2026} can lead to qualitatively similar stellar size evolution in CDM, signaling a potential degeneracy between FDM and these effects.} This effect also applies to subcomponents of dwarf galaxies (e.g., star clusters). Thus, \citet{Marsh:2018zyw} analyzed the central star cluster in Eridanus II to place a limit of $m_{\mathrm{FDM}}>0.6\times 10^{-19}~\mathrm{eV}$, based on the lack of observed stellar migration and an analytic calculation of FDM gravitational heating. Meanwhile, \citet{Teodori:2025rul} used tailored FDM simulations of Fornax, Carina, and Leo II (including tidal stripping) to show that FDM masses $5\times 10^{-22}~\mathrm{eV}<m_{\mathrm{FDM}}<5\times 10^{-21}~\mathrm{eV}$ are in tension with these dwarfs' observed half-light radii and velocity dispersions; \citet{Caputo:2026wmy} extended this analysis to include Leo I and Draco and showed that the same FDM mass range is disfavored when stellar self-gravity is also modeled. Considering lower-mass dwarfs, \citet{Dalal:2022rmp} used the observed sizes and velocity dispersions of Segue 1 and Segue 2 to derive $m_{\mathrm{FDM}}>3\times 10^{-19}~\mathrm{eV}$ using a quasiparticle-based calculation of FDM heating. \citet{May:2025ppj} extended this analysis to show that, if the Ursa Major III/UNIONS 1 system is a dwarf galaxy, the \citet{Dalal:2022rmp} bound strengthens to $m_{\mathrm{FDM}}>8\times 10^{-18}~\mathrm{eV}$.

Several potential caveats to these dynamical heating-based FDM bounds have been raised. In the case of Eridanus II, \citet{Chiang:2021uvt} claimed that constraints should be substantially weakened or nullified because the soliton oscillation timescale is larger than the orbital timescale of stars within the cluster, rendering FDM heating negligible. Other studies found that effects like tidal stripping or stellar self-gravity can reduce the FDM heating effect, which may weaken $m_{\mathrm{FDM}}$ bounds from both star clusters and entire dwarf galaxies (e.g., \citealt{Schive:2019rrw,DuttaChowdhury:2023qxg,Yang:2025bae,Caputo:2026wmy,Liu:2026uwd}; however, also see \citealt{May:2025ppj}). In the specific case of Ursa Major III/UNIONS 1, \citet{Cerny:2025} found that this system is likely a star cluster, which would nullify the constraint in \citet{May:2025ppj} from this system.

\emph{In situ} FDM physics can also impact dwarf satellite abundances; this signature has been explored in simulations but has not been used as often as dwarf dynamics to place FDM constraints. For example, \citet{Du:2018qor} found that satellites orbiting close to the center of the MW can only survive if $m_{\mathrm{FDM}}\gtrsim 2\times 10^{-21}~\mathrm{eV}$ by modeling the mass-loss rates of FDM solitons. Soliton disruption can also be accelerated by extensions of the FDM model that include repulsive self-interactions~\citep{Glennon:2022huu}, which would strengthen lower bounds on $m_{\mathrm{FDM}}$, and vice versa for models that include attractive self-interactions.

\textbf{Self-interacting Dark Matter}: Dwarf galaxy constraints on SIDM have been derived from distant, bright dwarfs---both at the population level (e.g., \citealt{Kaplinghat:2015aga,Kamada:2016euw,Ren:2018jpt,Roberts:2024uyw,Jia:2026ocr}) and using individual galaxies (e.g., \citealt{Leung:2020voe,ManceraPina:2024ybj,Zhang:2025bju})---and from faint, nearby dwarfs, and particularly MW satellites. This discussion focuses on the latter results; we refer the reader to \citet{Tulin:2017ara} and \citet{Adhikari:2022sbh} for broader reviews of SIDM limits from dwarf galaxies.

For low-amplitude cross sections, SIDM predicts that dwarfs inhabit cored DM halos, at least in systems where the stellar-to-halo mass ratio is small enough to prevent significant adiabatic contraction (e.g., \citealt{Vogelsberger:2012ku,Kaplinghat:2013xca,Elbert:2016dbb,Jiang:2022aqw}).\footnote{Note that baryonic feedback can also create cores in classical and bright dwarf galaxies; breaking this degeneracy requires more detailed density profile measurements and/or additional observables~\citep{BurgerSIDM,Straight:2025udg}.} Thus, the existence of cuspy dwarfs constrains these models. For example, \citet{Read:2018pft} combined Jeans modeling of the dwarf satellite Draco, which implies that it inhabits a cuspy DM halo, with analytic predictions for SIDM core sizes to derive $\sigma/m<0.57~\mathrm{cm}^2~\mathrm{g}^{-1}$ at $99\%$ confidence (also see \citealt{Valli:2017ktb}). This analysis effectively constrains the SIDM cross section at Draco's internal velocity scale ($\sim 20~\mathrm{km\ s}^{-1}$; \citealt{Robles:2020gqr}) and applies in the absence of gravothermal core collapse~\citep{Zavala:2019sjk,Sameie:2019zfo}.

Subsequent work extended these constraints to the population level. \citet{Hayashi:2020syu} combined UFD density profiles inferred from Jeans modeling with analytic predictions for SIDM core sizes to derive $\sigma/m<0.086~\mathrm{cm}^2~\mathrm{g}^{-1}$ and $\sigma/m<0.39~\mathrm{cm}^2~\mathrm{g}^{-1}$ using Segue 1 and Willman 1, respectively. Then, \citet{Silverman:2022bhs} compared subhalo density profiles from cosmological SIDM N-body zoom-in simulations to the densities of classical and ultra-faint satellites evaluated at their half-light radii, finding that velocity-independent cross sections of both $1$ and $5~\mathrm{cm}^2~\mathrm{g}^{-1}$ reduce satellites' inner densities too severely to be consistent with the data. More recently, \citet{Almeida:2025cee} derived $0.3~\mathrm{cm^2~g}^{-1}<\sigma/m<200~\mathrm{cm^2~g}^{-1}$ at $v\approx 5~\mathrm{km\ s}^{-1}$ by fitting SIDM models to the DM density profiles of UFDs as inferred from their stellar density profiles.

MW satellites' inner densities and pericentric distances have also been used to constrain SIDM. In particular, \citet{Kaplinghat:2019svz} and \citet{Andrade:2023fgr} found that MW satellites' inner densities, evaluated at $150~\mathrm{pc}$ from the center of each system, anticorrelate with their pericentric distances. These studies found that high-resolution CDM N-body simulations do not reproduce this anticorrelation (however, see \citealt{Cardona-Barrero:2023fwb} and \citealt{Kravtsov:2023oxa}). \citet{Ebisu:2021bjh} analyzed this anticorrelation using cosmological SIDM N-body zoom-in simulations to conclude that velocity-independent cross sections with $\sigma/m<3~\mathrm{cm}^2~\mathrm{g}^{-1}$ are favored in the absence of core collapse. In addition, \citet{Kim:2021zzw} found $\sigma/m\lesssim 0.1~\mathrm{cm}^2~\mathrm{g}^{-1}$ by comparing MW satellites' predicted velocity dispersion function without core collapse to completeness-corrected data.

These SIDM constraints have been extended to high-amplitude, velocity-dependent cross sections that yield both core-forming and core-collapsed subhalos. Specifically, \citet{Correa:2020qam} compared SIDM halo density profiles predicted from a gravothermal evolution model to the inferred inner densities of MW satellites. \citet{Correa:2020qam} concluded that classical satellites like Carina and Fornax prefer cross sections of $\sim 30$ to $50~\mathrm{cm}^2~\mathrm{g}^{-1}$, while UFDs prefer values up to $100~\mathrm{cm}^2~\mathrm{g}^{-1}$; in this way, they constrained the parameters of a velocity-dependent SIDM model described by a Yukawa potential. \citet{Slone:2021nqd} then placed complementary constraints by comparing semi-analytic predictions for both core-collapse and evaporation timescales to observed satellites' inner densities and orbital properties. This yielded constraints on both the amplitude and turnover scale of the velocity-dependent cross section in Equation~\ref{eq:xsec}, with $\sigma_0/m\gtrsim 40~\mathrm{cm}^2~\mathrm{g}^{-1}$ and $w\lesssim 200~\mathrm{km\ s}^{-1}$; \citet{Ando:2025qtz} performed a complementary analysis, reaching similar conclusions, while \citet{Fischer:2026ryr} found that most MW satellites need to be (mildly) core-collapsed to explain their observed inner densities, consistent with the simulation predictions from \citet{Yang:2022mxl}. These results imply that only a finite range of velocity-dependent SIDM parameter space remains viable if elastic DM self-interactions significantly impact small-scale structure; simulations of these viable cross sections produce subhalo populations that are broadly consistent with MW satellite data~\citep{Turner:2020vlf, Lovell:2022vzx,Yang:2022hkm,Nadler:2025jwh,Nadler:2025wra}.

Complementary SIDM constraints have been derived from MW satellite population analyses. In SIDM, subhalo--host halo interactions lead to DM evaporation, which accelerates subhalo disruption, reduces the abundance of surviving subhalos~\citep{Nadler:2020ulu}, and affects gravothermal evolution~\citep{Zeng:2021ldo,Klemmer:2026ivv}. For example, using cosmological SIDM zoom-in simulations, \citet{Vogelsberger:2012ku} found that a velocity-independent cross section of $\sigma/m=10~\mathrm{cm}^2~\mathrm{g}^{-1}$ is disfavored based on these effects. In addition, both \citet{Vogelsberger:2012ku} and \citet{Zavala:2012us} found that both velocity-independent cross sections of $\sim 1~\mathrm{cm}^2~\mathrm{g}^{-1}$ and velocity-dependent cross sections with amplitudes between $\sim 10$ to $100~\mathrm{cm}^2~\mathrm{g}^{-1}$ are compatible with observed satellite abundances.

\citet{Dooley:2016ajo} found that these viable velocity-dependent cross sections predict substantial stellar stripping in dwarf galaxies. Note that the velocity scale of the cross section probed by evaporation is determined by the orbital velocities of satellites with respect to the host, which \citet{Nadler:2020ulu} demonstrated using cosmological zoom-in simulations of velocity-dependent SIDM models. Several recent controlled and cosmological simulations have studied this effect in the context of high-amplitude velocity-dependent cross sections that yield core collapse (e.g., \citealt{Zeng:2021ldo,Kong:2025kkt,Nadler:2025jwh,Klemmer:2026ivv}).

Finally, individual dwarf galaxies with anomalous stellar velocity dispersions given their size or luminosity can be used to probe DM. For example, \citet{Borukhovetskaya:2021ahz} showed that Crater II has an extremely low velocity dispersion given its surface brightness, which is difficult to reproduce in CDM; on the other hand, SIDM can naturally lead to this configuration~\citep{Zhang:2024ggu}. Recent data also indicates that some UFDs may have lower inner densities than expected in CDM~(e.g., \citealt{Nguyen:2026gap,Li:2026ode}). This could result from significant tidal disruption or that alternative DM physics (e.g., self-interactions) produces cores in low-mass halos; the subsequent tidal stripping and disruption of cored systems is much more rapid than in CDM (e.g., \citealt{Errani:2022aru}).

\textbf{Decaying Dark Matter}: DDM reduces (sub)halos' central densities, lowering the enclosed \NEW{dynamical masses} in the inner regions of dwarf galaxies. Leveraging this effect, \citet{Peter:2010sz} constrained the DDM lifetime to $\tau>30~\mathrm{Gyr}$ by analyzing the \NEW{dynamical masses} of SDSS MW satellites, for DDM kick velocities spanning $20$ to $200~\mathrm{km\ s}^{-1}$. \citet{Wang:2014ina} then performed cosmological simulations to show that allowed DDM models can potentially alleviate the ``too-big-to-fail'' problem for MW satellites~\citep{Boylan-Kolchin:2011qkt}.

\citet{DES:2022doi} used cosmological DDM zoom-in simulations to predict the suppression of the SHMF in these models. By incorporating this suppression into the MW satellite modeling framework from \citet{DES:2020fxi}, they derived DDM constraints based on a Bayesian analysis of the MW satellite population detected by DES and PS1. At $95\%$ confidence, these authors inferred DDM lifetime limits of $\tau>18~\mathrm{Gyr}$ ($>29~\mathrm{Gyr}$) for kick velocities of $V_{\mathrm{kick}}=20~\mathrm{km\ s}^{-1}$ ($40~\mathrm{km\ s}^{-1}$) at $95\%$ confidence. This procedure assumed that subhalo disruption in the cosmological DDM simulations corresponds to complete disruption of satellites' stellar components; further work (e.g., using hydrodynamic simulations) is needed to characterize DDM satellite disruption in detail.

\subsubsection{Summary}

\NEW{Dwarf galaxy abundances currently provide among the most stringent constraints on \emph{ab initio} DM physics that suppresses small-scale structure. For thermal-relic WDM, the most robust limits are those derived from MW satellite population forward-modeling frameworks that marginalize over galaxy--halo connection uncertainties and the MW halo mass; these analyses consistently yield $m_{\rm WDM} \gtrsim 6~{\rm keV}$ and constrain a variety of other models that suppress small-scale power in a qualitatively similar fashion to WDM (e.g., DM--proton scattering IDM). Primary challenges for such analyses as they work toward $P(k)$ measurements (rather than conservative lower limits) include efficiently generating realizations of MW-like systems, folding in dwarf abundance data from satellite systems beyond the MW and field surveys, and reliably modeling the details of subhalo and satellite disruption.}

\NEW{WDM constraints from dwarf dynamics are complementary to but generally weaker than abundance measurements, with the exception of velocity dispersion function analyses. For FDM, constraints from dwarf abundances typically yield $m_{\rm FDM} \gtrsim 10^{-20}~{\rm eV}$, although these limits depend on FDM subhalo mass function modeling assumptions. Comparisons between dynamical measurements and soliton profiles yield preferred FDM mass ranges for some low-density dwarfs, but when considering dwarfs at a population level, dynamical measurements yield comparable FDM limits to dwarf abundances because FDM struggles to produce the observed diversity of high- and low-density dwarfs. Dynamical heating provides the most stringent reported FDM limits from dwarfs, with $m_{\mathrm{FDM}}\gtrsim 10^{-19}~\mathrm{eV}$, and involves distinct modeling uncertainties.}

\NEW{Meanwhile, dwarf galaxies serve as important probes of \emph{in situ} DM physics by testing the scale-dependent predictions of such models in low-mass halos. For SIDM, dwarfs are most sensitive to the amplitude and velocity dependence of the cross section at $v \sim 10$--$50~{\rm km\ s}^{-1}$. For low-amplitude cross sections that keep halos in the core-forming phase of gravothermal evolution, dynamical measurements of dwarf galaxies yield limits of $\sigma/m\lesssim 1~\mathrm{cm^2~g}^{-1}$ at this velocity scale. On the other hand, higher-amplitude cross sections of $\sigma/m\gtrsim 10~\mathrm{cm^2~g}^{-1}$ at these scales lead to a mixture of cored and core-collapsed halos that is allowed by current dynamical data. In addition, dwarf satellite abundances set complementary limits of $\sigma/m\lesssim 10~\mathrm{cm^2~g}^{-1}$ at typical orbital velocities of $v\sim 100~\mathrm{km\ s}^{-1}$ due to subhalo evaporation. Thus, as discussed in Section~\ref{sec:sidm}, only a finite range of velocity-dependent SIDM parameter space is viable if self-interactions significantly impact dwarf galaxies. We note that SIDM constraints from population-level dwarf analyses remain challenging due to modeling uncertainties associated with gravothermal evolution, its interplay with tidal effects and baryonic feedback, and the evolution of stellar populations in evolving SIDM (sub)halos.}

\NEW{Finally, dwarf abundance and dynamics place stringent constraints on the DDM lifetime of $\tau\gtrsim 20~\mathrm{Gyr}$ at kick velocities $V_{\mathrm{kick}}\lesssim 50~\mathrm{km\ s}^{-1}$ due to the coring and enhanced disruption of low-mass subhalos in these models.}

\subsection{Strong Lensing}
\label{sec:strong_lensing}

In strong gravitational lensing, a luminous astronomical source is lensed into multiple images, arcs, or an Einstein ring (see \citealt{Vegetti:2023mgp} for a review). The lensing signal depends on the properties of the source, lens, and intervening DM structure. Strong lensing is sensitive to low-mass DM structure through two complementary channels: perturbations to the flux ratios of multiply-imaged compact sources, which probe percent-level deviations from a smooth lens model, and perturbations to the surface brightness of extended lensed arcs, detected at milli-arcsecond precision with high-resolution imaging or interferometry. Together, these channels probe the abundance and densities of low-mass halos and thus place competitive constraints on DM models.

Two main techniques have been developed to translate strong-lensing observations to DM constraints:
\begin{enumerate}[label=(\roman*)]
    \item Statistical detection: The collective effects of DM (sub)structure on strong-lensing signal (e.g., flux ratios of a multiply-imaged compact source) are modeled, and population-level constraints are inferred from the data to inform DM models;
    \item Individual detection: The effects of individual DM (sub)halos on strong-lensing signal (e.g., perturbations along a lensed arc of an extended source) are modeled, and the existence or non-existence of perturbers with certain properties are used to inform DM models.
\end{enumerate}

\subsubsection{Galaxy Lensing}
\label{sec:galaxy_sl}

\textbf{Flux Ratio Anomalies}: Beginning with \citet{Mao:1997ek} and \citet{Metcalf:2001ap}, flux ratios of multiply-imaged quasars have been used as a probe of small-scale DM structure. \citet{Dalal:2001fq} applied this technique to a sample of seven quadruply-imaged quasars to infer that the DM substructure mass fraction in strong lens galaxies is $\approx 2\%$. This result broadly agrees with CDM predictions and disfavors DM models that reduce the abundance of low-mass subhalos (e.g., WDM), although such constraints were not quantified at the time. Subsequent work debated this conclusion (e.g., \citealt{Mao:2004iw,Xu:2009,Xu:2014dda}), and constraints on DM models using flux ratio anomalies remained both statistically and systematically limited for many years. Statistically, no new lenses with sufficiently sensitive radio or infrared coverage for DM substructure analyses were discovered. Systematically, compact continuum sources such as the optical/UV accretion disk of a quasar are small enough to be significantly affected by stellar microlensing, which contaminates the substructure signal; see the discussion in \citet{Vegetti:2023mgp}.

Two developments strengthened flux ratio anomalies as a DM probe before the sample of available lenses was expanded. First, improved modeling techniques were applied to existing quadruply-imaged quasar samples. These improvements included more accurate models for line-of-sight structure~\citep{Xu:2012,Despali:2017ksx} and the impact of baryonic structure in strong lens galaxies~\citep{Gilman:2016uit,Hsueh:2016aih,Hsueh:2017nlk,Hsueh:2017zfs}. \citet{Hsueh:2019ynk} combined these effects and an analytic model for the suppression of WDM subhalo abundances to predict flux ratio distributions in seven strongly-lensed quasars with radio or mid-infrared flux measurements. These authors derived $m_{\mathrm{WDM}}>5.58~\mathrm{keV}$ at $95\%$ confidence based on the amplitude of the observed flux ratio variance. In particular, warmer models produce less structure and thus predict less variable flux ratios, such that the observed ratios of the image brightnesses disfavor WDM. These WDM constraints are marginalized over nuisance parameters describing the lens and source mass and light profiles, along with unknown properties of the perturbing DM structure (e.g., the \NEW{stripped masses} and positions of individual (sub)halos). Note that \citet{Hsueh:2019ynk} did not model the suppression of halo concentrations in WDM, which can further weaken the predicted lensing signal; incorporating this effect would strengthen such limits.

Second, new measurements of narrow-line emission from strongly-lensed quasars were made \citep{Nierenberg:2014cga}. Unlike compact continuum sources, narrow-line emission regions are extended enough to be insensitive to stellar microlensing while remaining small enough to respond to flux perturbations from DM substructure, substantially reducing a key systematic that had limited earlier analyses. Specifically, \citet{Nierenberg:2019pdj} presented HST observations of eight quadruply-imaged quasars, doubling the sample available for DM analyses. \citet{Gilman:2019nap} analyzed these systems by combining the semi-analytic \textsc{Galacticus} (sub)halo population model~\citep{Benson:2010kx}, including the effects of WDM, with analytic lens and source models. These authors derived $m_{\mathrm{WDM}}>5.2~\mathrm{keV}$, at $95\%$ confidence, by applying the hierarchical Bayesian inference framework introduced in \citet{Gilman:2019vca}. These WDM constraints are marginalized over lens, source, and substructure nuisance parameters, and include the suppression of WDM (sub)halo concentrations. As demonstrated by \citet{Gilman:2019bdm}, the same quasar sample constrains the CDM mass--concentration relation in the \NEW{stripped mass} range of $\sim 10^6~M_{\mathrm{\odot}}$ to $10^{10}~M_{\mathrm{\odot}}$, assuming CDM (sub)halo populations. This implies that the WDM constraints in \citet{Gilman:2019nap} are partially driven by the reduced concentrations of WDM (sub)halos.

Subsequent studies extended the \citet{Gilman:2019nap} WDM constraints. In particular, \citet{Keeley:2023ive} forecasted flux ratio constraints on mixed WDM/CDM models, finding that $31$ quadruply-imaged quasars with JWST flux ratio measurements can distinguish a mixed WDM model with $f_{\mathrm{WDM}}=0.5$ versus a $100\%$ WDM model after marginalizing over $m_{\mathrm{WDM}}$ (also see \citealt{Kamada:2013sh}). \citet{Nierenberg:2023tvi} showed that JWST data probe $\sim 10^7~M_{\mathrm{\odot}}$ (sub)halos by resolving the warm dust emission regions of quasar sources. \citet{Keeley:2024brx} used flux ratio data from nine of these 31 JWST systems and the updated (sub)halo population model and statistical framework from \citet{Gilman:2019nap} to derive $m_{\mathrm{WDM}}>6.1~\mathrm{keV}$ at a posterior odds ratio of 10:1. \citet{Keeley:2025oig} then updated this analysis using $28$ of the $31$ lenses in the full sample and the subhalo evolution model from \citet{Du:2025xqi} to derive $m_{\mathrm{WDM}}>5.6~\mathrm{keV}$ at a 10:1 posterior odds ratio when assuming a prior on the SHMF from \textsc{Galacticus}; this limit strengthens to $m_{\mathrm{WDM}}>6.4~\mathrm{keV}$ when adopting a prior from the Symphony zoom-in simulations~\citep{Nadler:2022dvo}. Consistent with these results, \citet{Nierenberg:2026tma} placed upper limits on the \NEW{peak mass of the smallest} field halos and subhalos that contribute to this signal of $10^{8.3}~M_{\mathrm{\odot}}$ ($10^{8.2}~M_{\mathrm{\odot}}$) at a 10:1 odds ratio, when using SHMF priors from \textsc{Galacticus} (Symphony).

Several studies applied similar methods to probe other \emph{ab initio} DM scenarios. For example, \citet{Gilman:2021gkj} used eleven lenses (the eight studied in \citealt{Gilman:2019nap} plus an additional three) to place constraints on the primordial matter power spectrum $P(k)$. Specifically, this work implemented a running spectral index $n_s$ (and a running-of-the-running) in \textsc{Galacticus} to place constraints on $P(k)$ at wavenumbers $k>10~\mathrm{Mpc}^{-1}$. In the context of this power spectrum shape, they derived that $-0.4<\log_{10}(P(k)/P_{\mathrm{CDM}}(k))<0.5$ at $68\%$ confidence and also derived correlated constraints at higher $k$; models with less (more) small-scale power underpredict (overpredict) the observed flux ratio variance.

Information beyond image positions and flux ratios can also be used to constrain DM models in the statistical detection regime. For example, \citet{Birrer:2017rpp} modeled both the luminous arcs and image positions from HST observations of RXJ1131-1231 to derive $m_{\mathrm{WDM}}>2~\mathrm{keV}$ at $2\sigma$ confidence. \citet{Gilman:2024mcs} showed that simultaneously modeling lensed arcs and flux ratios can improve constraints on $M_{\mathrm{hm}}$ by up to a factor of $\sim 10$ for a mock sample of $25$ strong lenses with JWST flux ratio measurements. Thus, \citet{Gilman:2025fhy} analyzed the same sample of lenses studied in \citet{Keeley:2024brx} but used both flux ratios and extended arcs to constrain WDM models, yielding a 10:1 Bayes factor constraint of $m_{\mathrm{WDM}}>6.5~\mathrm{keV}$ ($m_{\mathrm{WDM}}>7.4~\mathrm{keV}$) when adopting the \textsc{Galacticus} (Symphony) SHMF prior. \citet{Gilman:2026uvq} cast these constraints as limits on the WDM free-streaming length, finding $\lambda_{\mathrm{fs}}<7.0~\mathrm{kpc}$ ($\lambda_{\mathrm{fs}}<6.0~\mathrm{kpc}$). 

Moving to \emph{in situ} DM physics, \citet{Laroche:2022pjm} derived FDM constraints by generalizing the \citet{Gilman:2019vca} framework to include the suppression of low-mass halo abundances, solitonic halo density profiles, and effects of wave interference in FDM models. These authors showed that FDM models with $m_{\mathrm{FDM}}<10^{-21.5}~\mathrm{eV}$ are disfavored by the same eleven lenses; in particular, the CDM-to-FDM likelihood ratio is $4:1$ at $m_{\mathrm{FDM}}=10^{-21.5}~\mathrm{eV}$ and increases for lower $m_{\mathrm{FDM}}$. Note that this constraint significantly strengthens if FDM density fluctuations on the de Broglie scale are not modeled, since these fluctuations increase the predicted flux ratio signal. \NEW{Thus, it is critical to self-consistently model both FDM suppression and density fluctuation effects to derive unbiased FDM constraints from flux ratio statistics, as done in \citet{Laroche:2022pjm}.} 

For SIDM, \citet{Gilman:2021sdr} forecasted flux ratio constraints by integrating a model for core formation and collapse into the \citet{Gilman:2019vca} framework. They found that a sample of $50$ lenses can be used to derive $\sigma/m\lesssim 11~\mathrm{cm^2\ g^{-1}}$ at a velocity scale of $20~\mathrm{km\ s}^{-1}$, and that an SIDM model with $\sigma/m=19.2~\mathrm{cm^2~g^{-1}}$ at this velocity scale can be distinguished from CDM with a $20:1$ likelihood ratio using the same data. \citet{Gilman:2022ida} then derived constraints on a family of SIDM scenarios, including models with velocity-dependent and resonant cross sections, using the same sample of eleven quadruply-imaged quasars as \citet{Gilman:2021gkj}. They found that SIDM models with cross sections of $\sigma/m\gtrsim 100~\mathrm{cm^2~g^{-1}}$ at a $50~\mathrm{km\ s}^{-1}$ velocity scale are disfavored relative to CDM because most low-mass (sub)halos core collapse in these scenarios, causing the predicted flux ratio signal to deviate from the data. \NEW{On the other hand, using the JWST quasar sample, \citet{Gilman:2026wul} found that SIDM models are \emph{favored} relative to CDM with a Bayes factor of at least $6:1$ unless subhalos and globular clusters are more abundant than standard CDM expectations. These results hint at departures from CDM in current flux ratio data.}

\NEW{Finally, we note that the detailed treatment of baryonic effects and subhalo tidal stripping is a key aspect of flux ratio modeling. In terms of stripping, \citet{Du:2025xqi} estimated that the SHMF evaluated using bound mass is suppressed by a factor of $\sim 20$ relative to the infall mass function, and that typical subhalos relevant for flux ratio signal lose $\sim 80\%$ of their mass since infall. As a result, the forward modeling frameworks above take the approach of marginalizing over the amplitude of the projected SHMF. Baryonic structures, including the central galaxy of the main lens, are predicted to substantially enhance subhalo stripping rates in MW-mass systems (e.g., \citealt{Garrison-Kimmel170103792}); at the strong lensing scale, hydrodynamic simulations show similar effects (e.g., \citealt{Fiacconi:2016cih,Richings:2020auv}), while some SAMs predict only a small change to the projected SHMF relevant for current flux ratio analyses~\citep{Gannon:2025nhr}. On the other hand, the effects of baryonic feedback within the low-mass subhalos probed by current flux ratio measurements is expected to be fairly minor, except for initially higher-mass systems that have been significantly stripped. As the precision of the data improves, more accurate modeling of these effects will be important to disentangle them from DM signatures.}

\textbf{Gravitational Imaging}: Detection of individual (sub)halos using strong lensing often uses the ``gravitational imaging'' technique~\citep{Koopmans:2005nr}. Here, the effects of individual (sub)halos on the lens potential are modeled as regularized pixellated corrections to an underlying smooth model. A Bayesian comparison between a smooth lens model and one including DM substructure is then used to detect individual perturbers or (in the case of non-detection) place limits on their abundance and other properties~\citep{Vegetti:2008eg}. This method---along with approaches that parametrically reconstruct the perturber mass profile---has been applied to ensembles of strong lenses, yielding a handful of individual detections~\citep{Vegetti:2009aa,Vegetti:2012mc,Hezaveh:2016ltk,Lange:2024pef,Amvrosiadis:2026wkp}. Some of these detections remain disputed (e.g., \citealt{Stacey:2025}), and a large number of non-detections have also been reported~\citep{Nightingale:2022bhh}.\footnote{Even in cases where a perturber has been confirmed by multiple independent groups, the inferred perturber \emph{properties} often vary depending on the lens, source, and perturber modeling assumptions (e.g., for the SDSSJ0946+1006 perturber, see \citealt{Minor:2020hic,Ballard:2023fgi,Enzi:2024ygw,Despali:2024ihn,Minor:2025,Tajalli:2025qjx,He:2025wco}; for the JVAS B1938+666 perturber, see \citealt{Sengul:2021lxe}).}

Population analyses of gravitational imaging detections and non-detections probe (sub)halo abundances and, when combined with a model for the selection function of objects in these data, have been used to constrain models like WDM. For example, \citet{Li:2015xpc} forecasted that $\sim 100$ strong lenses can distinguish a $7~\mathrm{keV}$ sterile neutrino model from CDM. \citet{Ritondale:2018cvp} then combined $17$ BELLS GALLERY and $11$ SLACS strong lenses with an analytic model for source, lens, and perturber populations to derive $m_{\mathrm{WDM}}>0.3~\mathrm{keV}$ at the $2\sigma$ level; similarly, \citet{Vegetti:2018dly} derived $m_{\mathrm{WDM}}>0.3~\mathrm{keV}$ at the $2\sigma$ level, using 11 SLACS lenses, and extended these limits to sterile neutrino models. \NEW{The two limits are comparable despite the different sample sizes because the SLACS sample includes a detection of a low-mass halo that strengthens the WDM constraint, whereas the BELLS GALLERY sample only contains non-detections.} Note that the constraining power of non-detections strongly depends on the sensitivity function (i.e., the probability that a given perturber could be detected as a function of its properties and the lens/source properties); for example, \citet{Ritondale:2018cvp} conservatively assumed a $10\sigma$ detection threshold. Thus, future datasets that better constrain this sensitivity function can yield more stringent DM constraints~\citep{ORiordan:2022qds}. In addition, \citet{ORiordan:2025zxl} showed that including the collective effects of low-mass (sub)halos below the individual detection limit can further increase the constraining power of gravitational imaging.

Gravitational imaging has also been applied to high-resolution interferometric data to constrain DM models. For example, \citet{Powell:2023jns} analyzed very long baseline (VLBI) measurements of the strongly-lensed radio jet MG J0751+2716 in an FDM scenario. By modeling FDM density granule fluctuations alone, they derived $m_{\mathrm{FDM}}>4.4\times 10^{-21}~\mathrm{eV}$ at a 20:1 posterior odds ratio. Subhalos were not included in this analysis because, given the $\lesssim 5~\mathrm{mas}$ resolution of the VLBI data, perturbations from FDM subhalos in the $m_{\mathrm{FDM}}$ range of interest are not significant compared to those from density granule fluctuations. More recently, \citet{Powell:2025rmj} found evidence for a $\sim 10^6~M_{\mathrm{\odot}}$ perturber in JVAS B1938+666 using VLBI data. This detection is consistent with CDM expectations for the number of detectable subhalos, but the perturber appears more centrally concentrated than expected at this mass scale~\citep{Vegetti:2026}.

Intriguingly, most perturbers currently detected via gravitational imaging show evidence for higher inner densities than naively predicted by CDM. This data is being used to inform beyond-CDM models; for example, \citet{Nadler:2023nrd} used cosmological SIDM--only zoom-in simulations of a strong lens analog to show that the velocity-dependent SIDM cross section from Equation~\ref{eq:xsec} with $\sigma_0/m=147.1~\mathrm{cm^2\ g}^{-1}$ and $w=120~\mathrm{km\ s}^{-1}$ can produce extremely compact subhalos analogous to the SDSSJ0946+1006 perturber. They also showed that CDM rarely produces such dense objects, consistent with other simulation results (e.g., \citealt{Minor:2020hic,Despali:2024ihn,Zhang:2026oui}). \citet{Yang:2024uqb} then derived similar SIDM predictions for a model with $\sigma_0/m=70~\mathrm{cm^2\ g}^{-1}$ and $w=120~\mathrm{km\ s}^{-1}$ by applying a parametric transformer model to CDM zoom-in simulations, and \citet{Kong:2025sqx} showed that zoom-in simulations of both models from the SIDM Concerto suite~\citep{Nadler:2025jwh} produce analogs of all known strong-lensing perturbers.

Recently, \citet{Li:2025kpb} revisited the SIDM explanation of the SDSSJ0946+1006 perturber, finding that core-collapse is only efficient for high-mass subhalos that are expected to host galaxies which violate upper limits on the perturber's luminosity (see \citealt{Kong:2025sqx} for further discussion). \citet{Tajalli:2025qjx} reanalyzed the same perturber to infer that an effective cross section $\sigma_{\mathrm{eff}}/m\gtrsim 300~\mathrm{cm^2~g}^{-1}$ at $v\sim 100~\mathrm{km\ s}^{-1}$ is required to produce sufficiently compact core-collapsed subhalos based on gravothermal fluid modeling. This analysis assumed a median-concentration halo; the lower bound on the cross section would be relaxed for a higher-concentration system. Finally, \citet{Yu:2025tmp} and \citet{Zhang:2026gur} considered the $\sim 10^6~M_{\mathrm{\odot}}$ JVAS B1938+666 perturber in CDM and SIDM scenarios, finding that its density profile is consistent with core-collapsed SIDM subhalos or with highly-stripped CDM subhalos that host an intermediate mass black hole. \citet{Oguri:2026csl} showed that the same system can be fit by a halo with an inner soliton and central black hole in an FDM model with $m_{\mathrm{FDM}}=4\times 10^{-21}~\mathrm{eV}$.

\NEW{Overall, at present, neither a CDM explanation (requiring highly stripped subhalos that host intermediate mass black holes) nor an SIDM explanation (requiring tuning of the cross section to produce sufficient core-collapse) of these compact (sub)halo perturbers is definitively favored. Resolving this will require larger samples of perturbers with well-measured densities and folding non-detections into population-level DM inference.}

\textbf{Lensing Power Spectrum}: A complementary approach to individual perturber detection is to characterize the collective effects of subhalos and line-of-sight halos through the power spectrum of the lensing convergence field. This method exploits the statistical signal of (sub)halo populations on lensed arcs, probing the SHMF even when individual subhalos fall below the detection threshold~\citep{DiazRivero:2017xkd,Cyr-Racine:2018htu}. In particular, the amplitude and slope of the convergence power spectrum on scales of $\sim 1~\mathrm{kpc}^{-1}$ depend sensitively on subhalo abundances and density profiles, making these quantities a powerful diagnostic of DM physics~\citep{Brennan:2018jhq}. CDM, WDM, FDM, and SIDM produce significantly different convergence power spectra at these scales; predictions for these scenarios in non-CDM models have been derived from cosmological simulations~\citep{DiazRivero:2018oxk,Kawai:2021znq}. In addition, the density slopes of individual perturbers have also been proposed as a DM model discriminator; for example, \citet{Sengul:2022edu} showed that this slope is measurable from HST-quality data and differs systematically between CDM, WDM, and SIDM.

In this context, an important modeling consideration is the contribution of line-of-sight (LOS) halos. For example, \citet{CaganSengul:2020nat} computed the effective convergence power spectrum from LOS halos and found that, for typical lens configurations, LOS halos and subhalos contribute comparably to strong lensing signal and must be modeled jointly to avoid biased DM inferences. \citet{Dhanasingham:2022nox} then showed that LOS halos imprint an anisotropic signature on the two-point function of the effective lensing deflection field, which statistically probes the impact of SIDM on subhalo populations~\citep{Dhanasingham:2023thg}. Meanwhile, ratios of multipole moments that are robust to the mass--sheet degeneracy may be measured by ELTs, further advancing galaxy strong lensing as a frontier DM probe.

\subsubsection{Cluster Lensing}
\label{sec:cluster_sl}

\textbf{Cluster Substructure}: Strong lensing of galaxy clusters is a promising probe of small-scale DM structure that complements galaxy--galaxy strong lensing measurements (see \citealt{Natarajan:2024iqm} for a review). Although cluster substructure modeling is challenging due to the large dynamic range and uncertain impact of baryonic physics processes involved, several recent studies have advanced this probe. For example, \citet{Meneghetti:2020yif} reported a significant excess of small-scale gravitational lenses in $11$ galaxy clusters. In particular, they found that the probability of galaxy--galaxy strong lensing (GGSL) events in these systems exceeded predictions from cosmological hydrodynamic CDM simulations by $\gtrsim 10\mathrm{x}$.
Follow-up studies argued that the GGSL cross section predicted in simulations is sensitive to resolution~\citep{Robertson:2021} and the baryonic feedback prescription~\citep{Bahe:2021bcs}. \citet{Meneghetti:2022apr} found that adopting a less efficient AGN feedback prescription brings CDM into better agreement with the data, but that such models form subhalos that are too massive given the observed Einstein radii of the strong lenses. Thus, \citet{Meneghetti:2023fug} concluded that a GGSL discrepancy of a factor of $\gtrsim 4\mathrm{x}$ persists.

Meanwhile, based on analysis of strong and weak lensing data for several galaxy clusters, \citet{Chiang:2025wup} found that the outer regions of cluster subhalos are compatible with CDM, while the inner regions favor more compact mass distributions; building on this result, \citet{Natarajan:2026ztd} argued that the spatial distribution of subhalos in several of these clusters is much more centrally concentrated than CDM predictions derived from the TNG-Cluster simulations.

These results have motivated studies of alternative DM models. For example, \citet{Yang:2021kdf} found that core collapse in SIDM models can increase the predicted GGSL probability and ease the discrepancy discussed above, even for a cross section of $\sigma/m=1~\mathrm{cm^2~g}^{-1}$ at the cluster substructure velocity scale of $\sim 250~\mathrm{km\ s}^{-1}$. This is particularly relevant since the observed strong lenses have small Einstein radii, implying more compact substructure than predicted by CDM. Indeed, \citet{Tokayer:2024wwo} found that refitting observed lenses with NFW profiles, including the effects of adiabatic contraction due to baryons, is insufficient to explain the discrepancy. Considering FDM, \citet{Kawai:2024zka} showed that de Broglie fluctuations can increase the predicted GGSL probability for $10^{-23}~\mathrm{eV}\lesssim m_{\mathrm{FDM}}\lesssim 10^{-20}~\mathrm{eV}$ by forming high-density solitonic cores, despite the fact that the SHMF is suppressed in this scenario. Finally, \citet{Ephremidze:2026ity} found that the residual power spectrum of multiply-imaged galaxies in cluster strong lenses can be used to statistically distinguish FDM and CDM on kiloparsec scales.

\textbf{Highly-magnified Stars}: Strong-lensing studies have recently discovered individual stars at cosmological distances strongly lensed by intervening galaxy clusters~\citep{Kelly:2017fps}. These objects are detectable due to their extreme magnification factors of $\sim 1000$, and thus probe smaller scales than the galaxy--galaxy lensing data discussed above. In particular, magnified stars in strong-lensing arcs probe both the underlying population of microlenses (e.g., lensed stars) and millilenses (e.g., DM substructure), as well as the macrolens model~\citep{Diego:2024qtk}. In this context, astrometric perturbations of the image positions can also be used to probe DM substructure with subhalo-to-cluster halo mass ratios of $\sim 10^{-7}$ to $10^{-9}$~\citep{Dai:2018mxx}.

\citet{Diego:2023qhp} applied this idea to derive DM constraints from JWST observations of Mothra, a highly-magnified star at $z=2.091$. They found that the lack of a particular counterimage in the lensing data cannot be explained by microlensing alone; instead, a perturber with \NEW{a stripped mass of} $M<2.5\times 10^6~M_{\mathrm{\odot}}$ is preferred. Given the angular scale probed, this perturber is compatible with CDM predictions when interpreted as a (sub)halo. In this way, \citet{Diego:2023qhp} derived $m_{\mathrm{WDM}}>8.7~\mathrm{keV}$ using their constraint on the number density of such perturbers. \citet{Perera:2024xdc} then reanalyzed the Mothra arc, finding that the perturber's \NEW{stripped mass} and core radius are respectively smaller than $\sim 10^6~M_{\mathrm{\odot}}$ and $17~\mathrm{pc}$.

\citet{Diego:2023qhp} also found that the FDM mass must be in the range $5\times 10^{-23}<m_{\mathrm{FDM}}<5\times 10^{-22}~\mathrm{eV}$ in order to produce density fluctuations compatible with the Mothra perturber, and \citet{Perera:2024xdc} updated this constraint to $m_{\mathrm{FDM}}\lesssim 3\times 10^{-21}~\mathrm{eV}$. Note that these analyses only model FDM density fluctuations and do not consider (sub)halos. On the other hand, \citet{Abe:2023hlf} found that CDM subhalos with \NEW{stripped masses} of $5\times 10^{7}~M_{\mathrm{\odot}}~h^{-1}<M<10^9~M_{\mathrm{\odot}}~h^{-1}$ can explain the anomalous position of Mothra, and that an extremely low mass of $m_{\mathrm{FDM}}\sim 10^{-24}~\mathrm{eV}$ is preferred in the FDM case. Meanwhile, \citet{Broadhurst:2024ggk} analyzed the ``Dragon Arc,'' finding that CDM subhalos in the \NEW{stripped mass range} $10^6~M_{\mathrm{\odot}}<M<10^8~M_{\mathrm{\odot}}$ produce strong-lensing features that are incompatible with the skewness and lensing structure of the data, while FDM with $m_{\mathrm{FDM}}\approx 10^{-22}~\mathrm{eV}$ provides a better fit. Given the extremely small scales probed and high sensitivity to DM physics, reconciling these measurements with each other and with other strong-lensing and small-scale structure DM constraints is an important area for future work.

\subsubsection{\NEW{Gravitational Wave Lensing}}
\label{sec:gw_lensing}

\NEW{A small fraction ($\sim 0.1$--$1\%$) of the gravitational wave (GW) transients detectable by ground-based interferometers are expected to be strongly lensed by intervening galaxies and clusters, producing multiple copies of the same GW signal arriving at the detector with characteristic time delays~\citep{Wang:1996as}. The number of such events and their time-delay distribution are sensitive to the abundance and properties of low-mass DM halos along the line of sight, providing a probe of small-scale structure that complements electromagnetic lensing measurements. The first strongly-lensed GW events are expected to be detected within the next few years~\citep{Haris:2018vmn}. Thus, while this probe is less established than optical lensing, it represents a promising future probe of DM physics.}

\NEW{For example, \citet{Jana:2024dhc} showed that the expected distribution of GW strong lensing events is sensitive to DM microphysics. In particular, these authors found that WDM and FDM both reduce the number of strongly-lensed GW events and reduce the number of events with short time delays relative to CDM due to the suppression of low-mass halos in these models. Based on the sensitivity of next-generation GW detectors such as the Einstein Telescope and Cosmic Explorer---which are expected to detect millions of binary black hole mergers---\citet{Maity:2026hcv} forecasted that GW strong lensing observations can constrain thermal-relic WDM masses up to $m_{\mathrm{WDM}}\sim 20~\mathrm{keV}$ and detect deviations from CDM if the WDM mass is $m_{\mathrm{WDM}}\sim 10~\mathrm{keV}$. These authors also demonstrated that the upcoming O5 and O6 LIGO--Virgo--KAGRA runs can set WDM constraints that are competitive with the strongest limits from current small-scale structure probes.}

\NEW{A separate GW lensing observable is the wave-optics diffraction pattern imprinted on the GW waveform when the Fresnel scale of the lens is comparable to its physical size. For halos with $10^3~M_{\mathrm{\odot}}\lesssim M \lesssim 10^5~M_{\mathrm{\odot}}$, this wave-optics effect is potentially detectable at LIGO frequencies, implying that frequency-dependent modulations in strongly-lensed GW signals can probe low-mass (sub)halos~\citep{Dai:2018enj,Caliskan:2023zqm}. For example, \citet{Tambalo:2022wlm} showed that advanced LIGO measurements will be sensitive to FDM masses of $m_{\mathrm{FDM}}\lesssim 10^{-17}~\mathrm{eV}$ and velocity-independent SIDM cross sections as low as $\sigma/m\sim 10^{-4}~\mathrm{cm^2~g}^{-1}$ due to the very small density cores in these models. Meanwhile, \citet{Ando:2026poq} studied wave-optics effects in CDM, finding that future LISA detections of strongly-lensed massive binary black hole mergers may probe substructure down to $\sim 10^4~M_{\mathrm{\odot}}$. Based on LISA's expected sensitivity to strongly-lensed GW events, \citet{Liu:2026xpi} forecasted that CDM, an FDM model with $m_{\mathrm{FDM}}=10^{-21}~\mathrm{eV}$, and an SIDM model that increases the concentrations of low-mass halos by a factor of $3\times$ due to core collapse are distinguishable with $\sim 100$ events. Finally, \citet{Singh:2025uvp} showed that weakly-lensed GWs detected by LISA are also sensitive to FDM through its alteration of low-mass halo profiles.}

\NEW{GW lensing has several attractive features relative to electromagnetic strong lensing as a small-scale structure probe. First, the intrinsic properties of GW waveforms are extensively studied, mitigating some of the usual degeneracy with source modeling. Second, multiple copies of a GW event can be accurately identified based on their waveforms, and the time delay between them can be measured with millisecond precision. The time-delay distribution of lensed events is sensitive to the integrated 
effect of all (sub)halos along the line of sight, and can therefore probe objects that are too small to be detected individually, akin to flux ratio anomalies. Current non-detections of strongly lensed GW events in LIGO and VIRGO data already constrain the abundance of compact DM clumps~\citep{Barsode:2024wda}, and the accumulation of such events over the coming decade are expected to probe the DM scenarios we have discussed.}

\subsubsection{Summary}

\NEW{Strong gravitational lensing is a powerful probe of small-scale structure that extends sensitivity to low-mass halos beyond the local universe regardless of their luminous content. Flux ratio anomalies from quadruply-imaged quasars currently provide the most stringent strong lensing constraints on \emph{ab initio} DM physics that suppresses structure, with the recent analyses of JWST data that combine flux ratios and extended arcs reaching $m_{\rm WDM} \gtrsim 7~{\rm keV}$. The same technique constrains a variety of other DM phenomenology, including wave interference in FDM, yielding $m_{\mathrm{FDM}}\gtrsim 10^{-21}~\mathrm{eV}$, and core collapse in SIDM, probing cross sections of $\sigma/m \gtrsim 10~\mathrm{cm^2~g}^{-1}$ at $v\sim 20~\mathrm{km\ s}^{-1}$. Systematic uncertainties for this technique include SHMF modeling in strong lens systems, with constraints differing at the factor-of-two level depending on whether semi-analytic or N-body simulation priors are assumed. In addition, current analyses rely on semi-analytic models because millions of realizations per lens are needed for inference; it will be important to benchmark these predictions against simulation-based analyses as the number of observed systems increases and observational sensitivity improves. In addition, disentangling substructure vs.\ line-of-sight halo contributions to the signal will be crucial to interpret any potential deviations from CDM.}

\NEW{Gravitational imaging has revealed several anomalously dense perturbers that are outliers relative to CDM predictions, and recent flux ratio analyses reach similar conclusions on a statistical level. This has motivated high-amplitude, velocity-dependent SIDM cross sections as a potential solution: models with $\sigma/m\sim 100~\mathrm{cm^2~g}^{-1}$ at low velocities drive core collapse in the relevant (sub)halos and produce sufficiently dense perturbers. Nonetheless, the interpretation of these systems is subject to modeling uncertainties, even in CDM, including the degree of tidal stripping and the contribution of baryons to perturber compactness. Combining perturber detections and non-detections from extended arcs can also set population-level constraints on WDM-like models; the central challenge for this technique is modeling the detection across lenses and perturber properties. Cluster strong lensing has likewise revealed an excess of small-scale lenses in clusters that are concentrated near the host center and not obviously compatible with CDM predictions on these scales. Thus, it is timely to develop targeted simulations of both galaxy and cluster lenses in a shared set of DM models and (particularly in the cluster lensing case) to develop modeling frameworks that can capture stripped subhalos in these systems with high fidelity and marginalize over the uncertain impact of baryonic feedback on these data.}

\NEW{Meanwhile, highly-magnified stars have emerged as a sensitive probe of both WDM-like models and FDM, with claimed sensitivity comparable to other small-scale structure probes. The detection and analysis of additional systems will be important to contextualize these results and further quantify DM sensitivity. Future observations of strongly-lensed gravitational wave events also promise remarkable sensitivity to both \emph{ab initio} and \emph{in situ} DM physics, potentially reaching WDM masses of $\gtrsim10$~keV and complementing the electromagnetic lensing techniques discussed above.}

\subsection{Lyman-$\alpha$ Forest}
\label{sec:lyman_alpha}

The Lyman-$\alpha$ forest is one of the most powerful probes of $P(k)$ and thus \textit{ab initio} effects from DM models (e.g., see \citealt{Weinberg:2003eg} for an early review). The Lyman-$\alpha$ forest is a qualitatively different cosmological tracer than the other small-scale structure observables discussed in this review. Rather than tracing the most overdense regions of the cosmic web (i.e., galaxies, clusters and their sub-structure), the Lyman-$\alpha$ forest signal traces only moderately over- and under-dense regions (i.e., filaments and voids). Thus, at any given wavenumber and redshift, the Lyman-$\alpha$ forest is less affected by nonlinear gravitational evolution than galaxies and clusters (e.g., $\Delta^2_{\mathrm{lin}}\approx 0.35$ at $k=0.7~\mathrm{Mpc}^{-1}$ and $z=3$; Figure~\ref{fig:k_z}). In this sense, the Lyman-$\alpha$ forest is a relatively clean probe of linear DM density fluctuations and thus the initial conditions of the DM. However, to interpret the Lyman-$\alpha$ forest, it is necessary to marginalize over the astrophysical state of the intergalactic medium (IGM) that sources the forest signal when constraining DM density fluctuations (e.g., \citealt{Garzilli:2015iwa,Garzilli:2018jqh}).

The Lyman-$\alpha$ forest is a series of absorption lines observed in the spectra of high-redshift (\(2 \lesssim z \lesssim 6\)) objects, in particular quasars and Lyman-break galaxies. When light is emitted from the source, it has a relatively smooth broadband electromagnetic spectrum with broad Lyman-$\alpha$ and Lyman-$\beta$ emission features. When the light passes through intervening neutral hydrogen gas in the IGM, Lyman-series absorption occurs (in particular, Lyman-$\alpha$ at {1216\AA} and Lyman-$\beta$ at {1026\AA} rest-frame wavelengths). Absorption always occurs at its rest-frame wavelengths, but previous absorption is continuously redshifted due to cosmological expansion, leaving the ``forest'' of lines that we observe. In a given spectrum, the spectral position and strength of an absorption line is a mapping from the density of neutral hydrogen along the line of sight to the background source. In turn, it is therefore a mapping from the underlying line-of-sight DM distribution. Observing many hundreds of thousands of Lyman-$\alpha$ forest spectra across the sky thus effectively builds up an (incomplete) three-dimensional map of the DM at high redshift. As a result, the Lyman-$\alpha$ forest can probe models that change the clustering and velocity distributions of the DM.

The Lyman-$\alpha$ forest is most accurately modeled by cosmological hydrodynamical simulations of the IGM (e.g., see \citealt{Cen:1994da,Meiksin:2001tn} for earlier works, \citealt{Lukic:2014gqa} for a systematic study and \citealt{Chabanier:2024knr} for a modern example). Lyman-$\alpha$ forest absorption is defined as originating from neutral hydrogen with column densities \(< 10^{17.2}\,\mathrm{atoms}\,\mathrm{cm}^{-2}\); the higher column density gas \citep[Lyman-limit systems and damped Lyman-$\alpha$ absorbers;][]{1986ApJS...61..249W} forms a contamination with absorption damping wings that need to be removed from the data \citep{Garnett:2016awq} and a residual that is modeled and marginalized in parameter inference \citep{Rogers:2017bmq,Rogers:2017eji}. The Lyman-$\alpha$ forest itself, sourced from low-density gas in the IGM, is only weakly sensitive to the details of galaxy formation and feedback \citep{Chabanier:2020uuh,Dong:2024,Tillman:2024pim}, meaning that Lyman-$\alpha$ forest simulations are less computationally expensive than hydrodynamical simulations focused on galaxy formation. Metal absorption lines must also be removed from the data and/or modeled and marginalized when inferring DM properties.

As mentioned above, the Lyman-$\alpha$ forest signal is sensitive not only to the density and velocity of the DM, but also to the heating and ionization rates of the UV background from hydrogen and helium reionization (e.g., \citealt{Meiksin:2024lso}). The full calculation of these effects requires computationally-expensive radiative transfer simulations \citep{Chardin:2015uza}. However, at the current statistical precision of data, the Lyman-$\alpha$ forest can be accurately modeled with a spatially-uniform UV background, radiative heating and cooling, adiabatic expansion and ideal gas pressure support. Nonetheless, though the physics of the IGM is well understood, there is significant parametric uncertainty on the timing and duration of reionization, as well as correlations between the IGM temperature, gas density and amplitude of the ionizing background. Sampling this high-dimensional parameter space with hydrodynamical simulations necessitates the acceleration and interpolation of simulations (e.g., \citealt{Villasenor:2022aiy}), and, more recently, the development of machine learning (ML) approaches like emulators and simulation-based inference (e.g., \citealt{Rogers:2018smb,Bird:2018efe,Rogers:2020cup,Pedersen:2020kaw}). Semi-analytic approaches have also been developed, including a halo model and the effective field theory of large-scale structure (e.g., \citealt{Irsic:2018hhg,Ivanov:2023yla,Ivanov:2024jtl}).

There are two main types of Lyman-$\alpha$ forest surveys. The first type covers large cosmological volumes using massively multiplexed spectroscopic instruments that collect hundreds of thousands of Lyman-$\alpha$ forest spectra \citep[e.g.,][]{DESI:2023pir}. These surveys have medium resolution (resolving power \(R \sim 5{,}000\)) and relatively lower signal-to-noise ratios. The large volume allows three-dimensional correlations in the transmitted flux to be detected, leading to the high-significance detection of the BAO feature \citep[separations \(r \sim 100\, h^{-1}\,\mathrm{Mpc}\), \(z \sim 2.3\);][]{DESI:2025zpo}. However, most of the \textit{small-scale} constraining power is in the line-of-sight modes. It is therefore typical to measure the one-dimensional Lyman-$\alpha$ forest flux power spectrum when constraining DM models \citep{Karacayli:2025svi,Ravoux:2025uik}, defined as the integral of the 3D flux power spectrum over modes transverse to the line of sight. For many cosmological and DM models of interest, after marginalizing over IGM parameters, most of the cosmological information in the 1D flux power spectrum measured from surveys like SDSS, the Baryon Oscillation Spectroscopic Survey (BOSS), and the Dark Energy Spectroscopic Instrument (DESI) can be compressed to the amplitude and spectral index of the linear matter power spectrum at \(k \sim 1\,h\,\mathrm{Mpc}^{-1}\) and \(z \sim 3\) (e.g., \citealt{SDSS:2004kjl,eBOSS:2018qyj,Pedersen:2019ieb,Pedersen:2022anu}).

The second type of survey targets a smaller number (\(\sim\) dozens) of bright quasars with very-high-resolution (\(R \sim 40{,}000\)) spectrographs on large-aperture telescopes like the HIRES spectrograph on Keck \citep{Vogt:1994}, the MIKE spectrograph on Magellan \citep[][]{2003SPIE.4841.1694B}, and the UVES spectrograph on the Very Large Telescope (VLT; \citealt{Dekker:2000}). The smaller number of spectra means that angular correlations between spectral pixels cannot be measured; however, the high spectral resolution resolves smaller-scale density fluctuations (\(k \lesssim 100\,h\,\mathrm{Mpc}^{-1}\)). Thus, high-resolution measurements of the 1D flux power spectrum with a small number of high-redshift quasars \citep[e.g.,][]{Boera:2018vzq} have proven to be a competitive DM probe. These measurements typically extend to redshifts \(z \lesssim 5.4\), such that metal contamination is less important but there is sensitivity to spatial inhomogeneities in hydrogen reionization \citep{Wu:2019sgk}. However, given the relatively few bright, high-redshift quasar targets, these analyses are still limited by statistical uncertainties (unlike recent DESI measurements). These considerations will change as the number of high-resolution spectra increases with high-resolution spectrographs on 30-meter-class telescopes (see Section~\ref{sec:observation_future} for further discussion).

\subsubsection{Constraints}

\textbf{Warm Dark Matter}: The connection between the statistics of the transmitted flux in the Lyman-$\alpha$ forest and the underlying small-scale structure was established through a series of hydrodynamical simulations  in the 1990s \citep[e.g.,][]{Cen:1994da,1996ApJ...457L..51H,Gnedin:1997td}. This connection was formalized by the fluctuating Gunn-Petersen approximation that analytically connects the Lyman-$\alpha$ forest flux to DM density perturbations \citep[e.g.,][]{1996ApJ...457L..51H,Croft:1997jf}. \citet{Croft:1998pe} inferred, from a sample of nineteen Lyman-$\alpha$ forest spectra, the amplitude and spectral index of $P(k)$ at \(k \sim 1\,h\,\mathrm{Mpc}^{-1}\) and \(z \sim 2.1\), finding consistency with the CDM model at the time.

\citet{Narayanan:2000tp} searched for the small-scale matter power spectrum cutoff from WDM in the Lyman-$\alpha$ forest, setting a lower limit on a WDM particle mass \(m_\mathrm{WDM} > 0.75\,\mathrm{keV}\) at the $3\sigma$ level. \citet{Viel:2005qj} combined high-resolution, high signal-to-noise spectra \citep{Croft:2000hs,Kim:2003qt} with CMB data from the Wilkinson Microwave Anisotropy Probe  \citep[WMAP;][]{WMAP:2003elm,WMAP:2003zzr} to constrain the mass of thermal-relic WDM \(m_\mathrm{WDM} > 0.55\,\mathrm{keV}\) and the Dodelson--Widrow sterile neutrino \(m_s > 2\,\mathrm{keV}\), both at the $2\sigma$ level; these authors also constrained early-decoupled thermal relics and light thermal gravitinos.\footnote{Throughout this subsection, $m_s$ refers to the particle mass of Dodelson--Widrow sterile neutrinos; however, we note that Lyman-$\alpha$ forest bounds have been extended to other sterile neutrino production scenarios (e.g., \citealt{Schneider:2016uqi,Parashari:2026dxo})}

In parallel to these early analyses of high-resolution spectra, SDSS increased the sample size of Lyman-$\alpha$ forests from dozens to \(> 3000\), leading to sub-percent precision measurements of $P(k)$ on small scales~\citep{SDSS:2004kjl}. \citet{Seljak:2006qw} used this measurement, the high-resolution spectra \citep{Kim:2003qt}, WMAP CMB data \citep{WMAP:2003elm,WMAP:2003zzr}, and SDSS galaxy clustering data \citep{2004ApJ...606..702T} to derive \(m_{\mathrm{WDM}} > 2.5\,\mathrm{keV}\) and \(m_s > 14\,\mathrm{keV}\) at $95\%$ C.L. Using almost the same data, \citet{Viel:2006kd} found \(m_{\mathrm{WDM}}> 2\,\mathrm{keV}\) and \(m_s > 10\,\mathrm{keV}\) at $95\%$ confidence. Then, \citet{Viel:2007mv} strengthened these limits using a combination of high-resolution HIRES spectra at $2.0<z<6.4$ and SDSS flux power spectrum measurements, deriving \(m_{\mathrm{WDM}}> 4\,\mathrm{keV}\) and \(m_s > 28\,\mathrm{keV}\) at $2\sigma$; their thermal-relic WDM limit weakened to \(m_{\mathrm{WDM}}> 1.2\,\mathrm{keV}\) when using the HIRES spectra alone. \citet{Boyarsky:2008xj} constrained a mixture of CDM and WDM using similar data, finding that \(m_{\mathrm{WDM}}< 1.1\,\mathrm{keV}\) is excluded at $95\%$ C.L. for WDM fractions down to $f_{\mathrm{WDM}}=0.4$.

Using the 1D flux power spectrum measurement from \(\sim 14000\) Lyman-$\alpha$ forest spectra from SDSS-III/BOSS DR9~\citep{BOSS:2013rpr, BOSS:2012dmf} and \textit{Planck} 2015 CMB data \citep{Planck:2015fie}, \citet{Baur:2015jsy} derived \(m_{\mathrm{WDM}}> 2.96\,\mathrm{keV}\) and \(m_s > 16\,\mathrm{keV}\) at $95\%$ C.L. The addition of \textit{Planck} CMB data weakened the Lyman-$\alpha$ forest-only limits owing to a difference in each dataset's inference of the spectral index of the primordial power spectrum; thus, when using Lyman-$\alpha$ forest data alone, \citet{Baur:2015jsy} found \(m_{\mathrm{WDM}}> 4.09\,\mathrm{keV}\) at $95\%$ C.L. Using BOSS DR9 and high-resolution data from HIRES, MIKE and XQ-100, \citet{Baur:2017stq} considered mixed CDM plus WDM, finding that $m_{\mathrm{WDM}}\geq 0.7~\mathrm{keV}$ at $95\%$ C.L.\ if thermal-relic WDM constitutes $\gtrsim 15\%$ of the DM abundance.

\citet{eBOSS:2018qyj} measured the 1D flux power spectrum from the \(\sim 44000\) highest-quality quasar spectra from the extended Baryon Oscillation Spectroscopic Survey (eBOSS) DR14 sample. \citet{Palanque-Delabrouille:2019iyz} used hydrodynamical simulations and these data to constrain cosmological and DM models, reporting a conservative limit of \(m_{\mathrm{WDM}}> 5.3\,\mathrm{keV}\) at $95\%$ C.L. In this work, a mild tension was reported when inferring the primordial spectral index from \textit{Planck} 2018 CMB and BAO data compared to the eBOSS Lyman-$\alpha$ forest. However, when projecting the data to parameters more directly measured by the Lyman-$\alpha$ forest, i.e., the amplitude and spectral index of $P(k)$ at \(k \sim 1\,h\,\mathrm{Mpc}^{-1}\) and \(z \sim 3\), \citet{Rogers:2023upm} measured a \(\sim 5 \sigma\) discrepancy between CMB and BAO data and the eBOSS Lyman-$\alpha$ forest data. They found that this tension is resolved by a DM fraction of \(\sim 5 \%\) with a thermal relic mass \(m_{\rm WDM} \sim 10\,\mathrm{eV}\). Through improved modeling of the effects of high column density absorbers \citep{Rogers:2017bmq,Rogers:2017eji} and metals, more recent DESI DR1 1D flux power spectrum data (using \(\sim 63000\) quasar spectra) have restored consistency with the CMB \citep{2026arXiv260121432C}.

Sample sizes of medium-resolution Lyman-$\alpha$ forest spectra from SDSS, BOSS, eBOSS, and DESI have grown dramatically, increasing the statistical precision of DM model constraints. In contrast, sample sizes of higher-resolution spectra have remained about the same (\(\sim\) dozens) limited by the number of high-\(z\) quasars that are bright enough (motivating the future use of 30 m-class telescopes). However, measurements of the flux power spectrum have been extended to smaller scales through a combination of a better control over instrumental systematics and improved hydrodynamical modeling. \citet{Viel:2013fqw} used a set of 25 Keck/HIRES and Magellan/MIKE spectra (combined with SDSS-I data) to derive \(m_{\mathrm{WDM}}> 3.3\,\mathrm{keV}\) at $2\sigma$, importantly marginalizing over the timing of hydrogen reionization. \citet{Yeche:2017upn} combined BOSS DR9 and HIRES/MIKE data with a new measurement of the XQ-100 flux power spectrum to place constraints of \(m_{\mathrm{WDM}}> 4.65\,\mathrm{keV}\) and \(m_s > 28.8\,\mathrm{keV}\) at $95\%$ C.L. \citet{Irsic:2017ixq} then combined HIRES/MIKE and XQ-100 data to set a limit of \(m_{\mathrm{WDM}}> 3.5\,\mathrm{keV}\), at the $2\sigma$ level, in their most conservative analysis that allowed for a non-smooth redshift evolution of the IGM temperature. In the same analysis, enforcing smooth IGM temperature evolution yielded \(m_{\mathrm{WDM}}> 5.3\,\mathrm{keV}\) at $2\sigma$, demonstrating the importance of marginalizing over IGM uncertainties. Using similar data and a general $P(k)$ parameterization, \citet{Murgia:2018now} obtained $m_{\mathrm{WDM}}>2.7~\mathrm{keV}$ in their most conservative analysis and $m_{\mathrm{WDM}}>3.6~\mathrm{keV}$ when enforcing a smoother IGM temperature history, both at $2\sigma$; we discuss this work further in the IDM subsection below.

Underscoring the importance of IGM and reionization history modeling, \citet{Garzilli:2019qki} derived $m_{\mathrm{WDM}}>1.9~\mathrm{keV}$ at $95\%$ C.L. by adopting a significantly colder set of reionization histories than previous work in an analysis of the Lyman-$\alpha$ forest data from \citet{Boera:2018vzq}. Meanwhile, \citet{Villasenor:2022aiy} analyzed similar data using a new suite of GPU-accelerated cosmological hydrodynamic simulations to obtain $m_{\mathrm{WDM}}>3.1~\mathrm{keV}$ at $95\%$ C.L. These authors point out that their likelihood weakly peaks at $m_{\mathrm{WDM}}=4.5~\mathrm{keV}$, hinting at a slight tension between CDM predictions and the \citet{Boera:2018vzq} data in their analysis. More recently, by exploiting the smallest-scale flux power spectrum measurements to-date, \citet{Irsic:2023equ} derived \(m_{\mathrm{WDM}}> 5.7\,\mathrm{keV}\) at $95\%$ C.L.; when limiting their analysis to comparable scales and thermal histories as previous studies, this limit weakens to \(m_{\mathrm{WDM}}> 4.1\,\mathrm{keV}\). \citet{Garcia-Gallego:2025kiw} used similar data to constrain mixed WDM and CDM, finding $f_{\mathrm{WDM}}<0.16$ for $m_{\mathrm{WDM}}=1~\mathrm{keV}$ at $2\sigma$, and weaker bounds on the fraction for increasing $m_{\mathrm{WDM}}$.

\textbf{Fuzzy Dark Matter}: Early Lyman-$\alpha$ forest analyses concentrated on thermal and non-thermal WDM as benchmark scenarios beyond CDM. However, with improvements to data quantity, precision, and accuracy, IGM modeling, and the introduction of precision and probabilistic ML emulator methods, it has become possible to distinguish different DM models by the effects of their \textit{ab initio} transfer functions on the Lyman-$\alpha$ forest.

We concentrate on constraints using hydrodynamical simulations where the physics of the DM model has been explicitly included, usually by modifying the initial conditions; these constraints are the most accurate in the literature. Since the first constraints on the amplitude \(\Delta_\star\) and spectral index \(n_\star\) of $P(k)$ from the Lyman-$\alpha$ forest \citep{Croft:1998pe}, the joint posterior of \(\Delta_\star\) and \(n_\star\) has nonetheless been used as an approximate compression to compare to dark matter model transfer functions, negating the need to run dedicated simulations. However, for much of the DM model space, this compression is highly lossy since the transfer functions can have strong local curvature that is not captured by a power-law approximation (e.g., power spectrum cutoffs and oscillations, see Section~\ref{sec:scenarios}). More recently, constraints on WDM-like transfer functions have been ``translated'' to other models. This translation is usually achieved by asserting that the DM model under consideration should suppress $P(k)$ more than the constrained WDM model at all observable wavenumbers. This method leads to approximate and usually conservative model bounds; \citet{Dienes:2021cxp} discuss the limitations of such translation approaches.

\citet{Amendola:2005ad} used compressed $P(k)$ likelihoods from Lyman-$\alpha$ forest analyses by \citet{Croft:2000hs} and \citet{SDSS:2004kqt} to limit the FDM particle mass \(m_\mathrm{FDM} > 10^{-23}\,\mathrm{eV}\). Using HIRES/MIKE and XQ-100 data with a dedicated suite of hydrodynamical simulations, \citet{Irsic1711903} then derived \(m_\mathrm{FDM} > 2\,\times\,10^{-21}\,\mathrm{eV}\) at $2\sigma$. \citet{Kobayashi:2017jcf} considered mixed CDM and FDM, limiting FDM to contribute no more than 30\% of the total DM abundance for \(10^{-23}\,\mathrm{eV} \lesssim m_\mathrm{FDM} \lesssim 10^{-21}\,\mathrm{eV}\) at $2\sigma$. Adding BOSS DR9 data, \citet{Armengaud:2017nkf} derived \(m_\mathrm{FDM} > 2.9\,\times\,10^{-21}\,\mathrm{eV}\) at $95\%$ C.L. \citet{Nori:2018pka} found that the inclusion of a fluid approximation to full FDM dynamics (though still neglecting full wave effects) does not change Lyman-$\alpha$ forest limits on \(m_\mathrm{FDM}\) given the precision of current data. On the other hand, \citet{Wang:2026wwq} found that wave dynamics may affect the flux power spectrum by modifying the DM velocity distribution, although this has not been verified using hydrodynamical FDM simulations. If confirmed, this result would imply that current FDM limits from the Lyman-$\alpha$ forest are conservative, and would be the first demonstration of \textit{in situ} FDM effects observable in the Lyman-$\alpha$ forest.

Several recent studies have applied emulator methods to the cosmological analysis of Lyman-$\alpha$ forest data~\citep{Rogers:2018smb,Bird:2018efe,Rogers:2020cup,Pedersen:2020kaw}. Unlike previous approaches using linear interpolation of simulated flux power spectra as a function of each model parameter individually, the use of Gaussian process regression captures the covariance of model parameters with a more flexible surrogate model and global uncertainty quantification. \citet{Rogers:2020ltq} used an emulator trained on hydrodynamical simulations and the smallest-scale HIRES/UVES flux power spectrum data available to limit \(m_\mathrm{FDM} > 2\,\times\,10^{-20}\,\mathrm{eV}\) at $95\%$ confidence. \citet{Rogers:2023upm} then considered the mixed CDM and FDM case using CMB, BAO, and eBOSS Lyman-$\alpha$ forest data, finding that the \(\Lambda\)CDM parameter tension is addressed by a FDM fraction of \(\sim 5\%\) with \(m_\mathrm{FDM} \sim 10^{-25}\,\mathrm{eV}\) \citep[however, see][for updated DESI flux power spectrum cosmological constraints]{2026arXiv260121432C}. Meanwhile, \citet{Liu:2026jkq} combined cosmological hydrodynamic simulations, a neural network emulator without the application of previous active learning methods, and 15 high-resolution HIRES and UVES spectra to FDM, finding $m_{\mathrm{FDM}}>1.9\times 10^{-21}~\mathrm{eV}$ in the pure case and $f_{\mathrm{FDM}}<0.07$, $0.12$, and $0.65$ for $m_{\mathrm{FDM}}=10^{-23}$, $10^{-22}$, and $10^{-21}~\mathrm{eV}$, respectively, at $95\%$ C.L.

\citet{Irsic:2019iff} used an interpolated grid of simulations originally designed for primordial black hole DM analyses~\citep{Murgia:2019duy} to constrain ultralight axions with a post-inflationary symmetry breaking that enhances the small-scale matter power spectrum. They found a symmetry breaking scale \(f_\mathrm{a} \gtrsim 10^{15}\,\mathrm{GeV}\) for axion masses \(m_\mathrm{a} \lesssim 10^{-17}\,\mathrm{eV}\) at $2\sigma$. \citet{Garcia-Gallego:2026phh} then searched for post-inflationary axion isocurvature enhancement to the small-scale matter power spectrum, finding in their most conservative analysis that \(m_\mathrm{a} > 1.73\,\times\,10^{-18}\,\mathrm{eV}\) at $95\%$ C.L. In related ultralight DM models with both a free-streaming suppression and Poisson-like enhancement in $P(k)$, \citet{Amin:2022nlh} placed a lower bound of $10^{-19}~\mathrm{eV}$ on the DM particle mass based on Lyman-$\alpha$ forest $P(k)$ constraints; \citet{Long:2024cak} then set limits on the DM mass as a function of its characteristic momentum $q_*$, finding that eBOSS Lyman-$\alpha$ forest $P(k)$ data exclude DM particle masses below $5\times 10^{-20}~\mathrm{eV}\times (q_*/10^3~h~\mathrm{Mpc}^{-1})$ at $90\%$ C.L.

\textbf{Interacting Dark Matter}: \citet{Rogers:2021byl} used the HIRES/UVES small-scale Lyman-$\alpha$ forest data and an emulator trained on dedicated simulations to limit DM--proton scattering IDM for the velocity-independent cross section case (\(n = 0\)) and DM particle masses \(10\,\mathrm{keV} \leq m_\mathrm{DM} \leq 100\,\mathrm{GeV}\); for example, these authors derived \(\sigma_{\mathrm{DM}-p} < 6\,\times\,10^{-30}\,\mathrm{cm}^{2}\) for \(m_\mathrm{DM} = 100\,\mathrm{keV}\) at $95\%$ confidence. Velocity-dependent DM-proton cross sections have only been constrained using the lossy $P(k)$ amplitude and spectral index compression \citep{Dvorkin:2013cea,Xu189710} or a WDM translation method based on $P(k)$ integrals (the ``area criterion''); in this way, \citet{Buen-Abad:2021mvc} limited \(\sigma_{\mathrm{DM}-p} < 4\,\times\,10^{-34}\,\mathrm{cm}^{2}\) for \(m_\mathrm{DM} \lesssim 100\,\mathrm{MeV}\) for \(n = -2\) at $95\%$ C.L. \citet{Slatyer:2018aqg} additionally limited DM--proton cross sections using IGM heating constraints from the Lyman-$\alpha$ forest \citep{Munoz:2017qpy}, although such constraints tend to be much weaker than using the small-scale structure as traced by the Lyman-$\alpha$ forest. \citet{Buen-Abad:2021mvc} also used the area criterion to probe velocity-dependent DM-electron cross sections to derive, e.g., \(\sigma_{\mathrm{DM}-e} < 3\,\times\,10^{-33}\,\mathrm{cm}^{2}\) for \(m_\mathrm{DM} \lesssim 100\,\mathrm{keV}\) for \(n = -2\) at $95\%$ C.L. \citet{Bose:2018juc} demonstrated that DAOs in $P(k)$ can manifest in the Lyman-$\alpha$ forest flux power spectrum, explicitly considering ETHOS transfer functions. \citet{Yuan:2026zlr} analyzed the impact of ETHOS-like IDM models on the Lyman-$\alpha$ forest using dedicated hydrodynamical simulations and a hybrid deep learning/Gaussian process emulator. These authors found that DM that produces DAOs in $P(k)$ cannot comprise more than 20\% of the DM, if the DAO peaks at \(k \sim 1\,h\,\mathrm{Mpc}^{-1}\), or more than 30\%, if the DAO peaks at \(k \sim 50\,h\,\mathrm{Mpc}^{-1}\).

The Lyman-$\alpha$ forest flux power spectrum is sensitive not only to a strong feature in $P(k)$ like a cutoff, but can also distinguish between transfer function shapes. In order to exploit this property of the Lyman-$\alpha$ forest, there is a body of literature that constrains general parameterizations of the transfer function that are tied to a family of DM models rather than specific scenarios. For example, \citet{Murgia:2018now} constrained the linear matter power spectrum through a generalization of the WDM transfer function (Equation~\ref{eq:transfer_wdm}): \(T(k) \equiv \left[1 + (\alpha k)^\beta\right]^\gamma\), where \(\alpha\) sets a suppression scale and \(\beta\) and \(\gamma\) set the shape of the suppression. They found, using HIRES/MIKE data and for the family of transfer functions represented by this parameterization, that \(\alpha < 0.03\,\mathrm{Mpc}\,h^{-1}\), equating to a suppression wavenumber limit \(k \gtrsim 33\,h\,\mathrm{Mpc}^{-1}\), at $2\sigma$ (note that this analysis does not use the most up-to-date smallest-scale HIRES/UVES data). \citet{Archidiacono:2019wdp} used the \((\alpha,\beta,\gamma)\) likelihood from this work to constrain DM--dark radiation interactions, relying on the translation methods discussed above to map from smooth to oscillatory transfer functions, and \citet{Miller:2019pss} used a similar approach to constrain DM produced nonthermally during an epoch of early matter domination. \citet{Hooper:2021rjc} used the same likelihood to probe DM--neutrino interactions, and \citet{Hooper:2022byl} extended this approach to account for mixed DM models by introducing an additional parameter to capture the fraction of non-cold DM in $T(k)$. Finally, \citet{Mosbech:2026xcc} placed the strongest current limits on DM--neutrino interactions using dedicated hydrodynamical simulations and the HIRES/Keck and UVES/VLT spectra from \citet{Boera:2018vzq}.

\textbf{Decaying Dark Matter}: \citet{Wang:2013rha} used the Lyman-$\alpha$ forest to probe the nonlinear power spectrum suppression induced by the non-relativistic kick to a daughter particle in two-body DDM. They found, given SDSS-I data, that the kick velocity \(V_\mathrm{kick} \lesssim 30 - 70\,\mathrm{km}\,\mathrm{s}^{-1}\) for decay lifetimes \(\tau \lesssim 10~\mathrm{Gyr}\) at $1\sigma$. The Lyman-$\alpha$ forest is sensitive to DDM not just through effects on the clustering of matter arising from the decay, but also from heat injection that can arise in the IGM, the temperature of which is constrained by Lyman-$\alpha$ forest observations (see discussion above). \citet{Liu:2020wqz} used a compendium of BOSS/XQ-100/HIRES/MIKE flux power spectra and higher-redshift Lyman-$\alpha$ transmission spike observations to constrain DM decays and annihilations by spurious IGM overheating. \citet{Capozzi:2023xie} extended such an analysis to sub-keV DM particle masses.

\subsubsection{Summary}

\NEW{The Lyman-$\alpha$ forest provides the most direct constraints on $P(k)$ (and thus \emph{ab initio} DM physics) among the established small-scale structure probes discussed in this review, with sensitivity to wavenumbers as large as $k \sim 40~{\rm Mpc}^{-1}$ at $z \sim 2$--$5$. For WDM, Lyman-$\alpha$ forest limits range from $m_{\mathrm{WDM}}\gtrsim 3$ to $6~\mathrm{keV}$, with the spread primarily driven by IGM thermal history modeling assumptions. For FDM, limits of $m_{\rm FDM} \gtrsim 10^{-20}~{\rm eV}$ from emulator-based analyses are among the tightest limits from any small-scale structure probe, and similar analyses yield stringent constraints on IDM models that produce cutoffs in $P(k)$.}

\NEW{Because of its more direct access to $P(k)$ than other small-scale structure probes, a particularly exciting aspect of Lyman-$\alpha$ forest analyses is their sensitivity to $P(k)$ shapes and features beyond a WDM-like cutoff. For example, Lyman-$\alpha$ forest constraints on fractional non-CDM components that suppresses power (e.g., WDM or FDM) probe non-CDM fractions down to $\sim 10\%$, whereas analyses of similar models using dwarf galaxies or strong lensing typically lose sensitivity below non-CDM fractions of $\sim 50\%$. Similarly, the effects of oscillatory features in $P(k)$ in IDM models leave imprints in the Lyman-$\alpha$ forest that are distinct from ``effective'' WDM models. As a result, upcoming Lyman-$\alpha$ forest analyses have the potential to not only place improved limits on $P(k)$, but to characterize any deviations from CDM in a way that can distinguish between different signatures of \emph{ab initio} DM physics.}

\NEW{The Lyman-$\alpha$ forest probes higher redshifts than other established small-scale structure probes, allowing it to test redshift-dependent effects of both \emph{ab initio} and \emph{in situ} DM physics. For WDM-like models, the delayed growth of structure enhances the constraining power of the Lyman-$\alpha$ forest compared to measurements on the same comoving scales at lower redshifts; this effect is built into the constraints discussed above. On the other hand, \emph{in situ} DM physics such as two-body decays can reverse this redshift dependence, such that higher-redshift structure is CDM-like and deviations increase as time progresses. The Lyman-$\alpha$ forest therefore plays a critical role in the search for new DM physics because it can more specifically characterize any detection of a deviation from CDM from lower-redshift probes on smaller comoving scales.  Vice versa, a Lyman-$\alpha$ forest detection of deviations in $P(k)$ from the CDM expectation would lead to specific predictions for structure at lower redshifts in each viable DM scenario that strong lensing and dwarf galaxies could then test.}

\NEW{Finally, the Lyman-\(\alpha\) forest measured on intermediate scales \(k \sim (1 - 10)\,\mathrm{Mpc}^{-1}\) is also complementary to smaller-scale probes. For example, the detection of a power spectrum suppression or enhancement at \(k \sim 100\,\mathrm{Mpc}^{-1}\) may not be able to adjudicate the origin between a dark matter model and spectral running in the primordial power spectrum. However, by combining probes across many decades in wavenumbers, the full transfer function can be measured.}

\section{Emerging Probes}
\label{sec:emerging}

This section summarizes current and forecasted DM limits from emerging probes of small-scale structure: stellar streams (Section~\ref{sec:streams}), high-redshift galaxies (Section~\ref{sec:high_z}), weak lensing (Section~\ref{sec:weak_lensing}), and 21-cm cosmology (Section~\ref{sec:21cm}). Constraints from these probes are rapidly improving; we discuss prospects for upcoming measurements and theoretical advances in Section~\ref{sec:outlook}.

\subsection{Stellar Streams}
\label{sec:streams}

Stellar streams---the disrupted remnants of globular clusters (GCs) and dwarf galaxies---have rapidly gained interest as a small-scale structure probe. Stellar streams orbiting the MW were first observed early in the era of large ground-based surveys \citep[e.g.,][]{Odenkirchen:2001, Majewski:2003}; stream observations have advanced rapidly with Gaia (see \citealt{Bonaca:2024dgc} for a review). Two main classes of stellar stream observables have been used to test DM physics:
\begin{enumerate}[label=(\roman*)]
    \item Individual detection: The phase-space distribution of stream stars can be used to search for perturbations from DM subhalos; in turn, the properties of these perturbers constrain DM physics. Perturbations in real space (e.g., density, track, and width variations) and velocity space (e.g., kinks, cocoons, and dynamically hot components) can be used as indicators of gravitational perturbations. 
    \item Statistical detection: The statistical properties of stream populations, including their mass function and radial distribution, constrain the underlying population of stream progenitors (i.e., dwarf galaxies and GCs); these progenitors' properties can probe departures from CDM.
\end{enumerate}

\subsubsection{Individual Perturbers}

As the most extensively studied kinematically cold stellar stream, GD-1 provides a benchmark for using stream perturbations to test DM physics. GD-1 was discovered by \citet{Grillmair:2006bd}. It spans $\sim$100~deg on the sky at a heliocentric distance that ranges from $\sim 8$--$10~\mathrm{kpc}$ \citep{deBoer2018GD1}. GD-1 is remarkably narrow ($\approx 70~\mathrm{pc}$ linear width) and kinematically cold ($\approx 1~\mathrm{km\ s}^{-1}$ velocity dispersion). These observations inspired projections for the number of perturbations due to subhalo impacts detectable with more precise data (e.g., \citealt{Yoon2010ClumpyStreams,Carlberg:2013gxa,Erkal2016Gaps}). With improved measurements and the advent of Gaia, GD-1 stars were localized and measured with sufficient precision to detect a prominent gap and spur~\citep{Price-Whelan:2018}.

By comparing these data to impact models based on releasing test particles from the Lagrange points of the disrupting progenitor, \citet{Bonaca181103631} argued that the GD-1 gap-and-spur feature is caused by an impact with a compact ($\lesssim 10~\mathrm{pc}$) perturber \NEW{with a stripped mass in the range} $\approx 10^5$ to $10^8~M_{\mathrm{\odot}}$. If the perturber is interpreted as a subhalo, these high densities are extremely rare for CDM systems based on current simulation results~\citep{Zhang:2026oui}. Using controlled N-body simulations of SIDM halos, \citet{Zhang:2024fib} argued that sufficiently dense perturbers are naturally produced in SIDM models that feature core collapse, i.e., with cross sections of $30$ to $100~\mathrm{cm^2~g}^{-1}$ at $V_{\mathrm{max}}\approx 10~\mathrm{km\ s}^{-1}$. The SIDM models shown in Figure~\ref{fig:xsec}, which were motivated by other small-scale probes, have comparable effective cross sections of $\sigma_{\mathrm{eff}}/m\approx 30$--$140~\mathrm{cm^2~g}^{-1}$ at this velocity scale; see \citet{Yu:2025tmp} for further discussion.

Studies that use specific perturbation features in streams like GD-1 to probe DM substructure face several challenges. First, baryonic structures (rather than DM subhalos) can also impact streams; for example, in the case of GD-1, \citet{deBoer2018GD1} argue that an interaction with the Sagittarius dwarf galaxy can create features similar to the observed spur. GCs are also sufficiently dense to create the GD-1 gap-and-spur features, although \citet{Bonaca181103631} argued that no known GCs are on orbits compatible with a GD-1 interaction, and \citet{Doke:2022jro} ran a suite of test-particle simulations confirming that the creation of a gap through a GC impact was unlikely. \NEW{Forecasts indicate that upcoming stream data can reveal a few significant gaps in several streams, probing subhalos with stripped masses down to $\sim 10^6~M_{\mathrm{\odot}}$ and WDM-like models similar to those currently constrained by MW satellite abundances~\citep{Menker:2024koc,Lu:2025qbp,Lu:2026zni}, underscoring the importance of controlling these systematics.}

\subsubsection{Statistical Constraints}

Beyond GD-1, a growing number of MW streams show evidence of perturbations. For example, Pal 5 hosts a well-studied system of gaps that were analyzed by \citet{Erkal:2017}. However, Pal 5's prograde orbit and proximity to the Galactic disk make it particularly susceptible to perturbations from the rotating bar~\citep{Pearson:2017}, limiting its DM perturber constraining power relative to more distant, retrograde streams. Meanwhile, the ATLAS-Aliqa Uma stream hosts a prominent kink whose morphology is consistent with a close passage of the Sagittarius dwarf galaxy \citep{Li:2021}, suggesting that higher-mass luminous substructure may be responsible for at least some observed stream features; this underscores the importance of accounting for known baryonic perturbers before attributing stream morphology to DM substructure. In contrast, the Jet stream (a cold, retrograde stream at large Galactocentric distance) exhibits a gap-and-spur morphology \citep{Ferguson:2022} and is expected to be relatively insensitive to disk and bar perturbations, making it a clean target for future DM studies. Thus, from the standpoint of DM constraints, the most powerful kind of stream may ultimately be a smooth one, since a stream with no detectable gaps, spurs, or other features places a robust upper limit on subhalo abundances and/or concentrations. In Section~\ref{sec:observation_future}, we discuss how upcoming wide-area surveys will extend stream discoveries to larger Galactocentric radii, substantially expanding the sample of MW streams well-suited to DM substructure analyses.

In addition, the dynamics of globular cluster disruption~\citep{Weatherford2025StreamsBHRetention} and interactions with both the MW galaxy and giant molecular clouds can significantly alter the structure of GC streams~\citep{Banik:2018pal,Arora2026NoStreamLeftUnscathed}, effectively constituting a ``noise background'' for subhalo perturbation measurements. On the other hand, interactions between streams and dwarf galaxies can be used to probe dwarfs' mass profiles, which are sensitive to DM properties; for example, \citet{Foote:2024ral} showed that the collision between the Cetus-Palca stream and the Segue 2 UFD yields significantly different predictions for the stream's velocity dispersion depending on Segue 2's inner enclosed mass, which will be tested by upcoming observations.

Meanwhile, \citet{Malhan2020AccretedGCStreams} argued that the widths and velocity dispersions of streams from GCs that evolved within dwarf galaxies before accreting onto the MW probe the DM density profiles of their parent dwarf galaxies' halos. \citet{Malhan:2022nfe} applied this idea to GC streams observed by Gaia, finding that their properties are consistent with evolution inside cored subhalos with \NEW{infall masses} of $\approx 10^8$ to $10^9~M_{\mathrm{\odot}}$, but not with evolution inside cuspy subhalos \NEW{with infall masses} above $\approx 10^9~M_{\mathrm{\odot}}$. These results can potentially constrain DM models like SIDM that alter the density profiles of low-mass subhalos.

Given the complexities associated with fitting individual stream features, an alternative strategy is to use statistical measurements, including density power spectra, velocity dispersion, and stream width, to infer the properties of stream perturbers. For example, \citet{Bovy:2016irg} developed a method based on linear perturbation theory for simulating the effects of subhalo impacts on stream power spectra, finding that the Pal 5 stream data can be explained by interactions with $10^{+11}_{-6}$ subhalos with \NEW{stripped masses} between $10^{6.5}$ and $10^9~M_{\mathrm{\odot}}$. \citet{Banik:2019cza} extended this method to include baryonic perturbers and jointly analyze the density power spectra of GD-1 and Pal 5. These authors inferred that a population of low-mass subhalos (\NEW{with stripped masses of} $\approx 10^7$ to $10^9~M_{\mathrm{\odot}}$) with an SHMF amplitude of $\approx 40\%$ relative to CDM-only predictions likely impacted these streams; at the distances of GD-1 and Pal 5, this SHMF amplitude is consistent with expectations for the suppression of the SHMF due to stripping by the Galactic disk~\citep{Garrison-Kimmel170103792,Nadler171204467,Kelley:2018pdy,2020MNRAS.499..116W,Wang:2024moc}. Furthermore, these authors derived WDM mass constraints $m_{\mathrm{WDM}}>3.6~\mathrm{keV}$ at $95\%$ confidence based on a joint analysis of GD-1 and Pal 5; this limit strengthened to $m_{\mathrm{WDM}}>4.6~\mathrm{keV}$ when analyzing GD-1 alone. Then, \citet{Banik:2019smi} used these SHMF measurements to constrain non-CDM models, deriving $m_{\mathrm{WDM}}>3.6~\mathrm{keV}$ and $m_{\mathrm{FDM}}>1.4\times 10^{-21}~\mathrm{eV}$ at $95\%$ confidence when combining GD-1 and Pal 5 data.
In this analysis, DM models that produce fewer subhalos than these ruled-out WDM and FDM scenarios cannot explain the high amplitude of these streams' density power spectra on degree scales, even when perturbations from baryonic structures are modeled.

A complementary approach leverages the collective effect of heating from subhalo impacts to constrain the underlying subhalo population and thus DM properties. For example, using WDM zoom-in simulations, \citet{Carlberg:2024jne} predicted that models with $m_{\mathrm{WDM}}=5.5$ and $7~\mathrm{keV}$ respectively produce exponential wings in the stellar radial velocity distribution of a mock stream of $3$ and $4~\mathrm{km\ s}^{-1}$, while CDM yields $6~\mathrm{km\ s}^{-1}$ wings. Similar ideas can be used to test FDM models with $m_{\mathrm{FDM}}\approx 10^{-22}~\mathrm{eV}$, which heat streams due to density fluctuations on the de Broglie scale~\citep{Dalal:2020mjw}. 
For example, \citet{Amorisco:2018dcn} obtained $m_{\mathrm{FDM}}>1.5\times 10^{-22}~\mathrm{eV}$ by placing an upper bound on this heating effect using pre-Gaia stream data.

Recently, \citet{Nibauer:2025ezn} reanalyzed GD-1 using the perturbative modeling framework for stream impacts from \citet{Nibauer:2024uue}. These authors found that the observed velocity dispersion of GD-1 can be reproduced as a result of numerous impacts with subhalos \NEW{that have stripped masses} $\lesssim 10^8~M_{\mathrm{\odot}}$, inferring a mass fraction in subhalos of $f_{\mathrm{sub}}=0.05^{+0.08}_{-0.03}$ at $68\%$ confidence, consistent with CDM-only predictions for substructure abundance at the distance of GD-1. However, these authors also argued that more compact subhalos than CDM are preferred by the data, consistent with the results based on the gap-and-spur structure from \citet{Bonaca181103631}. Complementary perturbative frameworks for stream analysis have been developed in \citet{Delos:2021ouc} and \citet{AnauMontel:2026kye}.

Statistical properties of stream populations can also probe DM models. For example, DM physics that enhances satellites' mass loss rates can affect dwarf galaxy stream properties. In this vein, \citet{Hainje:2025wtx} used tailored N-body simulations to show that the disruption of the Sagittarius dwarf galaxy differs between CDM and SIDM models with cross sections of $\approx 30~\mathrm{cm^2~g}^{-1}$ at Sagittarius' orbital velocity scale. Specifically, very little DM remains bound to the Sagittarius progenitor in the SIDM case due to subhalo--host halo interactions, whereas the CDM system retains a large amount of DM in its inner regions; the same physics is expected to increase the number of disrupted dwarf galaxies in SIDM compared to CDM (e.g., \citealt{Nadler:2025jwh}). Conversely, DM physics that reduces satellite mass loss rates causes streams to form later and retain more DM compared to CDM (e.g., \citealt{MandacaruGuerra:2026zqk}).

In this context, we note that predictions for satellite disruption and thus dwarf galaxy streams depend on baryonic physics modeling. Comparisons between hydrodynamic simulations and observed stream populations have recently been developed. For example, \citet{S5:2022zzx} showed that the mass function of observable streams from FIRE-2 simulations is consistent with current data, although FIRE-2 streams form at much larger pericenters than observed; \citet{Shipp2024AurigaStreamsII} reached similar results using Auriga. Thus, it will be important to further study stream predictions from hydrodynamic simulations, including their sensitivity to feedback modeling and resolution (e.g., see \citealt{Riley2024AurigaStreamsI}).

\subsubsection{Dynamical Signatures}

Beyond subhalo perturbers, stellar streams can also constrain the shape of the MW's DM halo. Stream tracks are sensitive to the flattening and triaxiality of the host halo potential due to orbital precession in non-spherical potentials. For example, \citet{Bovy:2016chl} used the phase-space tracks of Pal 5 and GD-1 to measure that the ratio of vertical to radial accelerations at the locations of these streams are consistent with a nearly spherical MW halo potential within 20 kpc. This result is in mild tension with CDM-only simulations, which generally predict mildly prolate or triaxial halos~\citep[e.g.,][]{Allgood:2005eu}, although adding baryons tends to produce more spherical shapes~\citep[e.g.,][]{Chua:2019}. Since SIDM can also isotropize the inner regions of halos~\citep{Peter:2012jh,Sameie:2018}, shape measurements from streams can potentially probe SIDM. Other dynamical probes of low-mass MW subhalos include (but are not limited to) pulsar accelerations~\citep{Chakrabarti:2025hfk}, wide binaries in dwarf galaxies~\citep{Olea-Romacho:2026pgn}, and the phase-space distribution of stars in the MW halo~\citep{Buschmann:2017ams}, or disk~\citep{Feldmann:2013hqa}.

A related consideration for stream-based DM analyses is the perturbation to the MW's halo structure from the recent infall of the LMC, which generates pronounced density and kinematic wakes in both the DM and stellar halos~\citep{Garavito-Camargo:2019kxw}. These wakes displace streams from their expected orbits in a smooth host potential, constituting an effective background that must be modeled when attributing stream features to low-mass DM substructure. At the same time, the amplitude and morphology of the MW halo's response to the LMC encodes information about the LMC's halo properties and the structure of its dynamical friction wake, which could in turn probe the nature of DM (e.g., \citealt{Foote:2023}).

\subsubsection{Summary}

\NEW{Stellar streams have rapidly emerged as a small-scale structure probe, primarily through analyses of stream gaps, density power spectra, and velocity dispersions. A key strength of streams is their potential sensitivity to MW subhalos below the galaxy formation threshold, which complements strong lensing measurements of similarly low-mass halos along cosmological lines of sight. The most quantitative DM constraints currently come from analyses of the GD-1 and Pal 5 streams, which reach $m_{\rm WDM} \gtrsim 4~{\rm keV}$ from streams alone and $\gtrsim 6~{\rm keV}$ when combined with MW satellite abundances. In addition, perturbation features in individual streams like the GD-1 gap and spur can constrain the properties of subhalo perturbers; current analyses of GD-1 prefer dense perturbers that can (when interpreted as subhalos) arise in SIDM models with $\sigma/m\gtrsim 30~\mathrm{cm^2~g}^{-1}$ at $v\sim 10~\mathrm{km\ s}^{-1}$.}

\NEW{A key systematic for DM inference using streams is the impact of baryonic perturbations, including the time-evolving MW potential, which is partially degenerate with subhalo impact signal. In addition, accurately modeling the low-mass subhalo population of the MW, particularly at small Galactocentric radii where subhalos experience significant tidal stripping, remains challenging and can be improved by combining higher-resolution cosmological simulations and SAMs calibrated to those simulations. Deriving robust DM constraints from upcoming stream data will require population-level analyses of streams with different intrinsic properties, orbits, and perturbation features to disentangle the statistical impact of baryonic versus DM structure.}

\subsection{High-$z$ Galaxies}
\label{sec:high_z}

When searching for \textit{ab initio} DM model effects, it is useful to find DM tracers close to the epoch at which the initial conditions are imprinted. In the future, it will be possible to trace DM fluctuations in the cosmic dark ages, before the first stars and galaxies form, by measuring the 21-cm signal from neutral hydrogen (Section~\ref{sec:outlook}). However, observations of the first galaxies with the James Webb Space Telescope (JWST) extending to \(z \sim 14\) and beyond are already beginning to constrain DM physics. According to the hierarchical model of structure formation, these galaxies sit in DM halos and so their abundance and spatial clustering is a tracer of the matter power spectrum at high redshift. At $z\gtrsim 6$, the comoving scales associated with these halos are still, on average, linear; the halos that host bright galaxies correspond to rare density peaks (Section~\ref{sec:small-scale}). While the clustering of galaxies has been detected over a range of scales at low \(z\), the lower number of observed galaxies at higher \(z\) means that the most precise statistic in constraining the nature of DM has so far been the UVLF, i.e., the number density of galaxies as a function of magnitude and redshift.

The UVLF is related to the abundance of DM halos, i.e., the halo mass function (HMF). The HMF, by the excursion set formalism, is related to an integral of $P(k)$ \citep{Sheth:1999mn}. To first order, a suppression (enhancement) in the small-scale matter power spectrum will lead to a suppression (enhancement) of lower-mass halos, in turn leading to a suppression (enhancement) of fainter galaxies. The primary astrophysical uncertainties are exactly how galaxies trace halos (relating stellar mass to halo mass) and the burstiness of star formation (important for relating luminosity to stellar mass). These questions can be addressed through cosmological hydrodynamical simulations, including both feedback and radiative transfer effects to model the ionizing background and its interplay with galaxy formation \citep[e.g.,][]{Kannan:2021xoz,Shen:2023lsf}. It is not currently feasible to run these simulations in sufficient number to span the full parameter space to test DM models even with the aid of ML emulator methods. The current state of the art is to use a probabilistic analytical model, calibrated to simulations and with astrophysical parameters that are marginalized when setting DM parameter limits. This approach has led to constraints on both DM models (discussed below) and on $P(k)$ itself (e.g., \citealt{2022PhRvD.105d3518S,Munoz:2026diy}; also see Figure~\ref{fig:p_k}). In practice, this approach means modeling a probability that a galaxy of a given UV luminosity arises from a halo of a given mass \NEW{at the relevant redshift}, with a mean relation that includes parametric uncertainty on star formation at the low-mass end of the HMF and gas feedback at the high-mass end. \NEW{In addition, Eddington bias corrections to the UVLF, which arise from asymmetric scatter due to the steepness of the underlying HMF, affect the inferred galaxy--halo connection and must be modeled to derive robust DM constraints~\citep{Krishnan:2025dti}.}

Complementary to the UVLF are measurements of the timing of cosmic reionization, in particular the optical depth \(\tau\), which is determined from both CMB and quasar spectra observations. An earlier reionization implies a population of star-forming galaxies at earlier times, in turn requiring a more abundant population of halos to host the galaxies and more power in the initial conditions. The number density of high-$z$ gamma-ray bursts can be used to place analogous constraints~\citep{Wei:2026}.

\subsubsection{Constraints}

\textbf{Warm Dark Matter}: \citet{Barkana:2001gr} modified the ePS formalism to account for the effects of WDM on the HMF. Then, using this formalism and considering constraints at the time that reionization completes by \(z \sim 5.8\), they set a limit on the thermal-relic WDM mass of \(m_\mathrm{WDM} \gtrsim 0.4\,\mathrm{keV}\) in their most conservative analysis of reionization timing. These authors also considered future measurements of the UVLF from the Next Generation Space Telescope (what is now JWST), forecasting sensitivity on \(m_\mathrm{WDM}\) up to 1 keV. In contrast, first data from the WMAP CMB satellite inferred \(\tau = 0.17 \pm 0.04\), implying a reionization redshift \(z = 17 \pm 5\) \citep{WMAP:2003ggs}. Such an early reionization inference heavily constrained DM models like WDM with small-scale suppression \citep{Somerville:2003sh}. Over time, the preferred value of \(\tau\) has dropped significantly \citep[\textit{Planck} CMB data infers \(\tau = 0.059 \pm 0.006\);][]{Pagano:2019tci} and so inference on the end of reionization has returned to \(z < 6\), also supported by quasar spectra observations.

\citet{Pacucci:2013jfa} used the first detection of two \(z \sim 10\) galaxies from the Cluster Lensing And Supernova survey with Hubble (CLASH) to limit the number density of high-\(z\) halos, thereby constraining \(m_\mathrm{WDM} > 0.9\,\mathrm{keV}\) at $99.9\%$ confidence. \citet{Schultz:2014eia} then combined DM-only \(N\)-body simulations with a SMHM relation to compare to galaxy counts from the Hubble Ultra Deep Field up to \(z \sim 8\), finding \(m_\mathrm{WDM} > 1.3\,\mathrm{keV}\) at $2.2\sigma$; these authors also considered optical depth inferences, finding that a $m_{\mathrm{WDM}}=2.6~\mathrm{keV}$ model requires a high escape fraction to match \textit{Planck} measurements. In a similar vein, \citet{Dayal:2014nva} applied a galaxy formation model to DM merger trees from cosmological simulations, finding that the UVLF in a \(m_\mathrm{WDM} =1.5\,\mathrm{keV}\) model is significantly suppressed and flattened compared to CDM.

Next, \citet{2016ApJ...818...90M} combined ePS merger trees and a galaxy--halo connection model to predict the \(z \sim 2\) UVLF, finding that \(m_\mathrm{WDM} > 1.5\,\mathrm{keV}\) is required to match HST UVLF measurements in their most conservative analysis; these authors also constrained Shi--Fuller sterile neutrino WDM, finding \(m_s > 4\,\mathrm{keV}\). \citet{Corasaniti:2016epp} then exploited the Hubble Frontier Fields UVLF for \(6 \lesssim z \lesssim 8\) to limit \(m_\mathrm{WDM} > 1.5\,\mathrm{keV}\) at $2\sigma$, using DM-only \(N\)-body simulations and halo abundance matching to build a probabilistic relation between star formation rates and halo masses. In addition, these authors constrained the redshift of non-thermal DM production, finding a formation redshift $z_\mathrm{T}>8\times 10^5$ at $2\sigma$. Finally, \citet{Rudakovskyi:2021jyf} used more explicitly Bayesian methods to limit \(m_\mathrm{WDM} > 2\,\mathrm{keV}\) at $95\%$ confidence, again using HST data.

The launch of JWST in 2021 has transformed observations of the first populations of galaxies. In particular, JWST is finding evidence of galaxies at unprecedentedly high redshifts; for example, there are claims of galaxy candidates up to \(z \sim 25\), although only photometric evidence exists to support this at the moment~\citep{Perez-Gonzalez:2025bqr}. Spectroscopic observations are needed to more conclusively determine redshift, e.g., to distinguish dusty lower-\(z\) interlopers \citep{2022arXiv220802794N,2023Natur.622..707A}. \citet{2024Natur.633..318C} use JWST Advanced Deep Extragalactic Survey (JADES) Near-Infrared Spectrograph (NIRSpec) data to spectroscopically confirm a galaxy at \(z = 14.32^{+0.08}_{-0.20}\). These observations have been used to infer the UVLF at \(z > 10\), although uncertainties arise also from the effects of dust attenuation and volume estimation in calculating number densities. There have been claims of challenges to the standard cosmological model from such an abundance of high-\(z\) galaxies, although this is closely related to both observational uncertainties (e.g., in derived quantities like stellar masses) and theoretical uncertainties (e.g., in how early galaxy formation proceeds). What is clearer is that an apparent \textit{excess} of early galaxies would limit DM models that suppress the initial conditions and thus \textit{delay} structure formation compared to standard CDM.

In this context, \citet{Maio:2022lzg} compared hydrodynamical simulations to early JWST data, finding that simulations with \(m_\mathrm{WDM} > 2\,\mathrm{keV}\) better match the observed UVLF. They argued that, while JWST finds the brightest galaxies at a given redshift, this fails to constrain the faint end of the UVLF that most probes the small-scale suppression from WDM. Next, \citet{Liu:2024edl} used a larger compendium of JWST data, in particular UVLF measurements for \(8.5 \lesssim z \lesssim 15.5\)~\citep{2024MNRAS.533.3222D}, to limit \(m_\mathrm{WDM} > 3.2\,\mathrm{keV}\) at $95\%$ C.L., using a modified version of the modeling approach introduced in \citet{2022PhRvD.105d3518S} and \citet{Sabti:2021unj}. Recently, \citet{Urrutia:2025fvp} considered photometric UVLF measurements up to \(z \sim 25\) but also introduced more uncertainty to their model considering Pop-III star formation, finding \(m_\mathrm{WDM} > 1.5\,\mathrm{keV}\) at $95\%$ C.L. At the same time, they found no preference for enhanced matter perturbations over standard CDM after marginalizing over astrophysical uncertainties. In addition to the stellar components of high-$z$ galaxies, the relation between black hole mass and galaxy mass at high redshifts can also probe DM models; for example, using a semi-analytic galaxy and black hole formation model, \citet{Ellis:2025dpw} found that JWST constraints on this relation disfavor WDM models with $m_{\mathrm{WDM}}<7.2~\mathrm{keV}$ at $95\%$ C.L.

\textbf{Fuzzy Dark Matter}: \citet{Bozek:2014uqa} used a modified halo model and abundance matching to limit \(m_\mathrm{FDM} > 10^{-22}\,\mathrm{eV}\) at \(3\sigma\). They compared their model to the Hubble Ultra Deep Field UVLF at \(6 \le z \le 10\) and the reionization optical depth inferred from WMAP CMB polarization, although their strongest bound relaxed as new \textit{Planck} reionization inferences lowered the optical depth. \citet{Schive:2015kza} quoted a limit of \(m_\mathrm{FDM} > 1.2 \times 10^{-22}\,\mathrm{eV}\) at $2\sigma$ based on a comparison of full wave-dynamical FDM simulations with HST UVLF data. \citet{Corasaniti:2016epp} then updated the results of \citet{Bozek:2014uqa} by using DM-only \(N\)-body simulations to perform abundance matching to the Hubble Frontier Fields UVLF, finding \(m_\mathrm{FDM} > 1.6 \times 10^{-22}\,\mathrm{eV}\) at $2\sigma$. Next, \citet{Menci:2017nsr} used the total number density of galaxies in the Hubble Frontier Fields to find \(m_\mathrm{FDM} > 8 \times 10^{-22}\,\mathrm{eV}\) at \(3 \sigma\). Despite the previous bounds, \citet{Leung:2018evj} argued that a deficit of low-luminosity galaxies in the cluster lensing fields of the Hubble Frontier Fields was a signature of FDM with \(m_\mathrm{FDM} \sim 10^{-22}\,\mathrm{eV}\). These authors thus predicted that JWST would observe almost no galaxies at \(z > 10\), which was not the case. More consistently with other works, \citet{Ni:2019qfa} used the faint end of the HST UVLF to limit \(m_\mathrm{FDM} > 5 \times 10^{-22}\,\mathrm{eV}\) at \(3 \sigma\).

\citet{Winch:2024mrt} recently carried out a full statistical analysis of UVLF data, including marginalization over a wide range of astrophysical uncertainties. Based on a probabilistic relation between the halo mass function and the HST UVLF from \citet{2022PhRvD.105d3518S}, these authors found \(m_\mathrm{FDM} > 2.5 \times 10^{-22}\,\mathrm{eV}\) at $95\%$ confidence, and that for \(10^{-26}\,\mathrm{eV} \leq m_\mathrm{FDM} \leq 10^{-23}\,\mathrm{eV}\), the fraction of DM in FDM must be \(< 22\%\). These authors also performed the first FDM consideration of spectroscopically-confirmed JWST UVLF data, finding that the data are consistent with CDM but not yet constraining enough to improve over HST UVLF FDM limits. Using a similarly probabilistic model and the faint-end UVLF from the Hubble Frontier Fields, \citet{Sipple:2024svt} found \(m_\mathrm{FDM} > 5 \times 10^{-22}\,\mathrm{eV}\) at $2\sigma$ in their most conservative analysis. Finally, \citet{Urrutia:2025fvp}, in their analysis of photometric JWST UVLF data, found \(m_\mathrm{FDM} > 5.6 \times 10^{-22}\,\mathrm{eV}\) at $95\%$ C.L. Finally, \citet{Ellis:2025dpw} extended their high-$z$ black hole mass--galaxy mass analysis to exclude FDM with $m_{\mathrm{FDM}}<2\times 10^{-20}~\mathrm{eV}$ at $95\%$ C.L.

\textbf{Interacting Dark Matter}. As the quantity and redshift reach of UVLF data have increased, accelerated by JWST, and as more robust statistical analysis frameworks have been developed, a wider range of DM models beyond WDM and FDM have been considered in UVLF analyses. For example, \citet{Lazare:2025gha} used the framework of \citet{2022PhRvD.105d3518S} to analyze HST UVLF data, including faint-end measurements from lensing fields, to constrain velocity-dependent DM-proton scattering IDM. These authors derived bounds for \(n=2\) (\(\sigma_{\mathrm{DM}-p} < 1.1 \times 10^{-25}\,\mathrm{cm^2}\)) and \(n=4\) (\(\sigma_{\mathrm{DM}-p} < 2.1 \times 10^{-22}\,\mathrm{cm^2}\)) at $95\%$ confidence; their $n=0$ bounds complement existing limits (see Figure~\ref{fig:idm}). \citet{Das:2025bnr} then considered both DM-proton and DM-electron scattering IDM with JWST UVLF data, improving bounds in the literature for \(n = -2\) DM-proton scattering for DM masses of \((1 - 500)\,\mathrm{MeV}\). Finally, \citet{Barron:2025dys} used the same framework to study the impact of ETHOS-like IDM models on UVLFs, finding that the first DAO peak must be at $k>50~h~\mathrm{Mpc}^{-1}$ unless the fraction of DM undergoing DAOs is under $7\%$.

\subsubsection{Summary}

\NEW{High-redshift galaxy UVLFs probe DM models that suppress structure formation at $z \gtrsim 6$, with JWST now extending these measurements to $z \sim 15$. For WDM, current UVLF limits reach $m_{\rm WDM} \gtrsim 3~{\rm keV}$; for FDM, they reach $m_{\rm FDM} \gtrsim 10^{-22}~{\rm eV}$. An important feature of UVLF measurements is that they probe the same comoving scales as other probes at different redshifts. Thus, similar to the Lyman-$\alpha$ forest, high-$z$ galaxy populations more directly constrain $P(k)$ compared to other small-scale structure probes, yielding sensitivity to features beyond a cutoff such as dark acoustic oscillations in (fractional) IDM models. The most important uncertainty for UVLF-based DM constraints is the high-redshift galaxy--halo connection, including the efficiency and burstiness of star formation in low-mass halos at $z \gtrsim 10$, along with the translation from UV luminosity, to stellar mass, and finally to halo mass. Marginalization over these uncertainties is critical to derive robust DM limits.}

\subsection{Weak Lensing}
\label{sec:weak_lensing}

In weak gravitational lensing, the lensing distortion of any individual source cannot generally be separated from intrinsic variation in the source properties. Although weak lensing typically functions as a precision probe of LSS or scales near $k_{\mathrm{nl}}$, a growing body of work applies weak lensing techniques to structure on nonlinear scales that are sensitive to DM microphysics.

\subsubsection{Galaxy Weak Lensing}

Galaxy weak lensing effects can be identified statistically by measuring the shapes (i.e., ellipticities) of a large population of galaxies. The \textit{cosmic shear} is the integrated lensing effect of all matter along the line of sight to a galaxy. Cosmic shear can be measured with high significance because the matter distribution between source and observer imposes a spatially correlated shear signal, while to zeroth order the intrinsic shapes and orientations of galaxies are uncorrelated.

In this sense, the galaxy shear correlation function is a tracer of the nonlinear matter correlation function (and thus power spectrum) at \(z \lesssim 1.5\) and for \(k \lesssim 10\,h\,\mathrm{Mpc}^{-1}\). In practice, there are intrinsic alignments in the orientations of galaxy shapes and this effect must be modeled and marginalized in cosmological analyses \citep{DES:2026fyc}, in addition to systematic effects of baryonic feedback \citep{Chisari:2018prw} on the matter power spectrum and uncertainties in the photometric redshift distributions of source galaxies. Unlike the spatial clustering of galaxies themselves, weak lensing is not directly sensitive to redshift-space distortions or how galaxies trace DM (\textit{bias}), making weak lensing a cleaner probe of the DM distribution in this regard.

Galaxy weak lensing has long been recognized as a potentially powerful probe of DM models that modify the matter power spectrum. For example, \citet{2011JCAP...01..022M} forecasted that a Euclid-like weak lensing survey, in combination with \textit{Planck} CMB data, could limit the WDM particle mass to be greater than 2.5 keV.  In the interim, generations of photometric surveys (e.g., the Canada-France-Hawaii Telescope Lensing Survey, the Kilo-Degree Survey, the Dark Energy Survey, the Hyper Suprime-Cam Subaru Strategic Program) have demonstrated that weak lensing most precisely constrains a combination of cosmological parameters \(S_8 \equiv \sigma_8 \left(\Omega_\mathrm{m}/{0.3}\right)^{1/2}\). Many studies \citep[e.g.,][]{Allali:2021azp,Rubira:2022xhb} have used constraints on \(S_8\) in the context of the \(\Lambda\)CDM model as an effective compression of weak lensing data to test DM models. Indeed, a \(2\sigma\) to \(3 \sigma\) discrepancy has emerged in the value of \(S_8\) inferred from CMB and weak lensing data assuming \(\Lambda\)CDM. This discrepancy could be a signature of unmodeled systematics, underestimation of baryonic feedback effects \citep{Amon:2022azi}, small-scale $P(k)$ suppression due to DM physics \citep[e.g.,][]{Rogers:2023ezo}, or a redshift-dependent effect from the dark sector \citep[e.g.,][]{Poulin:2022sgp}; \NEW{analyses of current data cannot distinguish these scenarios.}

However, \citet{Preston:2025tyl} demonstrated that, if \(S_8\) is inferred assuming \(\Lambda\)CDM, such a compression can bias the inference of DM parameters. Instead, to exploit weak lensing data correctly, it is more powerful to infer the full nonlinear matter power spectrum transfer function \citep[rather than just the amplitude parameter, \(\sigma_8\);][]{Preston:2024ggf}. By doing this, the nonlinear transfer function can then be compared to predictions from different DM models. The challenge is to model effects of different DM scenarios into the nonlinear regime using either simulations or semi-analytical approaches like the halo model. \citet{Dentler:2021zij} used a custom halo model accounting for FDM effects to analyze Dark Energy Survey (DES) Year 1 cosmic shear, in combination with \textit{Planck} CMB, to set a lower limit \(m_\mathrm{FDM} > 10^{-23}\,\mathrm{eV}\). The real power of the shear correlation function lies in precision measurements of the matter power spectrum that will be able to test not just sharp cutoffs but more subtle deviations from the CDM transfer function. \citet{Preston:2025tyl} forecast that the Vera C.\ Rubin Observatory's Legacy Survey of Space and Time (LSST) Year 1 cosmic shear dataset could limit FDM with \(m_\mathrm{FDM} = 10^{-25}\,\mathrm{eV}\) to be less than 5\% of the DM, probing up to \(m_\mathrm{FDM} = 10^{-21}\,\mathrm{eV}\) on the smallest accessible scales. Such sensitivity will increase further with the full ten-year LSST, as well as with Euclid and Roman.

Finally, recent photometric and spectroscopic measurements of dwarf galaxy weak lensing measure the mass profiles of $\sim 10^{10}$ to $10^{11}~M_{\mathrm{\odot}}$ halos~\citep{To:2025,Treiber:2025}. As discussed in Section~\ref{sec:observation_future}, these measurements have recently become feasible due to a combination of accurate dwarf galaxy identification and lensing measurements from imaging surveys along with confirmation and redshift measurements from spectroscopic surveys. These data have already been used to probe the faint-end SMHM relation and, in principle, can test DM physics that alters low-mass halo structure.

\subsubsection{CMB Weak Lensing}

CMB temperature and polarization anisotropies are distorted by weak lensing. In particular, the paths of CMB photons are deflected by an angle that is the gradient of a scalar \textit{lensing potential}. The angular power spectrum of this potential is an integral of the matter power spectrum across a wide range of redshifts (\(1 \lesssim z \lesssim 20\)). Through this integral, CMB lensing probes DM models that modify the matter power spectrum. Until recently, the lensing angular power spectrum was only measured precisely by CMB observatories like \emph{Planck} on linear scales. However, the advent of higher-resolution, ground-based observatories like the Atacama Cosmology Telescope (ACT) and the South Pole Telescope (SPT) has led to smaller-scale (\(k \lesssim 0.3\,h\,\mathrm{Mpc}^{-1}\)) lensing measurements that are sensitive to nonlinear clustering of matter, making them more powerful DM probes. For example, \citet{Lague:2026sbd} used a nonlinear halo model in a mixed CDM and FDM model to analyze a combination of \textit{Planck}, ACT DR6 and SPT-3G primary and lensing anisotropies, finding that FDM with \(m_\mathrm{FDM} = 10^{-25}\,\mathrm{eV}\) cannot be more than 8.8\% of the DM. Such constraints will increase in precision and extend to smaller scales with the upcoming Simons Observatory \citep{SimonsObservatory:2025wwn} and the proposed high-resolution, ground based CMB-HD experiment. CMB-HD is proposed to use lensing measurements to probe \(k \gtrsim 10\,h\,\mathrm{Mpc}^{-1}\), making it a competitive probe of the small-scale DM distribution~\citep{Nguyen:2017zqu,CMB-HD:2022bsz,MacInnis:2024znd,Cheslog:2026xyr}.

\subsubsection{\NEW{Astrometric Weak Lensing}}
\label{sec:astrometric_wl}

\NEW{DM subhalos moving through the MW halo weakly deflect the apparent positions of background sources. This weak lensing effect produces correlated, time-dependent shifts in the proper motions and accelerations of these sources~\citep{VanTilburg:2018ykj}. Unlike the galaxy and CMB weak lensing techniques discussed above, astrometric weak lensing probes the distribution of low-mass subhalos orbiting the MW. Similar to other types of lensing measurements, this technique is more sensitive to compact DM subhalos. Although there have been no subhalo detections using this method to date, current analyses indicate that astrometric weak lensing measurements can probe subhalos below the galaxy formation threshold if their inner densities exceed CDM predictions.}

\NEW{\citet{Mondino:2020rkn} first searched for astrometric weak lensing signatures in Gaia DR2 data by applying a matched filter to the proper motions of stars in the Magellanic Clouds. \citet{Mondino:2023pnc} extended this analysis to include Gaia DR3 data, excluding $\mathcal{O}(1)$ substructure fractions in halos with dynamical masses of $\approx 10^7$ to $10^9~M_{\mathrm{\odot}}$ and characteristic sizes $\lesssim 1~\mathrm{pc}$. Because CDM subhalos in this mass range are generally much more diffuse, these limits mainly constrain models of compact-object DM. However, the same authors forecasted that an acceleration-based search in Gaia DR4 data could probe substructure fractions of $\gtrsim 10^{-3}$ for subhalos with dynamical masses in the range $10$ to $3\times 10^3~M_{\mathrm{\odot}}$ and pc-scale sizes, approaching sensitivity to CDM subhalos. Meanwhile, \citet{Pardo:2021uzy} forecasted that Roman's microlensing survey of the Galactic bulge could probe DM subhalos with dynamical masses of $\lesssim 3\times 10^7~M_{\mathrm{\odot}}$ if they have NFW concentrations several times above the average CDM mass--concentration relation.}

\NEW{Astrometric weak lensing is therefore best viewed as a complementary probe of low-mass DM subhalos that is currently most relevant for scenarios that produce dense perturbers. Looking ahead, \citet{VanTilburg:2026eym} forecasted that astrometric weak lensing of multiply-imaged quasars measured with intensity interferometry could probe subhalos with dynamical masses in the range $3\times 10^{-6}$ to $10^2~M_{\mathrm{\odot}}$ and scale radii in the range $5\times 10^{-3}$ to $2~\mathrm{pc}$, near the fiducial CDM density profile prediction at these scales. No facility with the requisite sensitivity of sub-$\mu$as relative astrometry precision on arcsecond scales currently exists. Nonetheless, realizing this technique could dramatically improve the sensitivity of small-scale structure measurements; for example, by probing subhalos with sub-stellar masses, this technique could constrain WDM models with thermal-relic masses up to $\approx 400~\mathrm{keV}$.}

\subsubsection{Summary}

\NEW{Weak lensing is a precision probe of the matter power spectrum near $k_{\mathrm{nl}}$ and is becoming increasingly sensitive to DM properties due to improvements in data quality and modeling techniques. Current galaxy cosmic shear measurements constrain small deviations from the CDM power spectrum; for example, upcoming LSST measurements are forecasted to limit models down to FDM fractions of $\sim 5\%$. Thus, although weak lensing is less competitive for constraining full $P(k)$ cutoffs than those from the other probes discussed in this review, it measures small-scale clustering in a distinct redshift and wavenumber range, providing a crucial lever arm for testing scale-dependent effects of DM physics. Thoroughly characterizing the effects of baryonic feedback on $P(k)$ for the scales probed by upcoming data will be important to realize this potential. CMB weak lensing shares many of these advantages with cosmic shear, and currently probes FDM fractions of $\sim 10\%$; proposed high-resolution CMB measurements can potentially measure clustering on scales corresponding to $k\sim 10~\mathrm{Mpc}^{-1}$ if the impact of foregrounds can be controlled. On smaller scales, dwarf galaxy weak lensing is a newer technique that can measure halo profiles in the $10^{10}$--$10^{11}~M_\odot$ range, complementing dynamical measurements of Local Group dwarfs and bypassing the need for spectroscopic follow-up of individual stars. Finally, astrometric weak lensing can potentially probe extremely low-mass MW subhalos, and constraints from this technique are expected to improve rapidly with Gaia DR4 and Roman data.}

\subsection{21-cm Cosmology}
\label{sec:21cm}

Even as JWST peers further into the early Universe, the Universe becomes dark in the optical/infrared bands at sufficiently high redshifts, corresponding to the epoch before the formation of the first stars. As Figure~\ref{fig:k_z} illustrates, the so-called ``dark ages'' would provide a powerful probe of small-scale structure while it is still linear, and thus an excellent test of \emph{ab initio} DM physics. The neutral hydrogen in this epoch can emit and absorb photons in the 21-cm spin-flip transition. There is indeed a long-running program aiming to detect the 21-cm signal from neutral hydrogen in the IGM from the dark ages (\(30 \lesssim z \lesssim 200\)) through to ``cosmic dawn'' (\(6 \lesssim z \lesssim 30\)), the latter period probing the epoch of reionization.

Broadly speaking, 21-cm in the dark ages will be an excellent tracer of the high-\(z\) matter power spectrum with a close relation between the neutral hydrogen and the underlying DM distribution. 21-cm in the cosmic dawn, in an analogous manner to the Lyman-\(\alpha\) forest at lower redshifts, will become increasingly sensitive to the spatially-inhomogeneous process of reionization, which is itself highly sensitive to the DM model (see Section~\ref{sec:high_z} for discussion of how optical depth measurements have constrained DM models already). On the observational side, this field is poised to transform with the prospect of tomographic mapping of the neutral hydrogen across both these epochs from large arrays of radio telescopes like the Square Kilometre Array (SKA; see Section~\ref{sec:observation_future}).

There have already been observational constraints on the global monopole component of the 21-cm signal. After the first stars form, the UV radiation they emit penetrates the primordial hydrogen and alters the 21-cm excitation state such that background CMB photons are absorbed by the gas. The Experiment to Detect the Global EoR Signature (EDGES) claimed a detection of this radio spectral distortion with an amplitude of the absorption feature twice as strong as expected through standard galaxy formation physics in CDM~\citep{Bowman:2018yin}. However, \citet{Hills:2018vyr} pointed out a number of issues with the original analysis and concluded that the evidence for anomalously-strong absorption was not sufficient. In contrast, the Shaped Antenna Measurement of the Background Radio Spectrum (SARAS) experiment's antenna rests on water, reducing interference relative to the ground on which EDGES sits. SARAS claims a non-detection of the absorption feature at the same radio frequencies probed by EDGES~\citep{Singh:2021mxo,Bevins:2022ajf}. The upcoming Radio Experiment for the Analysis of Cosmic Hydrogen (REACH) may help adjudicate between these two results, while SKA will move beyond monopole measurements and map anisotropies in the 21-cm signal. In addition, the Hydrogen Epoch of Reionization Array (HERA) has recently set upper limits on the 21-cm power spectrum (see Section~\ref{sec:observation_future}).

\subsubsection{Global Signal}

A key physical mechanism through which DM microphysics can affect the 21-cm signal is the suppression of low-mass star-forming halos, which delays the buildup of Lyman-$\alpha$ coupling, X-ray heating, and reionization relative to CDM. These effects generally shift the global absorption trough to lower redshifts and narrow its width~\citep{Sitwell:2013fpa}. The timing and depth of the global 21-cm signal is therefore sensitive to both the amplitude and the shape of any modification to the matter power spectrum relative to CDM, probing a characteristic halo mass of $\sim 2\times 10^7~M_{\mathrm{\odot}}$ at $20\lesssim z\lesssim 40$~\citep{Park:2025phj}.

Several studies have used the timing and morphology of the global 21-cm signal in current data to constrain DM models, although the uncertain status of the EDGES detection complicates the interpretation of these results. For example, \citet{Schneider:2018xba} used an empirical galaxy--halo connection model to show that $m_{\mathrm{WDM}}>6.1~\mathrm{keV}$ for thermal-relic WDM and $m_{\mathrm{FDM}}>8\times 10^{-21}~\mathrm{eV}$ for FDM are favored by the timing of the EDGES absorption signal. \citet{Chatterjee:2019jts} similarly used the \textsc{Delphi} model of high-$z$ halo and galaxy populations to show that $m_{\mathrm{WDM}}>3~\mathrm{keV}$ is preferred based on both the timing and depth of the EDGES signal. However, \citet{Boyarsky:2019fgp} found that when astrophysical uncertainties are fully propagated, WDM constraints based on the EDGES signal timing weaken, such that both $m_{\mathrm{WDM}}=6~\mathrm{keV}$ and resonantly-produced $m_s=7~\mathrm{keV}$ sterile neutrino WDM are consistent with the data. Thus, degeneracies between DM physics and the efficiency of star formation in low-mass halos at high redshifts remain an important challenge for 21-cm analyses (e.g., \citealt{Hibbard:2022sng}), along with the proper extraction of DM signal from foreground systematics (e.g., \citealt{Rudakovskyi:2019cxt}).

For IDM, scattering between DM and baryons can lower the baryon temperature after CMB decoupling (e.g., \citealt{Tashiro:2014tsa}), although the magnitude (and even sign) of this effect depends on the DM--SM coupling and treatment of DM--baryon streaming velocity~\citep{Munoz:2015bca}. Based on this signature, \citet{Barkana:2018lgd} found that the hydrogen gas cooling implied by the EDGES result could be explained by a small fraction of the millicharged DM interacting with baryons.\footnote{This scenario is similar to \(n=-4\) IDM scattering, but only features interactions between DM and ions.} However, \citet{Kovetz:2018zan} showed that the viable parameter space for millicharged DM explanations of the EDGES anomaly is tightly constrained by CMB, BBN, and stellar cooling limits, restricting the viable IDM fraction to $\lesssim 10^{-4}$ to $4\times 10^{-3}$ of the total DM budget and particle masses to $0.5$ to $35~\mathrm{MeV}$. Moreover, \citet{Driskell:2022pax} demonstrated that the suppression of star-forming halos by DM--baryon scattering further reduces the viable parameter space, since 100\% millicharged DM cannot reproduce the EDGES signal once the effect on halo abundances is included.

Moving beyond EDGES, recent SARAS results have been analyzed in the context of DM--baryon scattering IDM models. In particular, \citet{Mittal:2026oyj} found that current SARAS data cannot distinguish CDM from IDM; however, these authors placed upper limits on the amplitude of the global signal in IDM. Forecasts based on upcoming global signal measurements indicate that even current instruments can improve existing bounds on Coulomb-like and velocity-independent DM--baryon scattering cross sections, and that these 21-cm constraints will be limited by degeneracies between DM physics and astrophysical uncertainties in the minimum halo virial temperature for star formation and the Lyman-Werner photon production rate~\citep{Rahimieh:2025fsb}. In addition, such interactions modify the evolution of baryon and DM temperature perturbations, which can further affect 21-cm and reionization signals~\citep{Short:2022bmm}.

For FDM, the width of the global absorption feature shrinks due to delayed and suppressed structure formation; this signal is relatively insensitive to the degeneracies discussed above. In particular, \citet{Nebrin:2018vqt} showed that the EDGES measurement implies $m_{\mathrm{FDM}}>6\times 10^{-22}~\mathrm{eV}$ based on the observed trough width, independent of the anomalous amplitude, and that a future detection of the 21-cm global signal at $z>14$ by interferometers such as the SKA could probe FDM models with $m_{\mathrm{FDM}}\lesssim 10^{-20}~\mathrm{eV}$.

\subsubsection{Power Spectrum}

The 21-cm power spectrum at cosmic dawn is sensitive to both the amplitude and the shape of any modification to the matter power spectrum at $20\lesssim z\lesssim 40$~\citep{Sikder:2025ykc}. For example, $P(k)$ suppression enhances large-scale 21-cm fluctuations because higher-mass (and thus more spatially biased) halos preferentially contribute to the signal in these models. This effect probes both WDM-like cutoffs in $P(k)$ and IDM models that imprint oscillatory features in $P(k)$.

For WDM, the large-scale enhancement of the 21-cm power spectrum noted by \citet{Sitwell:2013fpa} is predicted to be detectable with current interferometers for $m_{\mathrm{WDM}}\lesssim 3~\mathrm{keV}$. After reionization, 21-cm intensity mapping at $3\lesssim z\lesssim 5$ provides an independent constraint on DM models; for example, \citet{Carucci:2015bra} used hydrodynamic simulations to show that SKA1-LOW can rule out thermal-relic WDM with $m_{\mathrm{WDM}}=4~\mathrm{keV}$ at the $3\sigma$ level with 5000 hours of observation at $z>3$, while $m_{\mathrm{WDM}}=3~\mathrm{keV}$ can be ruled out at $>2\sigma$ with 1000 hours of observation at $z>5$. More recently, \citet{Zhang:2026mat} forecasted that a combination of DESI data and SKA 21-cm intensity mapping could limit $m_{\mathrm{WDM}}>7.1~\mathrm{keV}$ at $95\%$ C.L.\ based on large-scale ($k=0.05~\mathrm{Mpc}^{-1}$) effects on the Lyman-$\alpha$ opacity and 21-cm power spectrum at $4\lesssim z\lesssim 5.5$; these authors also forecasted that Stage-V spectroscopic surveys could improve these limits to the level of $m_{\mathrm{WDM}}>10$ to $40~\mathrm{keV}$.

Meanwhile, IDM models with dark acoustic oscillations in $P(k)$ imprint oscillatory features on the 21-cm power spectrum. Specifically, for IDM models within the ETHOS framework, \citet{Munoz:2020mue} ran a suite of hydrodynamic simulations to show that both the 21-cm global signal and power spectrum are sensitive to $P(k)$ suppression at $k\lesssim 300~h~\mathrm{Mpc}^{-1}$ even after marginalizing over astrophysical feedback. Crucially, ETHOS models with strong dark acoustic oscillations are distinguishable from pure WDM-like suppression through their oscillatory imprint on the 21-cm power spectrum. Forecasts indicate that HERA can separate ETHOS models from CDM for $P(k)$ suppression wavenumbers $k\lesssim 200~h~\mathrm{Mpc}^{-1}$ with 540 days of observations after marginalizing over astrophysical uncertainties~\citep{Verwohlt:2024efh}. Thus, the 21-cm power spectrum offers a powerful diagnostic of DM microphysics that can potentially discriminate between classes of DM models.

Finally, for FDM, \citet{Jones:2021mrs} modeled the 21-cm power spectrum across cosmic dawn and the epoch of reionization, finding that HERA can constrain the FDM particle mass to within $20\%$ at $2\sigma$ for $m_{\mathrm{FDM}}=10^{-21}~\mathrm{eV}$. \citet{Liu:2025cxq} extended this to include the full nonlinear FDM halo mass function, finding that SKA1-Low can potentially measure the FDM particle mass at the $\sim 10\%$ level with 1080 hours of observation, although these results will also be limited by foreground modeling assumptions~\citep{Sarkar:2022dvl}. For lower-mass FDM particles, \citet{Flitter:2022pzf} showed that HERA will be sensitive to FDM fractions as small as $1\%$ for $10^{-25}~\mathrm{eV}\lesssim m_{\mathrm{FDM}}\lesssim 10^{-23}~\mathrm{eV}$, a window not yet fully closed by other probes (see Figure~\ref{fig:fdm_limits}).

\subsubsection{Summary}

\NEW{21-cm cosmology has not yet delivered firm DM constraints: current results rely either on the contested EDGES global signal detection or are forecasts for upcoming 21-cm arrays. However, the physical basis for 21-cm cosmology as a DM probe has been well established. 21-cm global signal and power spectrum measurements at cosmic dawn will access redshifts up to $z \sim 30$ and wavenumbers up to $k\sim 100~\mathrm{Mpc}^{-1}$, corresponding to characteristic halo masses of $\sim 10^7~M_\odot$ at these redshifts. This reach provides the most direct window into the initial conditions for small-scale structure among all current small-scale structure probes. As a result, the 21-cm signal has the potential to discriminate between $P(k)$ shapes and thus DM scenarios; for example, WDM produces a smooth small-scale cutoff and large-scale bias enhancement in the 21-cm power spectrum, while dark acoustic oscillations in IDM models lead to a distinct signal that upcoming facilities like HERA can distinguish.}

\NEW{A challenge for 21-cm DM constraints is the astrophysical degeneracy between DM physics and the poorly constrained star formation efficiency in low-mass halos at high redshifts. The 21-cm power spectrum partially breaks this degeneracy by providing spatial information, and probe combination with the high-$z$ UVLF, optical depth measurements, and Lyman-$\alpha$ damping wings may provide additional constraining power. Thus, the combination of upcoming global signal and power spectrum measurements will sensitively probe a wide range of DM microphysics in the coming decade, with the particular advantage of accessing scales and redshifts inaccessible to any other current probe.}

\section{Probe Combination}
\label{sec:combination_constraints}

The small-scale structure probes discussed in Sections~\ref{sec:constraints}--\ref{sec:emerging} provide complementary windows into the small-scale DM distribution at different scales and redshifts. For example, dwarf galaxy and stellar stream observations constrain the abundance and internal structure of low-mass halos in the local universe, strong gravitational lensing is sensitive to substructure within and along the line-of-sight to (typically) low-redshift galaxies and clusters, and higher-redshift observations probe the first generations of galaxy and DM halo growth. These observables differ in scale and redshift and are subject to distinct theoretical and observational systematics.

We emphasize that these different small-scale structure probes are ultimately governed by the same underlying DM physics. Thus, rather than treating each probe independently, they should be viewed as different manifestations of the same (evolving) distribution of DM structure. In this view, the most powerful approach to small-scale structure analysis is to construct a joint inference framework that combines multiple probes, while consistently accounting for both their shared dependencies and distinct systematics, in order to break degeneracies and deliver robust DM constraints.

There have been several joint analyses in this spirit, some of which combine data from similar environments. For example, \citet{Banik:2019smi} showed that their $m_{\mathrm{WDM}}>3.6~\mathrm{keV}$ constraint derived from the GD-1 and Pal 5 stellar streams strengthens to $m_{\mathrm{WDM}}>6.2~\mathrm{keV}$, at $95\%$ confidence, when including classical MW satellite counts; their $m_{\mathrm{FDM}}>1.4\times 10^{-21}~\mathrm{eV}$ constraint likewise strengthens to $m_{\mathrm{FDM}}>2.2\times 10^{-21}~\mathrm{eV}$. Although these stream and satellite data both probe the MW subhalo population, their combination improves DM limits because classical satellites probe the high-mass end of the SHMF, constraining its normalization, while streams probe lower masses where the SHMF shape can change due to DM physics. Meanwhile, \citet{Schutz:2020jox} showed that WDM constraints from streams (plus classical MW satellite abundances) and strong lensing translate to a lower limit of $m_{\mathrm{FDM}}>2.1\times 10^{-21}~\mathrm{eV}$ based on an analysis of the underlying SHMFs.

In a benchmark example of small-scale probe combination, \citet{Enzi:2020ieg} derived WDM constraints by jointly analyzing a broader range of observables, including gravitational imaging~\citep{Vegetti:2018dly,Ritondale:2018cvp}, the Lyman-$\alpha$ forest~\citep{Murgia:2018now}, and MW satellites~\citep{Newton:2020cog} to derive $m_{\mathrm{WDM}}>6.048~\mathrm{keV}$ at $95\%$ confidence; these authors also constrained sterile neutrino models. This WDM constraint is slightly weaker than the strongest single-probe limit of $m_{\mathrm{WDM}}>6.989~\mathrm{keV}$ that entered the analysis. This follows because the joint constraint was derived by multiplying the marginalized posterior for the half-mode wavelength $\lambda_{\mathrm{hm}}$ from each probe, which slightly favors WDM over CDM in the \citet{Vegetti:2018dly} analysis.

In another pathfinder analysis, \citet{Nadler:2021dft} combined WDM constraints from MW satellites~\citep{DES:2020fxi} and strong-lensing flux ratios~\citep{Gilman:2019nap} to derive $m_{\mathrm{WDM}}>9.7~\mathrm{keV}$ at $95\%$ confidence. This analysis improved on the bounds from the individual probes by combining marginal likelihoods in the two-dimensional parameter space of SHMF normalization versus WDM half-mode mass. The resulting joint WDM constraint was stronger than either of the individual limits and than these limits' one-dimensional combination. This result highlights the importance of identifying shared degeneracies between small-scale probes and modeling them in a unified framework. \citet{Zelko:2022tgf} translated this limit to a variety of sterile neutrino WDM models, which yielded the strongest constraints on Higgs decay, GUT-scale production, the sterile neutrino minimal Standard Model, and the Dodelson--Widrow production mechanism.

Meanwhile, \citet{Lazare:2024uvj} combined high-redshift observables to constrain FDM. In particular, these authors analyzed HST UVLF measurements and HERA upper limits on the 21-cm power spectrum---along with constraints on the neutral hydrogen fraction and the optical depth to reionization measured by Planck---using an ML pipeline trained on semi-analytic models of these observables. Thus, \citet{Lazare:2024uvj} derived $f_{\mathrm{FDM}}<16\%$ for $m_{\mathrm{FDM}}=10^{-23}~\mathrm{eV}$ and $f_{\mathrm{FDM}}<1\%$ for $m_{\mathrm{FDM}}=10^{-26}~\mathrm{eV}$, both at $95\%$ confidence; in addition, they showed that these upper limits can strengthen to $f_{\mathrm{FDM}}\lesssim 1\%$ for $m_{\mathrm{FDM}}=10^{-23}~\mathrm{eV}$ if future 21-cm measurements detect the 21-cm power spectrum.

These results represent early demonstrations of DM constraints from small-scale probe combination; we discuss the outlook for such analyses in Section~\ref{sec:combination_future}.

\section{Challenges and Prospects}
\label{sec:outlook}

This section discusses the theoretical (Section~\ref{sec:theory_discussion}) and observational (Section~\ref{sec:observation_future}) outlook for small-scale structure studies. We organize this discussion by redshift range, as many modeling and observational considerations cut across probes in a given distance regime.

\subsection{Theory}
\label{sec:theory_discussion}

\subsubsection{Near-field and Low Redshift} \label{sec:low_z_theory}

\textbf{Near-field Cosmology}: Theoretical predictions for small-scale structure in the local universe need to accurately model the specific environment of the MW and, depending on the application, its surroundings (e.g., the Local Group or Local Volume). This situation presents several challenges. Computationally, it is difficult to engineer regions of the universe that resemble local structure in detail, such that a large number of realizations are required to generate appropriate samples. Yet, fluctuations in the small-scale matter distribution are large on the scale of the Local Group or smaller, implying that a large number of realizations are needed to robustly estimate sample variance. A growing number of constrained simulations aim to address these issues~(e.g., \citealt{Garrison-Kimmel:2013eoa,Carlesi:2016qqp,Sawala:2021npe,Wempe:2024rfj,Buch:2024ssx}), and constrained semi-analytic techniques have also been proposed~\citep{Nadler:2022kmy}. However, such simulations are only now starting to be used in forward-modeling pipelines to derive DM limits. Overall, it remains difficult to model small-scale structure in the local universe, at high resolution and across a representative range of DM models and baryonic feedback implementations, while accurately capturing the associated cosmic variance.

\textbf{Halo and Subhalo Populations}. Small-scale structure simulations relevant for both near-field cosmology and strong lensing face resolution challenges. To illustrate the dynamic range needed for frontier small-scale structure analyses, consider the recently-discovered low-mass B1938+666 strong-lensing perturber discussed in Section~\ref{sec:strong_lensing}. This perturber has an estimated \NEW{stripped mass} of $\sim 5\times 10^5~M_{\mathrm{\odot}}$ in its inner $\sim 10~\mathrm{pc}$~\citep{Vegetti:2026}, and it is likely a subhalo of a $\sim 10^{13}~M_{\mathrm{\odot}}$ host halo. To accurately model the assembly and evolution of this perturber in a cosmological setting, it is necessary to resolve it with $\gtrsim 10^4$ particles at infall (e.g., \citealt{Errani:2020wgn}), with sufficient spatial resolution to capture its inner regions where lensing data probes the enclosed mass. This translates to a requirement of $\sim 10^{12}$ particles in a cosmological volume, which is far higher than the resolution of most current cosmological simulations. Moreover, the perturber ideally needs to be simulated across a range of DM and feedback models.

For subhalo population modeling, the development of robust simulation codes and analysis algorithms in non-CDM models is a crucial area of ongoing work (e.g., see \citealt{Banerjee:2022qcb} for an overview). These modeling challenges vary across DM models; for example, SIDM codes can reach different results depending on the scattering implementation~\citep{Meskhidze:2022hwm,Ramos:2025lvk}, WDM simulations suffer from artificial fragmentation~\citep{Wang:2007he,Lovell:2013ola,Angulo:2013sza,Stucker:2021vyx}, and FDM codes take different approaches to solving the Schr\"{o}dinger-Poisson equations that have distinct strengths and weaknesses~(e.g., \citealt{Edwards:2018ccc,Nori:2018hud,Schwabe:2021jne,May:2021wwp,Johnston:2026qwx}). Most code comparisons and convergence tests to date have been performed in CDM, but a growing body of work shows that these issues need to be revisited in models like SIDM (e.g., \citealt{Mace:2024uze,Palubski:2024ibb,Fischer:2025rky}). It is also important to assess the performance of subhalo finder and merger tree algorithms---which remain an active area of development even in CDM~(e.g., \citealt{Han:2017lpe,Elahi:2019wap,Mansfield:2023prs,Kong:2025wig,Moreno:2025nps})---when analyzing simulations in non-CDM models~(e.g., see \citealt{Kong:2025kkt} for an example application to SIDM).

One promising avenue to address resolution challenges for dwarf galaxy, stellar stream, and strong lensing applications is the use of controlled N-body simulations. In this setup, subhalos of interest are initialized at high mass and spatial resolution and evolved in an analytic host potential. This technique has been used to generate libraries of subhalo evolution predictions in CDM (e.g., \citealt{Errani:2019sey,Errani:2020wgn,Ogiya:2019del,Aguirre-Santaella:2025oyz}), SIDM (e.g., \citealt{Sameie:2019zfo,Yang:2021kdf,Shirasaki:2022ttb,Zeng:2021ldo,Zeng:2023fnj,Zhang:2024ggu}), FDM~\citep{Du:2018qor,Glennon:2022huu}, and for subhalos with a range of inner density profiles~\citep{Errani:2022aru,Du:2024sbt}. Promising future directions for controlled simulation include using initial conditions from cosmological zoom-in simulation merger trees in a ``hybrid'' approach (e.g., \citealt{Zhang:2024fib}) and integrating controlled simulation results into SAMs~(e.g., \citealt{Benson:2022tzm,Du:2025xqi}).

Another increasingly popular modeling technique is to train ML emulators on cosmological simulations, across DM and baryonic feedback models, to rapidly predict (sub)halo population statistics in scenarios that have not been simulated. While this approach has been applied successfully for LSS modeling (e.g., \citealt{DeRose:2018xdj,CAMELS:2020cof}), applications to small-scale structure are relatively new. For example, ML emulators have been trained on the DREAMS simulations to predict WDM masses from density fields and galaxy properties~\citep{Rose:2023qbw,Lin:2024fyw,Costanza:2025yiz,Silvestrini:2026}, and to model the galaxy--halo connection~\citep{Nguyen:2024ndo}. Some studies have also trained models on processed simulation outputs---such as halo mass accretion histories or merger trees---to capture the dependence of summary statistics on parameters like the WDM mass~\citep{Nguyen:2023jzh,Nguyen:2025nip,Leisher:2025}. In a similar vein, some analyses have trained emulators on SAM predictions (e.g., \citealt{Lonergan:2025qdo}). Given recent improvements in cosmological simulations across DM models, emulators are poised to play a key role in the next generation of small-scale structure analyses.

Finally, there has also been rapid progress in training ML models directly on real or simulated data products rather than on cosmological simulations. For example, in strong gravitational lensing, ML models have been trained on mock or real lensing images to infer underlying DM structure and ultimately place DM constraints (e.g., \citealt{DiazRivero:2019hxf,Ostdiek:2020mvo,Montel:2022fhv,Zhang:2022djp,Zhang:2023wda,Filipp:2025qso,Tsang:2024agc,Wagner-Carena:2022mrn,Wagner-Carena:2024axc,Dhanasingham:2025tve,Filipp:2026hna}). Analogous techniques have been developed to infer subhalo perturber properties from stellar stream data (e.g., \citealt{Hermans:2020skz,LSSTDarkEnergyScience:2025zqb,Ma:2025srn}) or the density profiles of dwarf galaxies from kinematic data (e.g., \citealt{Nguyen:2022ldb,Nguyen:2026gap}). This class of ML techniques will become increasingly important as the complexity and volume of small-scale data increase in the coming years.

\subsubsection{Intermediate and High Redshift}

\textbf{Precision Probes}: High-precision probes of nonlinear structure like galaxy weak lensing face modeling challenges that are distinct from probes at the small-scale frontier. First, the nonlinear clustering of the underlying DM field, i.e., the nonlinear matter power spectrum up to \(k \sim 10\,h\,\mathrm{Mpc}^{-1}\) for \(z \lesssim 4\), must be computed efficiently across DM models. Second, the nonlinear matter power spectrum must be mapped to the observed galaxy shear correlation function (or equivalent statistics). The first step is currently primarily achieved either through a semi-analytical halo model approach or by training emulators on numerical simulations. However, there has been less work on the nonlinear matter power spectrum in beyond-CDM models, notwithstanding WDM \citep{Marsh:2016vgj} and FDM \citep{Vogt:2022bwy,Dome:2024hzq} halo models, the generalized DM model \citep{Sakr:2026jux} and WDM simulation emulators \citep{Parimbelli:2021mtp}. The main astrophysical uncertainty is the impact of baryonic feedback from active galactic nuclei, which appears to be much stronger than previously anticipated, and partially suppresses the matter power spectrum \citep[e.g.,][]{Amon:2022azi,DES:2024iny}. \citet{Preston:2025tyl} forecast that the effects of feedback and non-CDM models can be distinguished but only if improved hydrodynamical simulations in non-CDM models are combined with external measurements of feedback from the CMB Sunyaev-Zeldovich effect, X-ray cluster measurements and fast radio bursts \citep{Siegel:2025ivd}. The second step, mapping to the galaxy shear signal, needs hydrodynamical simulations to model the intrinsic alignments of galaxy shapes. Further, to extract more information from the billions of galaxy shapes next-generation wide-area surveys will provide, the robustness of these modeling approaches needs to be confirmed when estimating higher-order statistics and using simulation-based inference and field-level inference.

High-resolution CMB lensing does not currently probe as small scales as galaxy shear. But, the ongoing Simons Observatory and the future proposed CMB-HD experiment will probe equivalent scales, albeit at higher redshift (\(z \sim 2\)). CMB lensing is not sensitive to systematics like intrinsic alignments, although baryonic feedback will still affect the matter power spectrum. New, more powerful lensing estimators are being developed to extract the lensing signal from smaller angular scales in the CMB that are contaminated by foregrounds and other secondary anisotropies \citep[e.g.,][]{Chan:2023vye}.

\textbf{IGM and Reionization Physics}: DM model constraints from the smallest scales probed by the highest-resolution Lyman-\(\alpha\) forest are currently limited by statistical uncertainties from the small number of available spectra \citep[e.g.,][]{Boera:2018vzq}. It will be challenging to access much smaller scales through the Lyman-\(\alpha\) forest, as current data are already sensitive to the intrinsic pressure smoothing scale of the IGM (\(\sim 50\,\mathrm{kpc}\)) that erases smaller-scale DM features. However, there is the prospect of significantly increasing the existing sample size (Section~\ref{sec:observation_future}) and thus increasing the \textit{precision} of the matter power spectrum inference at \(k \sim 100\,h\,\mathrm{Mpc}^{-1}\) at \(z \sim 5\). To match this data precision, it will be necessary to improve our modeling of how the IGM density, temperature and ionization fields trace the DM.

Although hydrogen reionization has effectively completed by \(z \sim 5.5\), the Lyman-\(\alpha\) forest at \(4 \lesssim z \lesssim 5.5\) remains sensitive to the heat injection and ionization from this epoch. It is standard in Lyman-\(\alpha\) forest simulations to approximate the ultra-violet photo-ionizing background as spatially homogeneous \citep{Lukic:2014gqa}, which significantly speeds up simulations allowing the construction of emulators for data analysis \citep[e.g.,][]{Rogers:2020cup,Cabayol-Garcia:2023ygj}. However, radiative transfer (RT) simulations have demonstrated that there remain large-scale (\(\sim 40\,\mathrm{Mpc}\)) fluctuations in the temperature and ionization fields after reionization finishes \citep{Chardin:2015uza}. Although these scales are much larger than the small scales driving the most competitive DM limits, there is a secondary consequence from local variations in the mean opacity that has a \(\sim 10\%\) effect on the small-scale Lyman-\(\alpha\) forest flux power spectrum \citep{Wu:2019sgk}. This effect is below the sensitivity of current datasets, but this will change in the future. Modeling patchy reionization is challenging as the dynamic range is large (from tens of Mpc to tens of kpc), while full radiative transfer calculations are expensive. Solutions lie in building approximate models calibrated on fewer RT simulations \citep{Irsic:2023equ} or using multi-fidelity emulation techniques that combine larger, lower-resolution simulations with fewer smaller, higher-resolution simulations \citep{Fernandez:2022kxq}. For lower-\(z\) Lyman-\(\alpha\) forest (\(z \sim 3\)), e.g., probed by wide-area surveys like DESI, 4MOST, PFS, WEAVE or the future Spec-S5 and WST, further improvements in modeling patchy helium reionization are needed \citep{UptonSanderbeck:2020zla}.

Other astrophysical effects must also be controlled when extracting DM properties from the Lyman-\(\alpha\) forest. The bulk of the spectral absorption arises from low-density IGM, but most sightlines contain broadened lines from high column density absorbers \citep[column densities \(> 10^{17.2}\,\mathrm{atoms}\,\mathrm{cm}^{-2}\);][]{1986ApJS...61..249W}. The highest-density systems are identified and masked, while a model is marginalized for the residual contamination of Lyman-limit systems \citep{Rogers:2017bmq,Rogers:2017eji}. Again, as data precision increases, more robust simulations containing the co-evolution of the low-density Lyman-\(\alpha\) forest, high-density absorbers, metal contamination and feedback effects may be necessary.

\textbf{High-\(z\) Galaxy Formation}: As discussed in Section~\ref{sec:high_z}, our view of the high-\(z\) galaxy population, as summarized by the UVLF, is being transformed by JWST. Given the trajectory of HST, we may expect the impact of JWST to continue for a decade and more. The spectroscopic confirmation of galaxies forming by \(z \sim 14\) and photometric candidates up to \(z \sim 25\) must now be matched by a more comprehensive understanding of how the first population of stars and galaxies is formed. While there has been recent progress in building SAMs that map the halo mass function to the UVLF by probabilistic relations that allow uncertainties in, e.g., AGN and supernovae feedback effects to be marginalized, it is less clear at the moment how well this model extrapolates to \(z \gg 10\). Nonetheless, if those astrophysical uncertainties can be quantified robustly, they will yield stringent constraints on DM models that alter $P(k)$ relative to CDM, as even a small sample of very-high-\(z\) objects is very constraining for this purpose \citep{Winch:2024mrt}.

To achieve this goal, it will be necessary to run cosmological hydrodynamical simulations of early galaxy formation with radiative transfer effects \citep{Kannan:2021xoz}. As input to these simulations, we need, e.g., models of Pop-III star formation, the initial stellar mass function and feedback effects. Further, these simulations should be run \textit{with modified DM initial conditions} so that we understand the impact on the onset of hierarchical structure formation from a suppressed or enhanced linear matter power spectrum \citep{Rose:2024xcb}. Indeed, beyond astrophysical uncertainties, we do not yet have a complete understanding of the low-mass end of the halo mass function when the small-scale matter power spectrum is modified.

In addition to improvements in simulations, it will be useful to find complementary observations to the UVLF that can help break degeneracies between DM model and astrophysical parameters. \citet{Munoz:2023cup} point out that even measurements of the linear galaxy clustering amplitude (bias) can help achieve this goal. For SIDM, even existing HST UVLF data cannot disentangle astrophysical uncertainties from DM physics. Meanwhile, \citet{Wang:2026lly} find that combining the UVLF with determinations of the spatial topology of reionization from the 21-cm signal can help solve this problem.

The theoretical modeling challenges for 21-cm cosmology lie in understanding instrumental systematics (as demonstrated by the global monopole measurements above), Galactic foregrounds that dominate over the primordial signal in many regimes and a robust prediction for the distribution of neutral hydrogen from before (dark ages) and during (cosmic dawn) the epoch of reionization. For the latter, there is thus substantial overlap in the simulation requirements for UVLF and Lyman-\(\alpha\) forest analyses. The current workhorses for (forecast) 21-cm statistical analyses are SAMs like the \texttt{21cmFAST} code \citep{Mesinger:2010ne}. However, it is critical to ensure that the approximations made in such models are sufficiently close to more detailed simulations with gas hydrodynamics and radiative transfer. With the goal of constraining DM models, it is critical once again to understand the impact of modified initial conditions on the timing and duration of reionization and the high-\(z\) halo mass function, which are both currently relatively underexplored. Given the large number of astrophysical uncertainties, emulator and simulation-based inference methods are critical for efficiently sampling high-dimensional parameter spaces. Furthermore, due to the highly non-Gaussian nature of the 21-cm field, it will also be powerful to develop field-level inference methods that extract more information than the power spectrum alone.

\subsection{Observation} \label{sec:observation_future}

The coming decade will qualitatively shift our ability to test DM physics through astrophysical observations. This revolution will be driven by a combination of increasing scale in large astronomical survey projects and an increase in measurement precision. \NEW{We note two challenges that will accompany the increase in data volume across small-scale structure measurements. First, the \emph{volume of upcoming data} will present significant computational challenges. For example, many of the likelihood analyses discussed in this review involve expensive simulations, and scaling these pipelines to match the sample sizes anticipated from next-generation surveys will require investment in emulators, simulation-based inference methods, and simplified semi-analytic and empirical models. Second, \emph{reducing systematic biases} below the statistical error floor will become increasingly demanding. For example, the observational completeness functions that enter MW satellite analyses must be characterized at the $\lesssim 10\%$ level given the upcoming discovery of roughly hundreds of new dwarf galaxies. Analogously, the sensitivity functions for gravitational lensing surveys must be characterized. Overcoming these challenges will require close collaboration between project scientists and the broader community. With these challenges in mind, we now move to discuss several individual facilities.}

The Vera C.\ Rubin Observatory's LSST
\citep[][]{LSST:2008ijt} will fundamentally change how we view the dynamic night sky, and in the process, provide measurements of more astronomical objects than have been detected by all previous astronomical observations combined.
These observations will be delivered nearly in parallel with complementary wide-area, space-based imaging from ESA's Euclid mission \citep{Euclid:2024yrr}, NASA's Nancy Grace Roman Space Telescope \citep{Spergel:2015}, and the Chinese Space Station Telescope \citep[CSST;][]{Gong:2019yxt}, which will together provide exceptional spatial resolution and wider wavelength coverage into the infrared and ultraviolet.
Meanwhile, ground-based facilities are already in the midst of a spectroscopic revolution; building on the legacy of the highly successful SDSS \citep{SDSS:2000hjo}, BOSS \citep{BOSS:2012dmf}, and eBOSS \citep{eBOSS:2015jyv}, DESI \citep{DESI:2016fyo} has been blazing a trail for massively multiplexed ground-based spectrographs that are capable of collecting spectra for thousands of objects at a time, leading to surveys of tens of millions of stars and galaxies. A number of comparable wide-field massively multiplexed spectroscopic surveys are expected to deliver exciting results in the coming years including WEAVE \citep{Jin:2022ddg}, 4MOST \citep{2019Msngr.175....3D}, and PFS \citep{2014PASJ...66R...1T}.
In the near future, further advances in this domain are expected as the DESI instrument is reconfigured to target additional science cases \citep[including DM physics;][]{DESI:2022lza}, new devoted DM instruments such as Viaspec 
are constructed \citep{ViaCollaboration:2026}, and plans for future MOS instruments to deliver an order of magnitude increase in sensitivity are laid  \citep[e.g., MUST, Spec-S5, and WST;][]{Zhao:2024alp, Spec-S5:2025uom, WST:2024rai}.
At even higher redshifts, technology continues to advance toward performing wide area tomographic mapping from the epoch of reionization (LIM), the dark ages (21-cm), and back to the surface of last scattering (CMB).

\begin{figure*}[t!]
\centering
\includegraphics[width=0.95\textwidth]{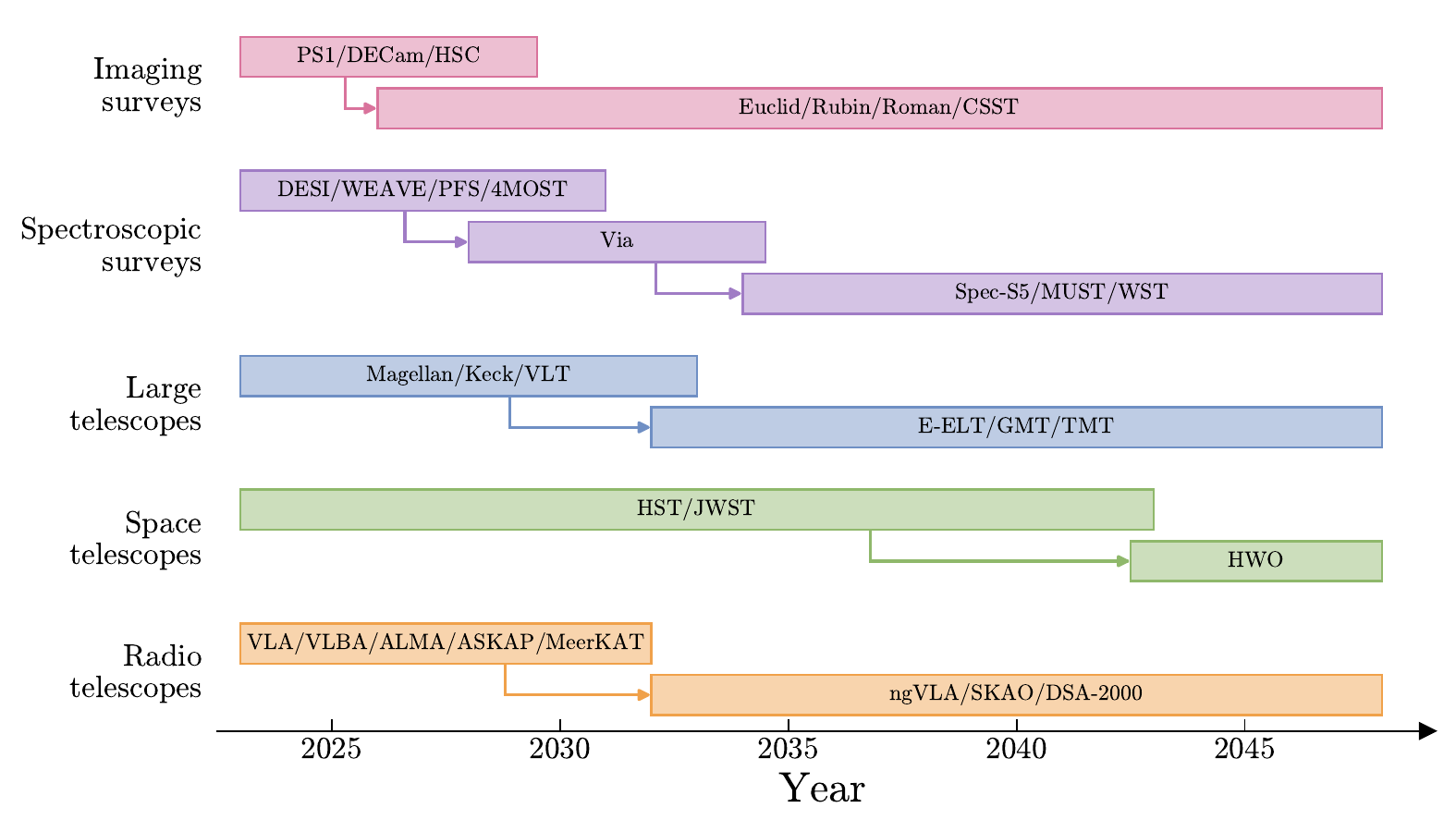}
    \caption{Schematic timeline for some of the major current and planned astronomical facilities to advance the study of dark matter on small cosmological scales. Facility timelines are difficult to predict, and the primary purpose of this figure is to serve as a visual guide for the discussion in Section~\ref{sec:observation_future}, which also describes the facilities in more detail.}
    \label{fig:obs_timeline}
\end{figure*}

These wide-area surveys are complemented by new instruments that are providing the unprecedented ability to perform precise measurements for individual objects. The Hubble Space Telescope (HST) has long been the exemplar in this domain, and is the foundation upon which our current understanding of small-scale structure from strong lensing resides. However, HST strong lensing small-scale structure measurements have quickly yielded to the power of the James Webb Space Telescope (JWST), which due to its combination of wavelength coverage and angular resolution is now allowing sensitivity to halos of comparable masses to those of the faintest dwarf satellites in the MW.
While these early observations are powerful, they have only targeted a few dozen lenses, and if HST is any indicator, the program to collect and analyze JWST lensing data will continue far into the next decade.
Furthermore, detailed single-object observations from the ground with VLBI have begun to revolutionize the sensitivity of gravitational imaging, leading to detections of anomalous features that can be interpreted as similar mass (but often anomalously dense) halos in nearby galaxies.
These measurements are an interesting example of the interplay between large surveys and targeted follow-up. The population of radio-loud lenses is currently limited by the lack of deep all-sky radio surveys, a situation that promises to change with the construction of SKA in the coming decade \citep{Weltman:2018zrl}; \NEW{some of these lenses could be followed up by an instrument like the Australian Long Baseline Array (LBA; \citealt{Petrov:2009})}. 
At the same time, the sensitivity of targeted radio observations promises to increase through the development of facilities such as ngVLA \citep{2018ASPC..517....3M}. 
Substantial advances in this domain are expected given the wide-area optical/NIR surveys from Rubin, Euclid, and Roman, and the deep high-resolution imaging and spectroscopy that is promised from the 30-m-class telescopes such as GMT \citep{2012SPIE.8444E..1HJ}, TMT \citep{2015RAA....15.1945S}, and E-ELT \citep{2007Msngr.127...11G}.
These ultra-sensitive, small field-of-view facilities are necessary to explore the most extreme satellites of the MW and high-resolution Lyman-$\alpha$ forest measurements.
While such detailed follow-up observations will necessarily target a sparse subset of the overall population, they will provide necessary calibration points needed for population-level modeling of survey data sets.

Below, we discuss some particularly exciting opportunities for improving measurements of small-scale structure in the low-, medium-, and high-redshift domains. To guide the discussion, Figure~\ref{fig:obs_timeline} provides a high-level overview of the timeline for current and upcoming observational facilities that will advance the small-scale structure probes focused on in this review. Note that we do not discuss microlensing measurements (e.g., \citealt{Croon:2020wpr}) or GW facilities (e.g., \citealt{Bertone:2019irm}) here.

\subsubsection{Near-field and Low Redshift ($z \lesssim 0.1$)}

\textbf{Dwarf Galaxies}: Over the last decade, DECam, HSC, and other ground-based surveys have led to a continuing phase of discovering fainter and more distant dwarf galaxies, and particularly MW satellites. The next major sensitivity gain will come from Rubin LSST, complemented heavily by the space-based angular resolution of Euclid, Roman, and/or CSST, which promises to be important for star/galaxy separation. Current surveys are already revealing a population of compact, ultra-faint systems, for which existing spectroscopic instruments on Keck, Magellan, and other large telescopes are unlikely to be able to measure resolved velocity dispersions. Upcoming multiplexed spectroscopic instruments such as Viaspec will push the systematic floor to $\sim$0.1 km/s and may enable velocity dispersion measurements in this regime~\citep{ViaCollaboration:2026}, while 30-m-class telescopes will provide the light-gathering power needed to obtain spectra of fainter member stars; however, resolving the sub-km/s dispersions expected for some of these systems will require instruments with sufficient spectral resolution, stability, and throughput.

The study of faint satellite populations is being extended to M31 and to more distant hosts, where space-based imaging can access resolved stellar populations in the ultra-faint regime; sensitivity estimates have been made for LUVOIR space telescope concepts~\citep{Luvoir}, and for the Habitable Worlds Observatory (HWO); for example, \citet{Doppel:2025} forecasted that HWO observations of Local Volume dwarf satellite populations could rule out thermal-relic WDM models with $m_{\mathrm{WDM}}\lesssim 13~\mathrm{keV}$. Nonetheless, the MW is expected to remain the most sensitive probe of the least luminous systems inhabiting halos near the galaxy formation threshold. In a future regime where resolved velocity dispersions become impossible to measure for the faintest objects around the MW and other nearby hosts, alternative indicators such as mean metallicity and metallicity dispersion will become more critical \citep{Cerny:2026pdr}. Furthermore, population-level inference will be essential for translating satellite abundances into constraints on DM subhalos without access to classification of individual systems. In parallel, DESI, PFS, Via, DESI-2, and Spec-S5 will kinematically probe the peripheries of faint dwarfs, providing critical information about their DM halos, the impact of tidal stripping, and potentially the role of early dwarf--dwarf mergers \citep[e.g.,][]{deason:2014}.

\textbf{Stellar Streams}: MW stellar streams are expected to be discovered in large numbers by Rubin LSST, assuming that star/galaxy contamination can be controlled (potentially through collaborative use of space-based imaging from Euclid and/or Roman). Photometric uniformity will be critical for measuring stream density variations, which mandates a program closer in character to LSS cosmology than to traditional Galactic surveys, and which can borrow tools and infrastructure from far-field cosmology. Spectroscopic follow-up is challenging for conventional and higher-resolution multi-object spectrographs but is well suited to massively multiplexed instruments. The trailblazing S$^5$ program on AAT/AAOmega has had an outsized impact, while DESI is now beginning to systematize these studies within the cosmology community. Gaia DR4 (late 2026) and the final DR5 (expected no sooner than 2030) will be valuable both for efficient target selection and for measuring the kinematic signatures of subhalo impacts in stellar proper motions. Distance measurements will remain a limiting factor in the outer halo, although spectrophotometric observations may improve this situation, and stream membership and bulk properties could potentially be used to set priors that reduce per-star distance uncertainties. The ultimate goal of obtaining full 6-D phase-space information on a star-by-star basis may be possible through some streams with Via, and eventually all streams with Spec-S5. 

\subsubsection{Intermediate Redshifts ($0.1 \lesssim z \lesssim 5$)}

\textbf{Halo Profiles in Small Galaxies}: Current constraints on the inner density profiles of low-mass galaxies are limited in part by the small (175-galaxy) SPARC sample from Spitzer \citep{Lelli:2016}. Efforts are ongoing to expand this sample by more than an order of magnitude through H\,{\sc I} observations from public telescope archives~\citep{Haubner:2024}, and SKA pathfinders (as well as SKA itself) promise an even larger sample in the future.
Optical/near-infrared IFUs on 30-meter-class telescopes (e.g., HARMONI on E-ELT, IRIS on TMT, GMTIFS on GMT) are expected to deliver stellar (and ionized-gas) kinematics at $\sim$10–50 pc resolution in Local Group and Local Volume dwarfs. 
However, the high oversubscription rates expected on these facilities will make it difficult to assemble a very large sample of galaxies.
Weak lensing offers a complementary observational route to measure DM halo profiles in this mass regime, with a natural division of labor across upcoming facilities: imaging surveys (e.g., DECam, Euclid, Rubin, Roman) provide target identification, massively multiplexed spectroscopic surveys (e.g., DESI, PFS, Spec-S5) will deliver redshifts and spectroscopic confirmation, and the lensing measurements themselves come from the same deep imaging. Some initial demonstrations of this technique have been performed with existing data sets \citep{Treiber:2025, To:2025}. Future programs have the potential to extend mass-profile measurements down to masses of $\sim 10^9\,M_\odot$ \NEW{for field halos}, which could provide insight into the cusp--core and diversity of rotation curve problems. However, this is still in the regime where baryonic feedback is expected to modify inner halo profiles, so disentangling DM physics from baryonic feedback will remain a central challenge.

\textbf{Strong Lensing}: Constraints on DM substructure from both flux ratio anomalies in lensed quasars and gravitational imaging of extended arcs will follow a discovery, confirmation, characterization pipeline, though the facilities and techniques used in each stage differ. 
For flux ratio anomalies, Rubin, Euclid, and Roman are projected to grow the galaxy-galaxy lens sample to $\sim 10^5$ and to expand the lensed quasar population by an order of magnitude over the current inventory \citep{Oguri:2010ns, Collett:2015roa}. Spectroscopic discovery and characterization (i.e., source/lens redshifts) will come from massively multiplexed surveys, with the DESI strong-lensing program already demonstrating this capability \citep{2026ApJS..282...41H, 2026ApJ...999..198K} with future facilities on the horizon.
Contamination from stellar microlensing of compact source regions has driven the field toward extended emission regions in the mid-IR (warm dust) and optical/NIR narrow lines, accessible to JWST MIRI/NIRSpec and eventually ELT-class IFUs. Together, the larger samples and improved per-system measurements should move the field beyond current population-averaged constraints toward halo mass function sensitivity across the $\sim$ $10^7$--$10^9\,M_\odot$ regime.

For gravitational imaging, the most sensitive single-perturber detections have come from VLBI observations of lensed AGN with spatially extended radio emission, where milliarcsecond-scale resolution enables sensitivity to individual ${\sim} 10^{6}$--$10^{8}\,M_\odot$ perturbers \citep{Powell:2025rmj}. Expanding the lens target set from the CLASS-era sample requires wide-field radio surveys (e.g., VLASS, LOFAR, MeerKAT, DSA-2000, and ultimately SKA) to enlarge the population for VLBI follow-up. 
Global VLBI campaigns, augmented by the ngVLA long‑baseline array, will provide expanded access to sub‑milliarcsecond resolution. Eventually, 30-meter-class telescopes with adaptive optics will scale high-resolution gravitational imaging to the much larger optical/NIR-bright lens population uncovered by Rubin, Euclid, and Roman, where radio-loud counterparts are rare. While the optical/near-infrared measurements are not expected to compete with VLBI in angular resolution, they will expand the sample and provide multi-wavelength information.
In both cases, the relevant signal includes both subhalos of the main deflector and the line-of-sight halo population. \NEW{In addition, the ongoing ALMA wideband sensitivity upgrade~\citep{Gonzalez:2022zaf} will improve strong lensing studies in radio bands akin to \citet{Hezaveh:2016ltk}.}

\subsubsection{High Redshift ($z \gtrsim 5$)}

\textbf{Lyman-$\alpha$ Forest}: Constraints on the small-scale matter power spectrum from the Ly-$\alpha$ forest at $z < 5.4$ currently set strong bounds on suppression of the matter power spectrum from the linear and quasilinear regime \citep[e.g.,][]{Viel:2013fqw, Irsic:2017ixq, Rogers:2020ltq, Villasenor:2022aiy}. Observationally, these measurements are limited by the sample of $z \gtrsim 5$ quasars that are bright enough for high-resolution ($R \gtrsim 40{,}000$) echelle spectroscopy with current 8--10 m facilities (Keck/HIRES, VLT/UVES and ESPRESSO, Magellan/MIKE). Progress over the coming decade will come from two complementary directions. First, the bright quasar sample will grow substantially as Rubin, Euclid, and Roman identify rare luminous quasars at $z \gtrsim 6$, while NIR-selected candidates from Euclid and Roman will be particularly important for $z \gtrsim 7$ (where Lyman dropout selection from optical surveys becomes inefficient). Second, the flux limit of high-resolution spectroscopy will be extended by 30-meter-class facilities and their planned high-resolution instruments (e.g., ANDES on the ELT, G-CLEF on GMT, and MODHIS on TMT) extending the usable sample to the more numerous, intrinsically fainter quasars and enabling higher S/N on existing bright targets. The combination should substantially increase the number of independent sightlines at $z \sim 5$ where the small-scale flux power spectrum carries the strongest leverage on the DM free-streaming scale.

The dominant systematic for DM inference from high-resolution Lyman-$\alpha$ forest measurements is the thermal and ionization state of the IGM near reionization due to significant degeneracies between suppression of the matter power spectrum and a warmer/more pressure-supported IGM. Thus, future constraints depend not only on an expanded observational sample, but also on parallel progress in constraining the IGM thermal history, possibly through joint analyses with reionization probes (CMB optical depth, 21-cm, Lyman-$\alpha$ damping wings in high-$z$ quasar and galaxy spectra).

\textbf{High-$z$ Galaxies}: The faint end of the rest-frame UV luminosity function (UV LF) at $z \gtrsim 6$ is a powerful probe of the low-mass halos that host the faintest UV-emitting galaxies. 
Constraints from HST and JWST already disfavor thermal-relic WDM masses below a few keV under standard assumptions about the galaxy--halo connection (e.g., \citealt{Menci:2017nsr, Corasaniti:2016epp, Rudakovskyi:2021jyf, Liu:2024edl, Sabti:2021unj}). 
In the coming decade, JWST will extend this along two complementary avenues: deeper and more uniform NIRCam imaging that will push the LF to higher redshifts ($z \gtrsim 12$) and fainter magnitudes ($M_{\rm UV}$), and surveys of galaxy cluster lenses (UNCOVER, BEACON, GLIMPSE, and successors) that probe intrinsically fainter sources through gravitational magnification, at the cost of smaller volume and magnification-dependent systematics. 
In parallel, Roman's High-Latitude Wide-Area Survey will provide the increased area to control cosmic variance and to characterize the bright end and characteristic magnitude across $z \sim$ 7--10, complementing JWST's depth with statistical power. 
The contribution of 30-meter-class telescopes will be primarily spectroscopic: HARMONI on the ELT, IRIS on TMT, and instruments such as MOSAIC and MANIFEST will deliver the spectra of faint, JWST-selected galaxies that are needed for secure redshifts, stellar-population properties, and emission-line diagnostics that constrain the galaxy--halo connection mediating between the UVLF and the underlying halo mass function. The dominant systematic for DM inference is expected to remain the degeneracy between the suppression of low-mass halos by DM physics and the suppression of star formation in the galaxies that occupy those halos. These systematics will be reduced as we learn more about feedback, photoionization, and reionization in $z \gtrsim 10$ galaxies revealed by JWST.

\textbf{21-cm Cosmology}: The redshifted 21-cm signal opens a window on the matter power spectrum at $z \sim 6$--30, which is otherwise inaccessible. Sensitivity to the small-scale power spectrum scales favorably with redshift because the relevant halo masses are smaller and the IGM is more nearly neutral. 
We focus on two distinct observational avenues that promise sensitivity to DM physics. 
First, statistical measurements of the 21-cm power spectrum from cosmic dawn and the Epoch of Reionization (HERA, LOFAR, MWA, and ultimately SKA-low) constrain the abundance of the low-mass halos hosting the first galaxies. Existing HERA upper limits already place model-dependent bounds on power-spectrum suppression through their impact on early heating and ionization \citep{HERA:2021noe, HERA:2022wmy}.
SKA-LOW is expected to detect the 21-cm power spectrum across cosmic dawn, with corresponding leverage on the matter power spectrum at $k \sim 10$--$100\,h\,\text{Mpc}^{-1}$ under standard assumptions about star formation in low-mass halos \citep{Munoz:2019hjh}.
The response to the reported EDGES absorption feature at $z \sim 17$ \citep{Bowman:2018yin} demonstrates the potential power of these measurements; however, the SARAS-3 non-detection has challenged these results~\citep{Singh:2021mxo}. 

Second, the 21-cm forest absorption against bright high-$z$ radio sources by neutral hydrogen in the IGM and in intervening minihalos offers a fundamentally different probe, in principle delivering line-of-sight constraints on the small-scale matter power spectrum at $z \gtrsim 6$. The technique is currently sample-limited rather than sensitivity-limited: it requires bright radio-loud AGN at $z \gtrsim 6$, of which only a handful are known, and detection of the absorption features at $\sim$kHz resolution is feasible for a handful of the brightest targets prior to SKA-low. Anticipated wide-area radio surveys with LOFAR, MeerKAT, ASKAP, and ultimately SKA itself are expected to expand the high-$z$ radio-loud population substantially, but progress will depend equally on foreground subtraction and instrumental calibration at a level that has not yet been demonstrated. Uncertainties in the poorly-constrained astrophysics of early star formation, X-ray heating, and reionization are the dominant systematic for 21-cm probes of DM physics. This offers a powerful opportunity to combine 21-cm observations with observations of Lyman-$\alpha$ damping, the UV LF, and the CMB optical depth to reduce these systematics and jointly probe DM physics at high redshift.

\subsection{Probe Combination}
\label{sec:combination_future}

As discussed in Section~\ref{sec:combination_constraints}, combining small-scale structure probes provides powerful constraints on DM physics because different observables access complementary aspects of the underlying matter distribution and have distinct systematics. Recent analyses have demonstrated the potential of this approach for constraining WDM (e.g., \citealt{Enzi:2020ieg,Nadler:2021dft}), showing that degeneracies in individual probes can be broken by incorporating additional observables that depend differently on the same underlying physical parameters.

More generally, joint inference approaches can be understood within a unified statistical framework in which all observables are modeled as functions of a shared set of latent variables (e.g., the SHMF). In this picture, the joint likelihood takes the schematic form
\begin{equation}
    \mathcal{L}(\vec{D}|\vec{\theta}) = \prod_{\mathrm{probe\ } i}\mathcal{L}_i(D_i|\vec{\theta}_{\mathrm{DM}},\vec{\theta}_{\mathrm{nuisance,}i}),
\end{equation}
where $\vec{\theta}_{\mathrm{DM}}$ denotes the  DM model parameters of interest and $\vec{\theta}_{\mathrm{nuisance,}i}$ represents the (astrophysical) nuisance parameters for probe $i$, which has an associated data vector $D_i$. The challenge is to construct accurate and computationally efficient models to marginalize over $\vec{\theta}_{\mathrm{nuisance,}i}$, while consistently accounting for shared uncertainties and correlations across probes.

The recent advances in theory and observation highlighted above are making this program increasingly tractable. In particular, emulators trained on suites of simulations provide a means to rapidly evaluate predictions across high-dimensional parameter spaces, enabling joint analyses that would otherwise be computationally prohibitive. Looking ahead, combining multiple small-scale structure probes within a unified inference framework will be essential for fully exploiting upcoming data. By leveraging the complementary information encoded in different observables, such analyses have the potential to significantly improve DM constraints and robustly disentangle new physics from astrophysical systematics.

\section{Summary and Conclusions}
\label{sec:conclusion}

\subsection{Where Do We Stand?}
\label{sec:current_status}

\NEW{Over the past two decades, measurements of small-scale structure have evolved from potential challenges to the CDM paradigm into a powerful probe of the fundamental nature of DM. For example, the combination of increasingly complete observations of faint galaxies and stellar streams, advances in strong lensing, and the development of flexible, empirically-grounded modeling frameworks has enabled robust inference that connects DM microphysics to data at the small-scale frontier. As a result, these probes now provide among the most stringent constraints on many DM properties, complementing LSS analyses and terrestrial experiments. At the same time, our understanding of small-scale structure remains incomplete, and new tensions have emerged on highly nonlinear scales, many of which are related to the inner densities of low-mass subhalos.}

\NEW{Based on our survey of the literature, the main DM scenarios we have reviewed stand as follows:}

\begin{enumerate}
    \item \NEW{\textbf{Warm dark matter}: Dwarf galaxy abundances, strong lensing flux ratios, and the Lyman-$\alpha$ forest now robustly constrain thermal-relic WDM based on its \emph{ab initio} $P(k)$ suppression, yielding $m_{\rm WDM} \gtrsim 6~{\rm keV}$ from multiple independent groups. The most constraining probe combination analyses push this limit to $\gtrsim 10~{\rm keV}$. Variations in WDM bounds \emph{within} each probe are mostly driven by different astrophysical assumptions (e.g., the galaxy--halo connection for dwarf galaxies or the IGM thermal history for the Lyman-$\alpha$ forest), rather than WDM modeling uncertainties or discrepant data. The field is converging on forward modeling frameworks that marginalize over these uncertainties, which should lead to convergent limits within each probe. Variations in WDM constraints \emph{across} small-scale probes mainly reflect the distinct scales and redshifts that they measure. Upcoming data for both established and emerging probes should substantially improve WDM sensitivity and reach well into the $m_{\mathrm{WDM}}\gtrsim 10$ keV range.}
    \item \NEW{\textbf{Fuzzy dark matter}: Probes of \emph{ab initio} $P(k)$ suppression due to FDM, and particularly its impact on quasilinear structure measured by the Lyman-$\alpha$ forest and the (sub)halo mass function as probed by dwarf galaxies and strong lensing, now consistently yield lower limits of $m_{\rm FDM} \gtrsim 10^{-20}~{\rm eV}$. Meanwhile, the \emph{in situ} effects of FDM on (sub)halo density profiles, which were originally a key motivation for FDM, generally require masses that are inconsistent with \emph{ab initio} bounds without additional model building; in addition, temporal fluctuations of the FDM field heat stars in dwarf galaxies, which leads to bounds of $m_{\mathrm{FDM}}\gtrsim 10^{-19}~\mathrm{eV}$. Thus, the simplest single-field FDM models face substantial tension from small-scale structure measurements. The development of fractional and/or self-interacting FDM models, along with production mechanisms beyond the standard misalignment scenario, is underway to meet this challenge. Unlike WDM, substantial modeling uncertainties remain for FDM, mainly due to approximations introduced to avoid the computational expense of full Schrödinger-Poisson simulations. Addressing these uncertainties will be important to move from conservative bounds to searches for unique signatures of FDM.}
    \item \NEW{\textbf{Interacting dark matter}: Interactions between DM and Standard Model particles have been tightly constrained by dwarf galaxies and the Lyman-$\alpha$ forest based on their \emph{ab initio} $P(k)$ suppression. These limits complement direct detection searches for the same physics by reaching both lower masses and higher cross sections than many terrestrial experiments probe. IDM models generally produce $P(k)$ features beyond a WDM-like cutoff, such as dark acoustic oscillations, which measurements of high-redshift structure including high-$z$ galaxies can potentially distinguish. The \emph{in situ} effects of DM--SM interactions are less explored and have only been used to limit IDM in a handful of analyses to date and will benefit from the development of simulation codes that reliably model such interactions.}
    \item \NEW{\textbf{Self-interacting dark matter}: Over roughly the last decade, SIDM has regained interest as a rich DM scenario that is pushing the interpretation of small-scale structure data and modeling into new territory. This is largely due to the fact that SIDM can produce both cored and core-collapsed halos, owing to the underlying distribution of halo formation histories, which set initial concentrations and thus core-collapse timescales. The mixture of these systems that arises in velocity-dependent cross sections with $\sigma/m\gtrsim 10~\mathrm{cm^2~g}^{-1}$ at $v\lesssim 50~\mathrm{km\ s}^{-1}$ is consistent with current measurements of dwarf galaxy dynamics, while lower-amplitude cross sections that mainly produce cored halos are tightly constrained, at the level of $\sigma/m\lesssim 0.1~\mathrm{cm^2~g^{-1}}$ at $v\sim 1000~\mathrm{km\ s}^{-1}$ and $\sigma/m\lesssim 1~\mathrm{cm^2~g^{-1}}$ at $v\sim 10~\mathrm{km\ s}^{-1}$. Recent detections of anomalously dense perturbers in strong lensing and stellar stream data---along with the growing population of high-redshift supermassive black holes and little red dots---have further motivated high-amplitude, velocity-dependent SIDM cross sections, and models have been proposed that simultaneously fit these observables. More accurate simulations and semi-analytic models are needed to robustly predict SIDM (sub)halo populations across a range of scales, environments, and baryonic feedback models. These tools will enable SIDM analyses to move from individual systems to statistical tests using larger samples of dwarf galaxies, stellar streams, and gravitational lenses delivered by upcoming data.}
    \item \NEW{\textbf{Decaying dark matter:} In the case of two-body DM decays, small-scale structure data and particularly dwarf galaxy abundances and dynamics imply that the DDM lifetime must be greater than the age of the universe across a range of mass splitting values. More complex models (e.g., specific decay channels or fractional scenarios) may evade these constraints. An important aspect of \emph{in situ} DDM effects is that their impact on structure generally grows with time, which is distinct from \emph{ab initio} WDM-like models and highlights the importance of combining small-scale probes across a wide redshift range.}
\end{enumerate}

\subsection{Where Are We Headed?}

\NEW{The search for new DM physics using small-scale structure is approaching an exciting confluence of powerful new measurements, robust simulation and modeling techniques, and theoretical development of DM scenarios that produce distinct astrophysical signatures. In this landscape, we expect constraints on all of the DM scenarios discussed above to substantially improve, shrinking viable DM parameter space and providing critical guidance to future cosmological and terrestrial experiments. Realizing this potential will require both observational and theoretical advances. On the observational side, continued investment in facilities that increase the statistical power of established probes and enable measurements of structure from emerging probes will be crucial. On the theoretical side, progress will depend on accurately modeling nonlinear structure across DM scenarios, robustly marginalizing over the impact of baryonic physics on small-scale structure, and combining data from multiple probes in a unified framework.}

\NEW{Moving beyond constraints, analyses of next-generation small-scale structure data have the potential to discover a departure from the CDM paradigm, which would constitute a major discovery and step toward identifying the fundamental nature of DM. At present, we are seeing tantalizing hints that a number of small-scale structure probes in the late-time universe prefer more compact low-mass substructure than predicted by CDM. These measurements include the inner densities of subhalos in gravitational imaging, flux ratio, and cluster strong lensing data, the properties of stellar stream perturbers, and the dynamical properties of dwarf satellite galaxies. Analyses of upcoming data that robustly characterize observational selection effects and combine information across these probes will be crucial to confirm or refute this potential preference for dense low-mass subhalos.}

\NEW{A key milestone in the coming years will be the detection (or robust exclusion) of DM halos below the galaxy formation threshold. Establishing the existence of completely dark halos would open a new observational window into structure formation and enable the most incisive small-scale structure tests of DM physics to date.  On the other hand, an absence of such systems would constitute strong evidence against CDM and point to new DM physics beyond gravity. More broadly, measuring the statistics and properties of dark halos will establish whether DM physics beyond gravity has any observable impact on galaxy formation. Either outcome will mark an important transition in our understanding of DM.}

\NEW{We predict that small-scale structure will play a central role in the broader search for DM physics. In the presence of a terrestrial detection, cosmological confirmation will be essential to show that the detected particle(s) are the cosmological DM, and small-scale structure provides a powerful avenue to achieve this goal. Conversely, robust evidence for departures from collisionless CDM inferred from nonlinear structure would offer critical guidance for direct detection and collider experiments by narrowing the viable particle DM parameter space. Thus, small-scale structure bridges cosmology, astrophysics, and particle physics, creating interdisciplinary opportunities at the interface of astrophysical data and particle theory. Together, these developments point toward a future in which small-scale structure enables precision tests---and perhaps discovery---of fundamental DM physics.}


\section*{Acknowledgments}

We are grateful to Kensuke Akita, Kim Boddy, Keith Bechtol, Adrienne Erickcek, Vera Gluscevic, Doddy Marsh, Michael Martinez, Julian Muñoz, Masamune Oguri, Sikder Sudipta, and Haibo Yu for comments on the manuscript, and we thank Sandip Roy for helpful discussions related to this work. KKR is supported by an Ernest Rutherford Fellowship from the UKRI Science and Technology Facilities Council (grant no. ST/Z510191/1).

\NEW{Claude Opus 5.5 was used to proofread the manuscript and unify plotting scripts during the review stage; all output was verified by the authors.}

\appendix

\section{Summary of Thermal-relic WDM Constraints}
\label{sec:wdm_summary}

Table~\ref{tab:wdm} lists the thermal-relic WDM constraints shown in Figure~\ref{fig:wdm_timeline}.

\begin{table*}[t!]
\caption{Thermal-relic WDM mass lower limits shown in Figure~\ref{fig:wdm_timeline}. The first column shows the probe type, the second column lists the reference, and the third and fourth columns list the $m_{\mathrm{WDM}}$ lower limit and statistical interpretation. When two $m_{\mathrm{WDM}}$ limits are shown, the first (second) value indicates the more conservative (less conservative) constraint. Blank entries in the final column indicate results that are not explicitly derived from likelihood analyses. \NEW{Studies that use the \citet{Vogel:2022odl} transfer function are marked by a $\dagger$}. All constraints listed here assume $f_{\mathrm{WDM}}=1$. We do not include limits based on the EDGES global 21-cm signal~\citep{Schneider:2018xba,Chatterjee:2019jts}, the status of which is uncertain (Section~\ref{sec:21cm}).}
\begin{ruledtabular}
\begin{tabular}{llcc}
Probe & Reference & $m_{\rm WDM}$ lower limit [keV] & Interpretation \\
\hline
\multicolumn{4}{l}{\textit{Lyman-$\alpha$ forest}} \\
  & \citet{Narayanan:2000tp} & $0.75$ & $3\sigma$ \\
  & \citet{Viel:2005qj} & $0.55$ & $2\sigma$ \\
  & \citet{Seljak:2006qw} & $2.5$ & $95\%$ C.L. \\
  & \citet{Viel:2006kd} & $2.0$ & $95\%$ confidence \\
  & \citet{Viel:2007mv} & $1.2$, $4.0$ & $2\sigma$ \\
  & \citet{Viel:2013fqw} & $3.3$ & $2\sigma$ \\
  & \citet{Baur:2015jsy} & $2.96$, $4.09$ & $95\%$ C.L. \\
  & \citet{Irsic:2017ixq} & $3.5$, $5.3$ & $2\sigma$ \\
  & \citet{Yeche:2017upn} & $4.65$ & $95\%$ C.L. \\
  & \citet{Murgia:2018now} & $2.7$, $3.6$ & $2\sigma$ \\
  & \citet{Palanque-Delabrouille:2019iyz} & $5.3$ & $95\%$ C.L. \\
  & \citet{Garzilli:2019qki} & $1.9$ & $95\%$ C.L. \\
  & \citet{Villasenor:2022aiy} & $3.1$ & $95\%$ C.L. \\
  & \citet{Irsic:2023equ} & $4.1$, $5.7$ & $95\%$ C.L. \\
\multicolumn{4}{l}{\textit{Dwarf galaxies (abundances)}} \\
  & \citet{Maccio:2009isa} & $1.0$ & -- \\
  & \citet{Polisensky:2010rw} & $2.3$ & $2\sigma$ \\
  & \citet{Lovell:2013ola} & $1.5$ & -- \\
  & \citet{Kennedy:2013uta} & $2.0$, $3.3$ & $95\%$ confidence \\
  & \citet{Jethwa161207834} & $2.9$ & $95\%$ confidence \\
  & \citet{Nadler:2019zrb} & $3.3$ & $95\%$ confidence \\
  & \citet{DES:2020fxi} & $6.5$ & $95\%$ confidence \\
  & \citet{Newton:2020cog} & $2.0$, $4.0$ & $95\%$ confidence \\
  & \citet{Dekker:2021scf} & $3.6$, $5.1$ & $95\%$ C.L. \\
  & \citet{Nadler:2025fcv}$^{\dagger}$ & $5.9$ & $95\%$ confidence \\
  & \citet{Liu:2025vhk}$^{\dagger}$ & $6.2$ & $95\%$ confidence \\
\multicolumn{4}{l}{\textit{Dwarf galaxies (dynamics)}} \\
  & \citet{Dalcanton:2000hn} & $0.7$ & -- \\
  & \citet{Boyarsky:2008ju} & $0.5$ & -- \\
  & \citet{Alvey:2020xsk} & $0.4$ & $95\%$ C.L. \\
  & \citet{Kim:2021zzw} & $6.0$ & -- \\
  & \citet{Bezrukov:2025ttd} & $0.3$, $0.5$ & $95\%$ C.L. \\
  & \citet{Delos:2025ees}$^{\dagger}$ & $5.8$ & $95\%$ confidence \\
\multicolumn{4}{l}{\textit{Strong lensing (gravitational imaging)}} \\
  & \citet{Vegetti:2018dly} & $0.3$ & $2\sigma$ \\
  & \citet{Ritondale:2018cvp} & $0.3$ & $2\sigma$ \\
\multicolumn{4}{l}{\textit{Strong lensing (flux ratios)}} \\
  & \citet{Birrer:2017rpp} & $2.0$ & $2\sigma$ \\
  & \citet{Hsueh:2019ynk} & $5.58$ & $95\%$ C.L. \\
  & \citet{Gilman:2019nap} & $5.2$ & $95\%$ C.L. \\
  & \citet{Keeley:2024brx} & $6.1$ & 10:1 odds ratio \\
  & \citet{Keeley:2025oig}$^{\dagger}$ & $5.6$, $6.4$ & 10:1 odds ratio \\
  & \citet{Gilman:2025fhy}$^{\dagger}$ & $6.5$, $7.4$ & 10:1 Bayes factor \\
\multicolumn{4}{l}{\textit{High-$z$ galaxies}} \\
  & \citet{Barkana:2001gr} & $0.4$ & -- \\
  & \citet{Pacucci:2013jfa} & $0.9$ & $99.9\%$ confidence \\
  & \citet{Schultz:2014eia} & $1.3$ & $2.2\sigma$ \\
  & \citet{2016ApJ...818...90M} & $1.5$, $1.8$ & -- \\
  & \citet{Corasaniti:2016epp} & $1.5$ & $2\sigma$ \\
  & \citet{Rudakovskyi:2021jyf} & $2.0$ & $95\%$ confidence \\
  & \citet{Maio:2022lzg} & $2.0$ & -- \\
  & \citet{Liu:2024edl} & $3.2$ & $95\%$ C.L. \\
  & \citet{Urrutia:2025fvp} & $1.5$ & $95\%$ C.L. \\
  & \citet{Ellis:2025dpw} & $7.2$ & $95\%$ C.L. \\
\multicolumn{4}{l}{\textit{Stellar streams}} \\
  & \citet{Banik:2019cza} & $3.6$, $4.6$ & $95\%$ confidence \\
  & \citet{Banik:2019smi} & $3.6$, $6.2$ & $95\%$ confidence \\
\multicolumn{4}{l}{\textit{Combinations}} \\
  & \citet{Enzi:2020ieg} & $6.0$ & $95\%$ C.L. \\
  & \citet{Nadler:2021dft} & $9.7$ & $95\%$ confidence
\label{tab:wdm}
\end{tabular}
\end{ruledtabular}
\end{table*}

\bibliography{references}

\end{document}